\documentclass[twocolumn,english,superscriptaddress,amsmath,floatfix,longbibliography,floats,prb,aps,10pt]{revtex4-2}

\usepackage[table]{xcolor}
\usepackage{array,ragged2e}
\usepackage{grffile}
\usepackage{multirow}
\usepackage{algpseudocode}
\usepackage{epstopdf}
\usepackage{booktabs}
\makeatletter

\usepackage{amssymb}
\usepackage{amsfonts}
\usepackage{graphicx}
\usepackage{xcolor}
\usepackage{dcolumn}
\usepackage{bm}
\usepackage{dsfont}
\usepackage{appendix}
\usepackage{placeins}
\usepackage{bbold}
\usepackage{comment}
\usepackage{hyperref}
\usepackage{ifthen}

\newboolean{raster}

\begin{document}

\setboolean{raster}{false}
\ifthenelse{\boolean{raster}}{\graphicspath{{./Figures/}}}{\graphicspath{{./Figures/}}}

\title{Detuning the Floquet anomalous chiral spin liquid}

\author{Matthieu Mambrini}
\affiliation{Universit\'e Toulouse, CNRS, Laboratoire de Physique Th\'eorique, Toulouse, France}

\author{Nathan Goldman}
\affiliation{Laboratoire Kastler Brossel, Coll\`ege de France, CNRS, ENS-Universit\'e PSL, Sorbonne Universit\'e, 11 Place Marcelin Berthelot, 75005 Paris, France}

\author{Didier Poilblanc}
\affiliation{French-American Center for Theoretical Science,
CNRS, KITP, Santa Barbara, CA 93106-4030, USA}

\date{\today}


\begin{abstract}

At high-frequency a periodically-driven quantum spin-1/2 system can emulate a chiral spin liquid (CSL) described by an effective static local chiral Hamiltonian. In contrast, at low-frequency these settings realize "Swap" models exhibiting {\it anomalous} CSL phases, in which one-way spin transport occurs at the edge although the bulk time-evolution operator over one period is trivial.
In this work we explicitly construct a family of Floquet quantum spin-1/2 models on the square lattice implementing Swap models to investigate the stability of the anomalous CSL under frequency detuning and 
the transition to the high-frequency regime. We used the average-energy spectrum on finite-size torus and cylinders to unfold the Floquet quasi-energy spectrum over the whole frequency range and obtain the geometrical Berry phases. This enabled us to identify three regimes upon increasing detuning: i) a finite-size regime (with no folding of the Floquet spectrum), ii) an intermediate (narrow) regime with folding and very few resonances and iii) a regime with an increased density of resonances suggesting heating. At small detuning, edge modes are revealed by spectroscopic tools and from the Streda response of the system giving access to the anomalous winding number. 
The analysis of all the data suggests that the anomalous CSL is not continuously connected to the high-frequency CSL. We also discuss the possible occurrence of a long-lived prethermal anomalous CSL. 

\end{abstract} 

\maketitle

\section{Introduction}

In non-interacting lattice systems, applying a time-periodic drive can induce topological band structures. In the high-frequency regime, the drive can open topologically non-trivial gaps that support chiral edge states. A prominent example is graphene illuminated with circularly polarized light~\cite{Karch2010}, where such effects have been theoretically predicted and explored. As first demonstrated by Kitagawa et al.~\cite{Kitagawa2010}, lowering the drive frequency introduces resonances that can generate additional topological gaps and chiral edge modes. This low-frequency regime is typically referred to as ``anomalous Floquet topology'', and is characterized by a breakdown in the conventional bulk-edge correspondence: the Chern numbers of Floquet bands no longer fully account for the presence of edge states~\cite{Rudner2013}.

An intriguing avenue of exploration concerns the transposition of Floquet topological concepts to genuine many-body systems, such as periodically driven spin lattices. Within this context, two distinct dynamical regimes have been identified: (i) High-frequency regime: Here, the driven spin system can emulate a chiral spin liquid (CSL) supporting chiral edge modes~\cite{Chen2016}. This regime mirrors the high-frequency limit studied by Kitagawa~\cite{Kitagawa2010}, and finds a concrete realization in the driven Kitaev honeycomb model~\cite{Fulga2019,Sun2023,Kalinowski2023}. 
(ii) Low-frequency regime: In contrast, the drive effectively implements simple ``swap'' gates~\cite{Po2016,Zhang2021}. This latter scenario parallels Rudner’s anomalous Floquet regime~\cite{Rudner2013}, in which the bulk time-evolution operator over one period is trivial (i.e. the Floquet Hamiltonian vanishes), yet robust chiral edge modes exist.
These regimes highlight the rich landscape of driven spin systems and open new perspectives on realizing topological phases in interacting quantum matter.

Another outstanding issue in the field concerns whether one can interpolate between the high-frequency and low-frequency regimes of driven spin systems within a unified framework, and, crucially, elucidate the nature of the transition between them. 

In this work, in order to address the two fundamental issues described above, we have adopted the following outline. In Section~\ref{sec:tuned} we first review properties of Swap models~\cite{Po2016,Zhang2021} on a two-dimensional (2D) square lattice. We explicitly derive an infinite discrete set of periodic drives of XXZ Hamiltonians realizing Swap models for specific tuned values of the drive frequency. On cylinder geometries, breaking of discrete-time invariance in the stroboscopic time evolution of edge states is outlined, showing similarities with Time Crystals~\cite{Khemani2019}. The investigation of the effects of detuning the frequency is carried out in Section~\ref{sec:detuning}. A careful analysis of Floquet quasi-energy spectra and average-energy spectra on torus and cylinder geometries reveals different regimes upon increasing detuning: (i) a finite-size regime with unfolded Floquet spectrum and no energy absorption; 
(ii) an intermediate regime with folded Floquet spectrum but only few resonances which may signal the occurrence of a prethermal regime~\cite{Abanin2015,Luitz2020}, even in the thermodynamic limit;
(iii) a regime where the proliferation of resonances suggests heating. In such an analysis, the average-energy concept~\cite{Schindler2024}, by providing a mathematical (but still physical) tool to unwind the Floquet spectrum, and the associated geometrical Berry phases, have been very helpful to characterize the various regimes. When a clear Floquet low-energy ``ground-state'' could be identified, signature of edge modes and anomalous spectral flow has been observed. 
In Section~\ref{sec:interpol} of this work we introduce a controlled detuning of the Swap model, which smoothly connects to a dynamical chiral spin liquid phase in the high-frequency limit~\cite{Mambrini2024,Poilblanc2024}. Remarkably, the transition between these regimes is found to be ``chaotic'', marked by the breakdown of topological protection and the disappearance of chiral edge modes due to the proliferation of resonances. This behavior stands in stark contrast to the non-interacting lattice case~\cite{Kitagawa2010}, where the edge states persist across the frequency domain under well-defined conditions.

\section{Tuned Swap models}
\label{sec:tuned}

{\it Drive Hamiltonians --} Let us first introduce Swap models which, for precisely tuned drive frequencies, implement an exact swap circuit of spins-1/2 (or hard-core bosons) on the square lattice~\cite{Po2016,Zhang2021}. 
Here the $T$-periodic sequence involves four discrete steps in which $1/4$ of the bond interactions are successively turned on and off for a time $T/4$. 
In a system with open boundaries, particle/spin numbers remain stable in the bulk while one-way transport of quantum information occurs at the edge(s), characterizing an {\it anomalous CSL}. 

\begin{figure}
	\centering
	\includegraphics[width=\columnwidth]{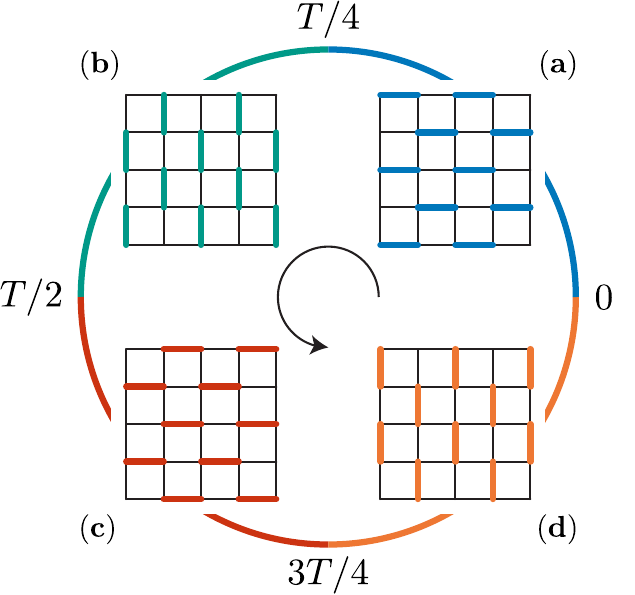}
     \caption{Schematic representation of the drive Hamiltonian $H_{\rm drive}(t)$ involving four XXZ nearest-neighbor couplings $H_a$, $H_b$, $H_c$ and $H_d$ acting successively during a time $T/4$ on the 4 sets of bonds labelled by $a$, $b$, $c$ and $d$.}
	\label{fig:model}
\end{figure}

We start by constructing an infinite set of XXZ drive Hamiltonians realizing, over a complete period, the same swap circuit on the square lattice but characterized by different micromotions. These Swap models of arbitrary low frequencies $\omega_0/(2p+1)$, where $p$ is integer, can be viewed as descendants of a parent Swap model at maximum frequency $\omega_0$. 

We start with a XXZ spin-1/2 model (with U(1) spin-rotation symmetry) on a square lattice
\begin{equation}
H = J \sum_{\langle i, j \rangle} S_i^x S_j^x +S_i^y S_j^y + \Delta S_i^z S_j^z 
\label{eq:XXZ}
\end{equation}
which is decomposed in four parts $H_a, H_b, H_c$ and $H_d$ acting on four disconnected sets of bonds, according to the setup described in figure~\ref{fig:model}.
For each time period $T$, these four (time-independent) Hamiltonians are applied successively during a time $T/4$, defining the time-periodic drive Hamiltonian $H_{\rm drive} (t)$.  Note that, on the square lattice, $\Delta$ can be changed into $-\Delta$ by a simple gauge transformation (so the sign of $\Delta$ is irrelevant).

We are interested in the unitary time evolution $U(t_1,t_0)={\cal T}_t\exp{\left( -i\int_{t_0}^{t_1}H(t)\, dt \right )}$ in the case of a sequence of constant steps, $H(t)=H_{\rm drive}(t)$. 
Since during each $T/4$ time interval the Hamiltonian is static, the Floquet unitary operator $U_F[t_0]=U(T+t_0,t_0)$ after a time period takes the simple product form (for $t_0=0$),
\begin{equation}
U_F = U_d^{\Delta}(T/4)\, U_c^{\Delta}(T/4)\, U_b^{\Delta}(T/4)\, U_a^{\Delta}(T/4)\, .
\label{eq:UnitaryGateProduct}
\end{equation}
Furthermore, since each term of the Hamiltonian is made of independent commuting terms, further simplification occurs for each of the above unitaries,
\begin{equation}
U_{\cal B}^{\Delta}(T/4) = \bigotimes_{l\in{\cal B}} u_l (\omega,\Delta)\, 
\label{eq:UnitaryGate}
\end{equation}
where $\cal B$ stands for one of the sets of (disconnected) bonds, $a$, $b$, $c$ or $d$. 

At a 2-site level, the Hamiltonian reads (taking $J=1$),
\renewcommand{\arraystretch}{1.5}
\begin{equation}
h(\Delta) = 
\begin{pmatrix}
 \frac{\Delta }{4} & 0 & 0 & 0 \\ 
 0 & -\frac{\Delta }{4} & \frac{1}{2} & 0 \\
 0 & \frac{1}{2} & -\frac{\Delta }{4} & 0 \\
 0 & 0 & 0 & \frac{\Delta }{4}
\end{pmatrix}_{\!\!\!l},
\label{eq:XXZmatrix}
\end{equation}
and the corresponding 2-site unitary gate acting on bond $l$ reads
\renewcommand{\arraystretch}{1}
\begin{align}
u_l (\omega,\Delta) &= e^{-i (T/4)h_l} = e^{-\frac{i \pi  \Delta }{8 \omega }}
\begin{pmatrix}
  1 & 0 & 0 & 0 \\
 0 & \alpha & \beta & 0 \\
 0 & \beta & \alpha & 0 \\
 0 & 0 & 0 & 1 \\
\end{pmatrix}_{\!\!\!l}\\
 &\stackrel{\text{def}}=  \Omega \;\tilde{u}_l (\omega,\Delta)
\label{eq:UnitaryGateMatrix}
\end{align}
with $\omega=2\pi/T$, $\Omega=e^{-\frac{i \pi  \Delta }{8 \omega }}$ and
\begin{align*}
\alpha &= \frac{1}{2} \left(1+e^{\frac{i \pi }{2 \omega }}\right) e^{\frac{i \pi  (\Delta -1)}{4 \omega }} \\
\beta &= -\frac{1}{2} \left(-1+e^{\frac{i \pi }{2 \omega }}\right) e^{\frac{i
   \pi  (\Delta -1)}{4 \omega }}
\end{align*}

\begin{figure}
	\centering
	\includegraphics[width=\columnwidth]{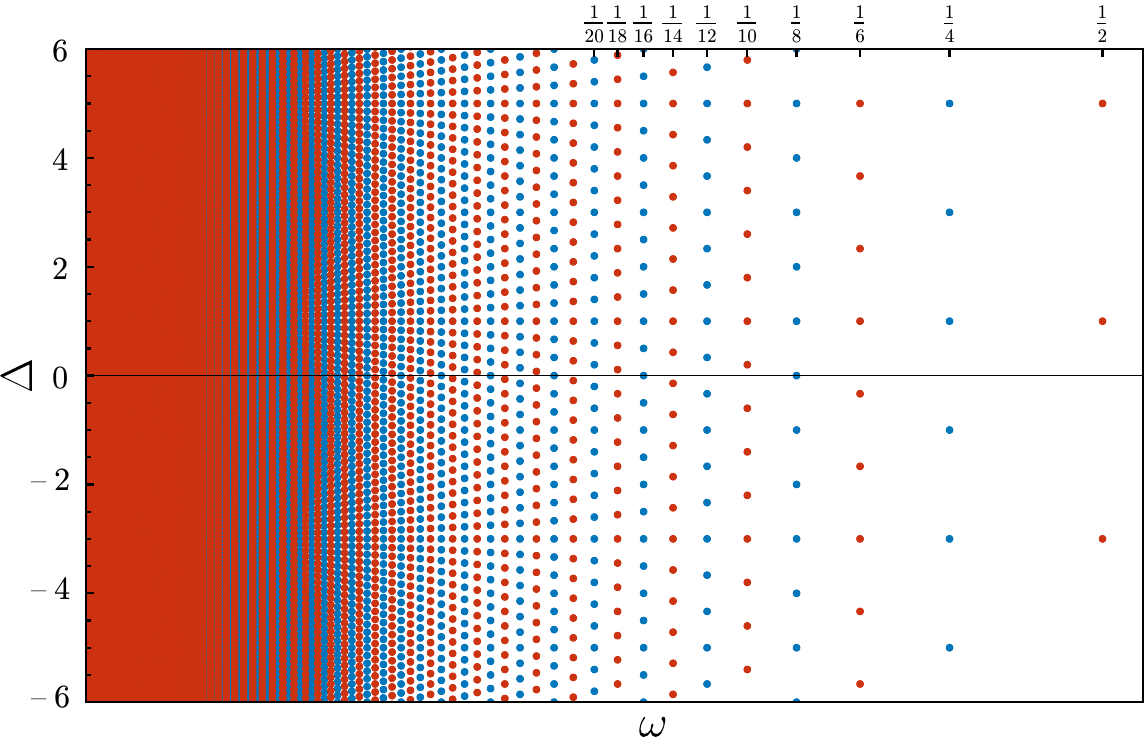}
     \caption{Fined-tuned gates map in the $(\omega,\Delta)$ plane. Trivial gate instances are represented as blue dots and swap gates as red dots. These two cases are denoted respectively as trivial points and swap points in the text. Note that, due to a gauge symmetry on bipartite lattices, $(\omega_s,-\Delta_s)$ give another equivalent set of swap gates (not shown).}
	\label{fig:FineTuned}
\end{figure}

{\it Fine-tuning --}
Interestingly, for some (discrete) values of $\omega$ equation~(\ref{eq:UnitaryGateMatrix}) takes a particularly simple form.
The first case, denoted as trivial, corresponds to $(\alpha_t,\beta_t)=(1,0)$ for which $\tilde{u}_l (\omega,\Delta) = \mathbb{1}$. The set of $\Delta$ and $\omega$ parameters corresponding to this solution writes,
\begin{align*}
\Delta_t & = 1 + 2\frac{q}{p}, \\
\omega_t  & = \frac{1}{4p},
\end{align*}
with $p\in\mathbb{N}^*$ and $q\in\mathbb{Z}$. 

More interestingly, for $(\alpha_s,\beta_s)=(0,1)$ $\tilde{u}_l$ in Eq.~(\ref{eq:UnitaryGateMatrix}) takes the form of a permutation operator or swap gate, denoted as
\begin{equation}
{\cal S}^{(0)}_l = \begin{pmatrix}
1 & 0 & 0 & 0 \\
 0 & 0 & 1 & 0 \\
 0 & 1& 0& 0 \\
 0 & 0 & 0 & 1 \\
\end{pmatrix}_{\!\!\!l}
\label{eq:Swap0}
\end{equation}
This occurs whenever 
\begin{align}\label{eq:Deltas}
\Delta_s & = \frac{4 q-2 p+1}{2 p+1} \\
\omega_s  & = \frac{1}{4p+2}\label{eq:Omegas}
\end{align}
with $p\in\mathbb{N}$ and $q\in\mathbb{Z}$. The two sets $(\omega_t,\Delta_t)$ and $(\omega_s,\Delta_s)$ are represented in figure~\ref{fig:FineTuned}. Trivial points do not deserve particular attention since $U_{\cal B}(T/4) = \Omega^{N/2} \;\mathbb{1}$. Hence $U_F  = \Omega^{N_{\rm bonds}} \; \mathbb{1}$ for any finite or infinite system, with either periodic (in such a case $N_{\rm bonds}=2N$) or open boundary conditions.
On the other hand, for swap points, the unitary operator over one period of time is $U_F =  \Omega^{N_{\rm bonds}} \; \mathbb{1}$ (in the bulk) while $U_\alpha(T/4)$ {\em is not} proportional to $\mathbb{1}$, hence allowing for non-trivial micromotion on the edge, as we further discuss below.

The largest possible frequency of the Swap models is $\omega_0=1/2$ (in units of $J$) for $p=0$. In our construction Swap models are generically of the Ising-type except for $q=0$, ..., $p-1$ giving $p$ different XY Swap models for $p\neq 0$.  Interestingly, by choosing $q=p$ one gets a series of SU(2)-symmetric Swap models originating from the $p=q=0$ ``parent'' model at $\omega=\omega_0$.

\begin{figure}
	\centering
	\includegraphics[width=\columnwidth]{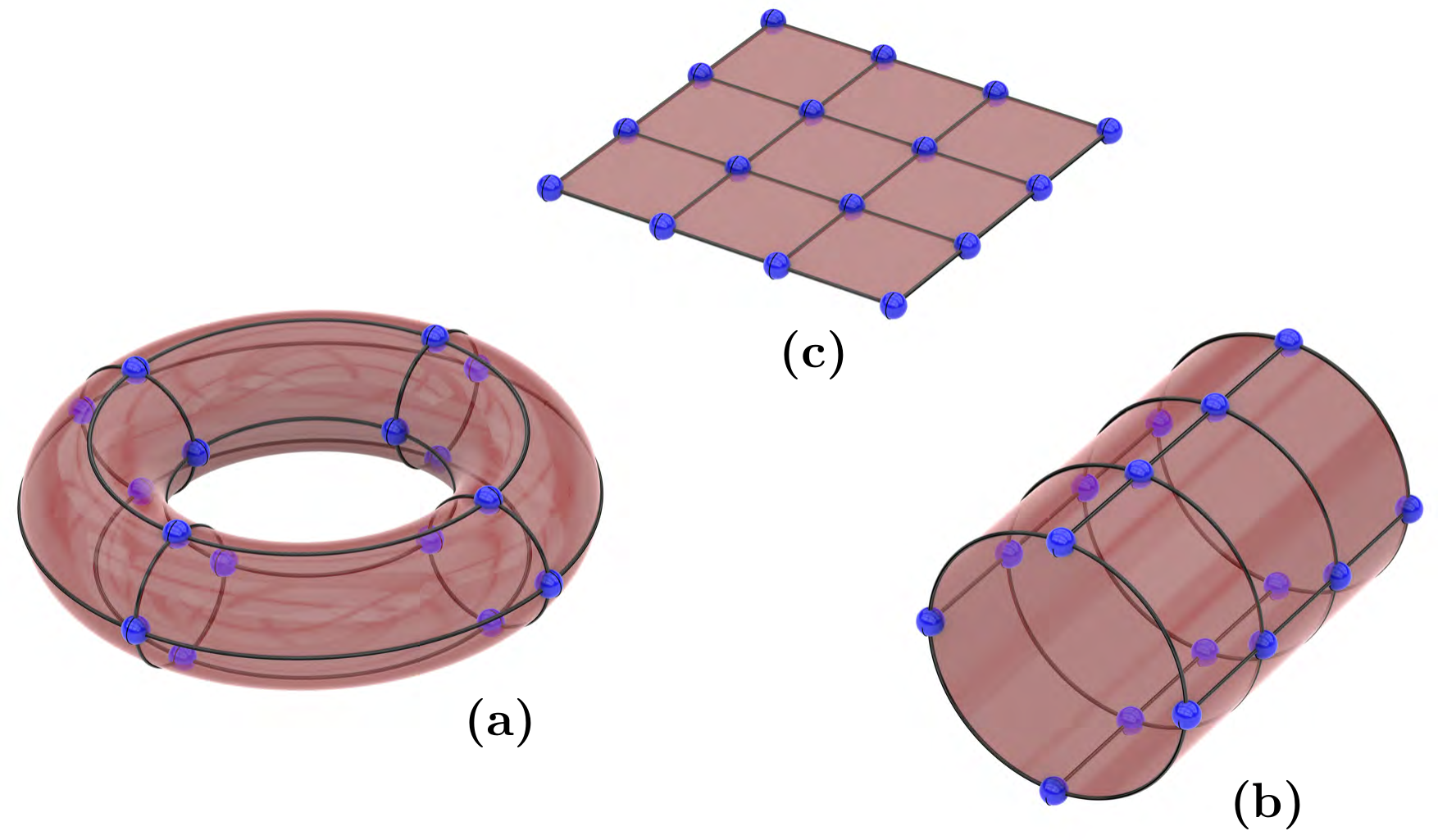}
     \caption{$4\times4$ square lattice cluster with {\bf (a)} periodic boundary conditions along $x$ and $y$ axes (torus), {\bf (b)} along $x$ axis (cylinder), and {\bf (c)} open boundary conditions.}
	\label{fig:geometries}
\end{figure}

{\it Stroboscopic motion on different geometries --} The periodicity of the motion in the Swap models depends on the topology of the system manifold. Here we shall investigate the time evolution on the various geometries shown in figure~\ref{fig:geometries}. While $U_F= \mathbb{1}$ on the torus, we find that on $L_x\times L_y$ cylinders (periodic e.g. along $x$, $L_x$ even) or on $L_x\times L_y$ systems with open boundaries, discrete time symmetry is broken like in Time Crystals~\cite{Khemani2019}, i.e. $U_F\neq \mathbb{1}$, and $U(T_{\rm geo})= (U_F)^{{\cal L}_{\rm geo}}=\mathbb{1}$ with new periodicities $T_{\rm geo}={\cal L}_{\rm geo} T$ where the integer multiplicative factors are given by:
\begin{eqnarray}
    {\cal L}_{\rm cyl} &=& (L_x/2)  \\
    {\cal L}_{\rm open} &=& (L_x+L_y-1)  \, ,
    \label{eq:lengths}
\end{eqnarray}
for a cylinder and an open (planar) geometry, respectively (while in contrast ${\cal L}_{\rm torus}=1$). In other words, when edges are present the time-periodicity is multiplied by an integer scaling like the size of the edge. This is true for all Swap models, independently of the $p$ and $q$ integers and $S_z$ total magnetization. This is to be associated to the chiral edge mode characteristic of the anomalous chiral spin liquid~\cite{Po2016,Zhang2021}. After one period, particles/spins move on the boundary by a finite length and an integer number ${\cal L}_{\rm geo}$ of periods is necessary to complete a full loop. 

\begin{figure}
	\centering
	\includegraphics[width=0.9\columnwidth]{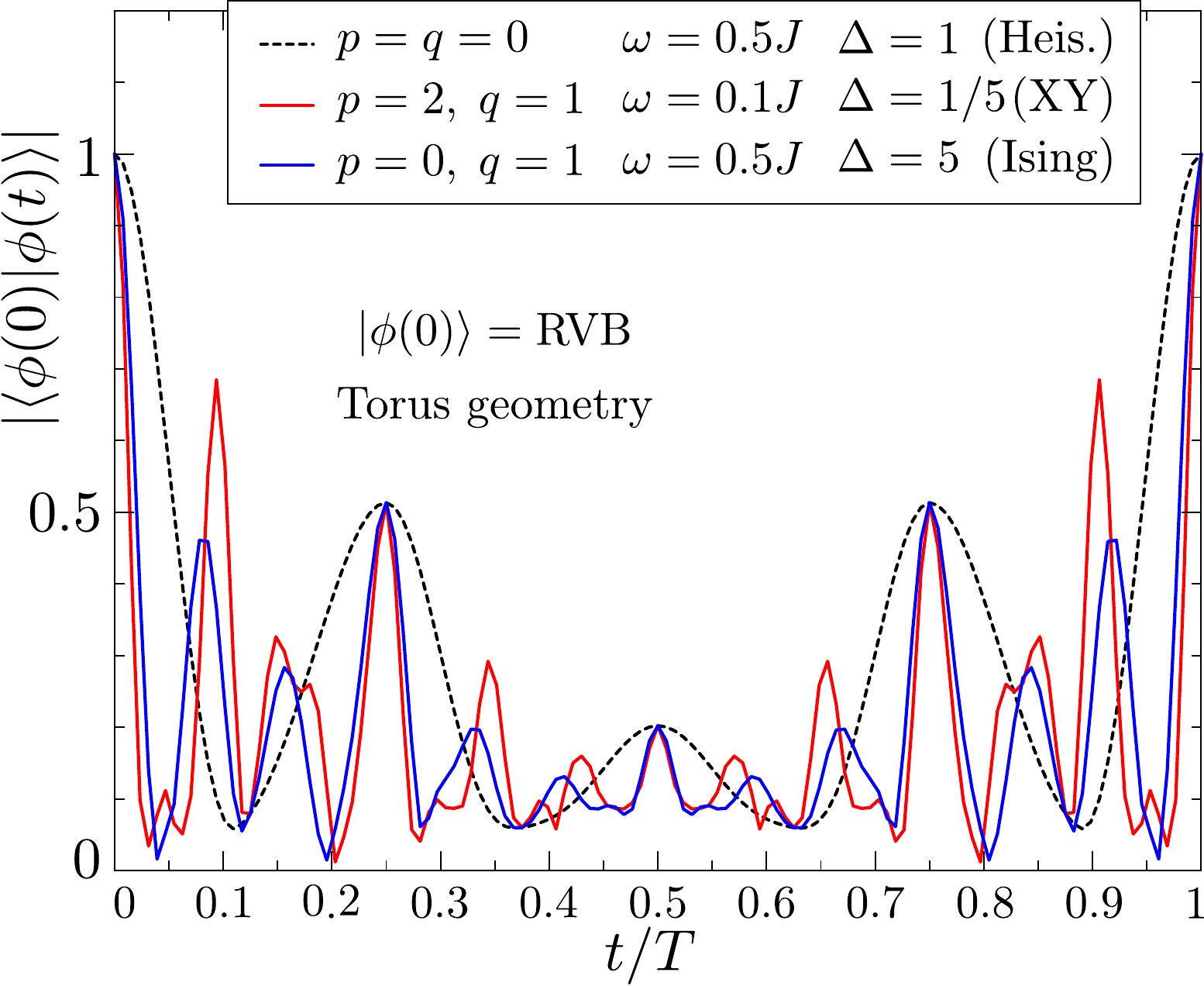}
     \includegraphics[width=0.9\columnwidth]{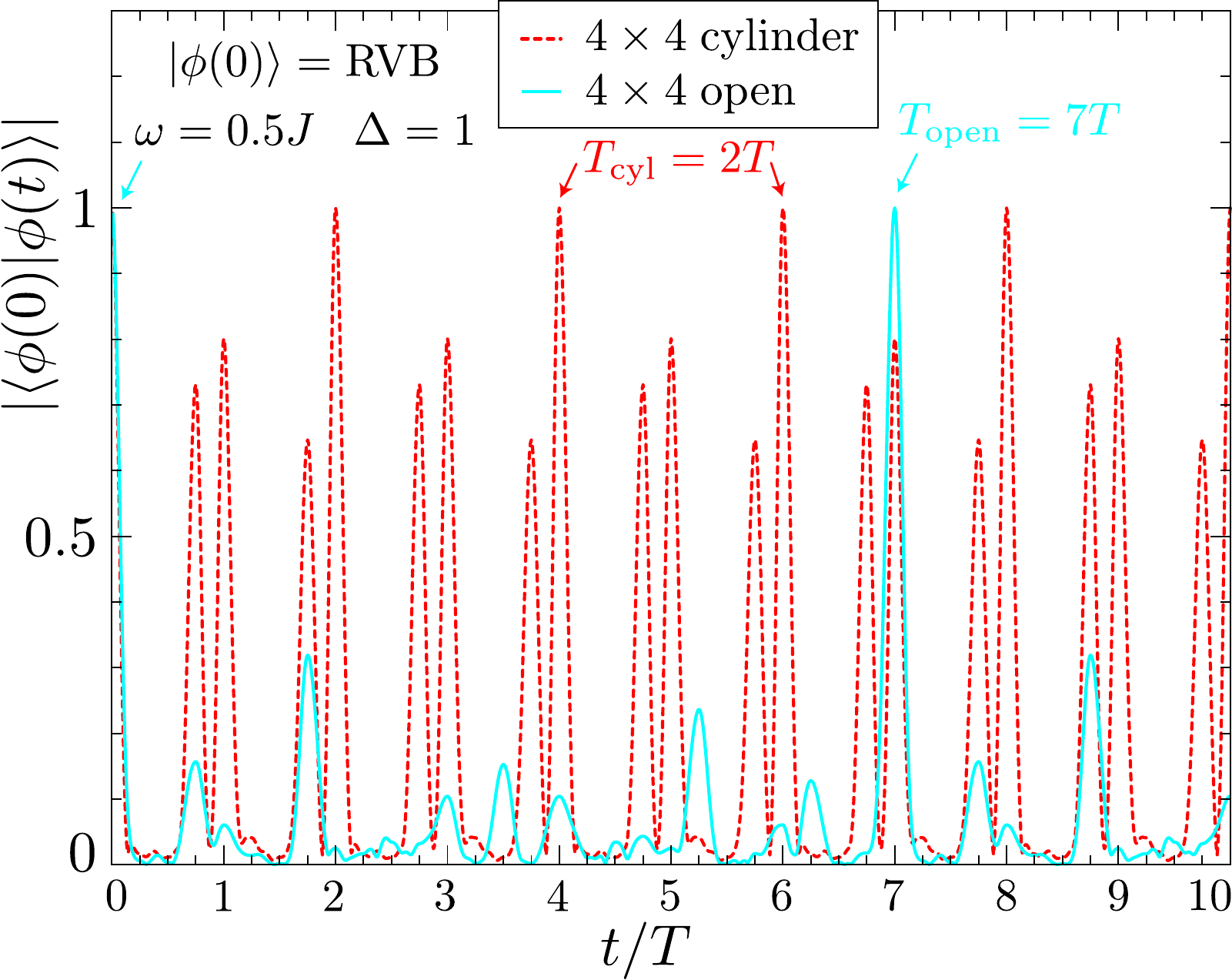}	
     \caption{Periodic micromotions in fine-tuned models on a $L\times L$ ($L=4$) system -- The fidelity $|\langle\phi(0)|\phi(t)\rangle|$ at $n=1/2$ boson density (the initial state being the RVB state, see text) is plotted vs reduced time. Top: torus geometry. Heisenberg, XY and Ising drives are considered. Bottom: cylinder and open geometries showing emerging $T_{\rm cyl}=(L/2) T$ and $T_{\rm open}=(2L-1)T$ time-periodicities (shown between arrows), respectively.   }
	\label{fig:overlaps}
\end{figure}

{\it Micromotion --} 
In contrast to the stroboscopic motion above, the micromotion (i.e. the time evolution within a time period) depends strongly on the integers $p$ and $q$ defining the model and on the (hard-core) boson density (i.e. the total magnetization $S_z$). Examples are shown in figure~\ref{fig:overlaps} for Heisenberg, XY and Ising drives applied to a zero-magnetization spin liquid initial state $|\phi(0)\rangle$. For simplicity we have chosen a Resonating Valence Bond (RVB) state defined as an equal-weight superposition of all valence bond coverings~\cite{Anderson1973}. In practice, the initial RVB spin liquid state can be constructed using tensor network techniques~\cite{Poilblanc2012} and is a linear superposition of many Floquet (singlet) eigenstates. 

Interestingly, we observe that the time-evolved states coincide at multiple of $T/4$ in all Swap models (on the same geometry).  This is expected as the latter are given recursively by $|\phi_{\cal B}\rangle=\bigotimes_{l\in{\cal B}}{\cal S}^{(0)}_l |\phi_{\cal B'}\rangle$, where $({\cal B},{\cal B'})=(a,0)$, $(b,a)$, $(c,b)$, $(d,c)$, which do not depend on the parameters of the model ($|\phi_{d}\rangle=|\phi_{0}\rangle$ on a torus geometry). 

\begin{figure*}
\begin{center}
	\includegraphics[width=1.95\columnwidth]{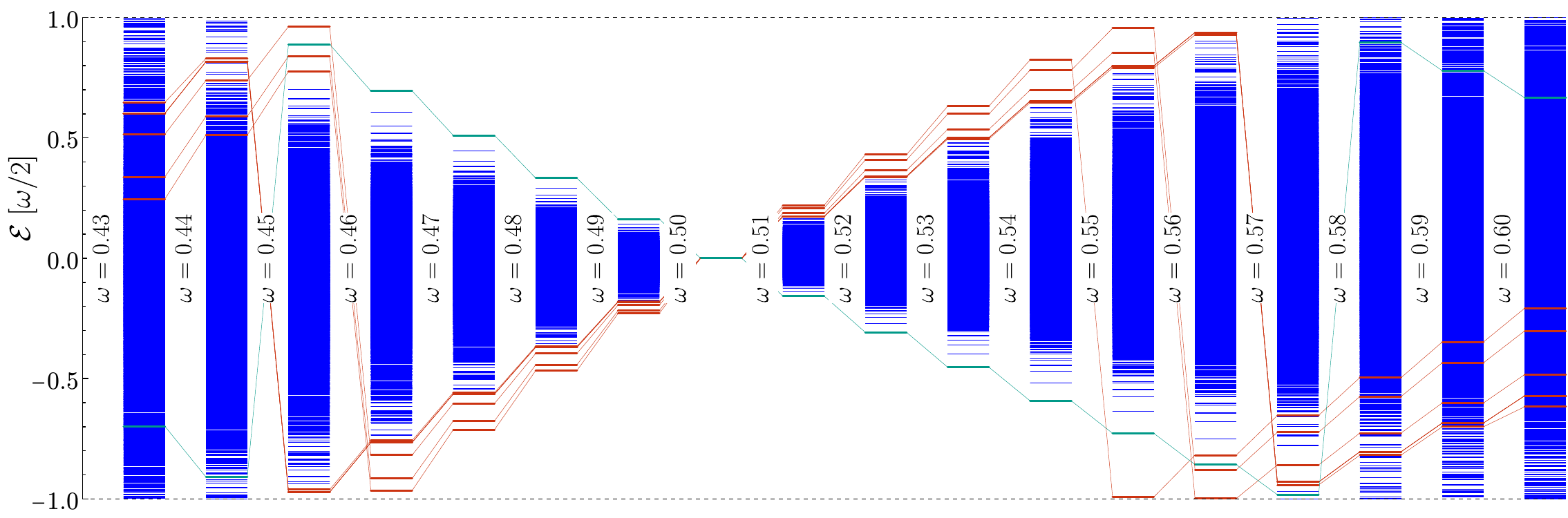}
 	\includegraphics[width=1.95\columnwidth]{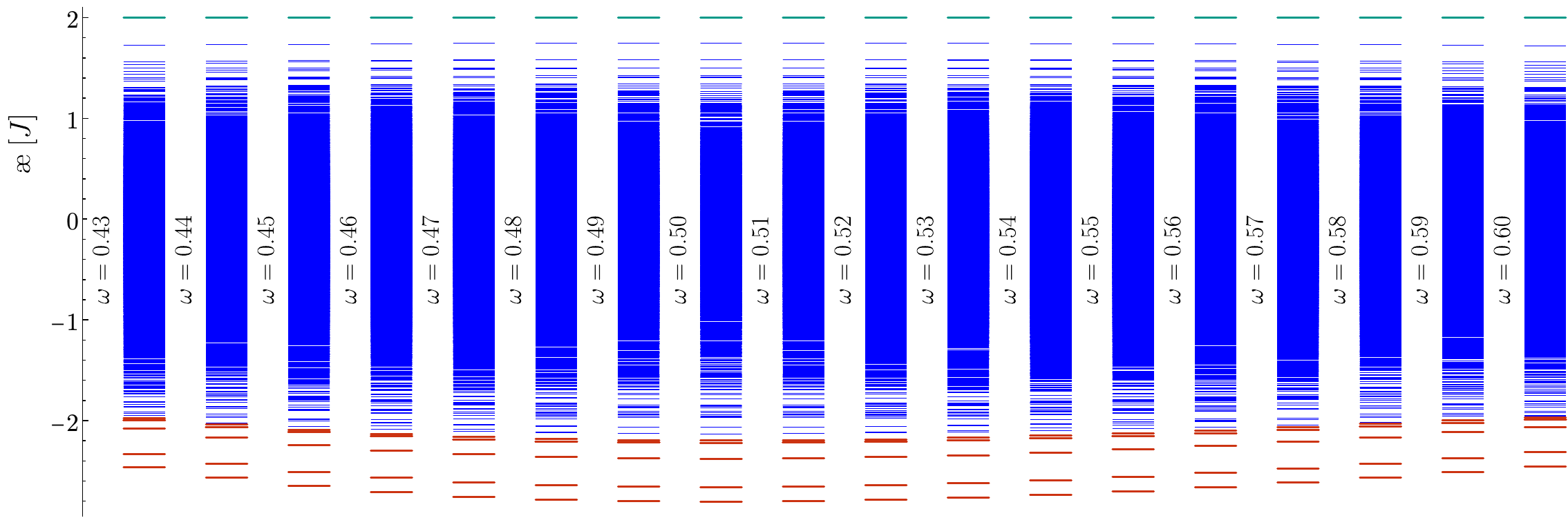}
\hspace*{-0.01\columnwidth}\includegraphics[width=1.95\columnwidth]{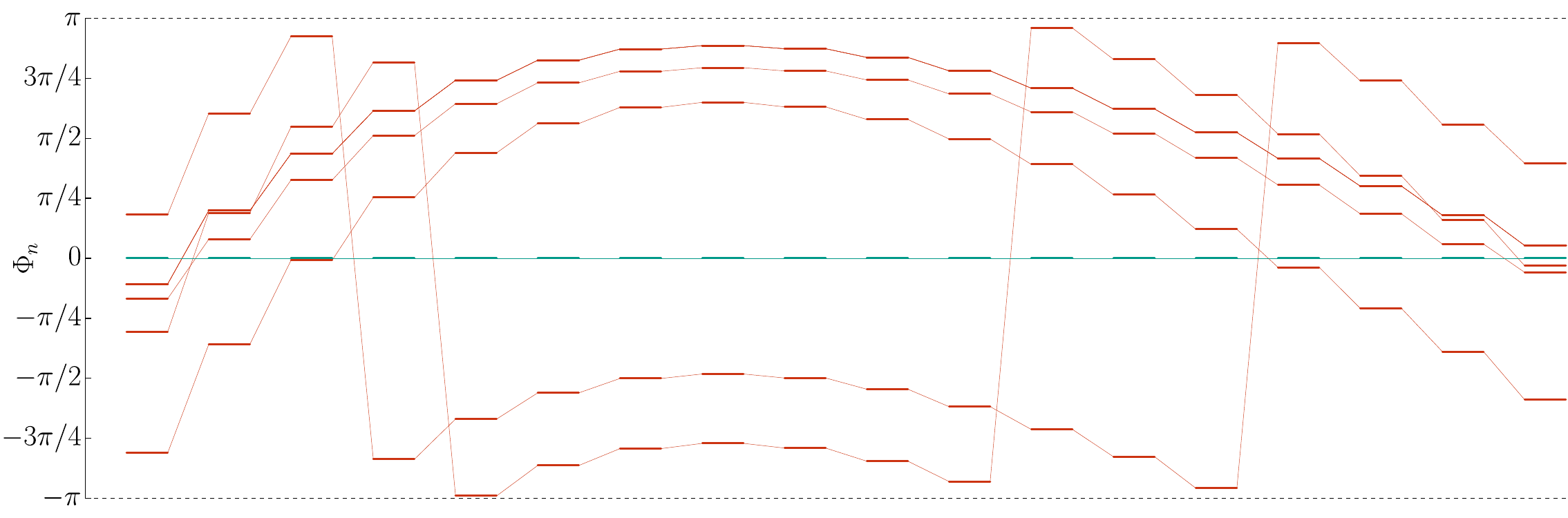}
 \end{center}
  \caption{Energy spectra on a $4\times 4$ torus around the tuned frequency $\omega_s=0.5$ (for $p=q=0$) in the finite-size unfolded regime $|\delta\omega|\lesssim \delta\omega_{\rm fold}(N)$ and slightly beyond,  i.e. $\omega\in [0.43,0.6]$. 
  Top row: quasi-energy spectra in units of $\omega/2=\pi/T$ (the Floquet energies being defined modulo 2 in $\omega/2$ units, ${\cal E}=\pm 1$ must be identified). Middle row: average-energy spectra in units of $J$. 
  The eight lowest average-energy states (some of them being degenerate belonging to related momentum sectors) and the ferromagnetic state are represented by red lines and green lines, respectively, in both the Floquet and average-energy {\AE} spectra. Note that for $\omega=0.5$ the average-energy spectrum obtained by taking the limit $\delta \omega\rightarrow 0$ (in practice $\delta \omega=10^{-6}$) does not see the massive degeneracy of the Floquet spectrum. 
  Bottom row: geometric Berry phases $\Phi_n= T({\cal E}_n - {\text{\ae}}_n)\;\;{\text{mod}} [2 \pi]$ of the nine states mentioned above for the corresponding values of $\omega$.
 }
	\label{fig:FloquetSpectraT}
\end{figure*}

\section{Detuning the frequency}
\label{sec:detuning}

We now consider detuning the Swap models and characterize the effects. Detuning can occur by deviations of the $\Delta$ parameter and the frequency $\omega$ from their exact tuned values. Changing $\Delta$ (only) has only mild effects as shown in Appendix~\ref{app:Delta_detuning} so we shall hereafter focus on frequency detuning. In particular, we investigate the modification of the bulk and edge quasi-energy spectra and provide a perturbative expansion of the Swap Hamiltonian.  We shall focus here on the SU(2)-symmetric case but we believe conclusions are generic. 

\subsection{Quasi-energy spectrum and effective Floquet Hamiltonian}
\label{sec:QuasiEnergies}

As a first step we use numerical calculations on finite size systems to obtain a heuristic understanding of the role of detuning. We assume a detuned frequency $\omega=\omega_s +\delta\omega$ around $\omega_s=\omega_0$, keeping $\Delta=\Delta_s=1$. 
The Floquet many-body spectrum is obtained from the exact diagonalization of the Floquet unitary operator $U_F$ using lattice translation symmetry (see Appendix~\ref{app:sym}). By identifying $U_F[t_0]$ with $\exp{(-iH_F[t_0]T)}$ we extract the spectrum of the Floquet hamiltonian $H_F[t_0]$ (independent of $t_0$) defined modulus $\omega=2\pi/T$. We show in the top panel of figure~\ref{fig:FloquetSpectraT} the evolution within the Floquet-Brillouin zone (FBZ) of the energy spectrum on a $4\times 4$ torus, as a function of the detuning frequency.  
As seen
the bandwidth increases linearly with $\delta\omega$. This defines a finite-size cross-over detuning frequency $\delta\omega_{\rm fold}(N)\propto J/N$ above which the winding of the Floquet spectrum in the quasi-energy FBZ occurs. For $N=16$ sites we find that $\delta\omega_{\rm fold}\simeq 0.05$.  The frequency interval $\omega\in [0.45,0.56]$ can then be considered as a ``finite-size'' regime characterized by a limited extension of the Floquet spectrum within the quasi-energy FBZ. 

\begin{figure*}
	\centering
    \includegraphics[width=0.8\columnwidth]{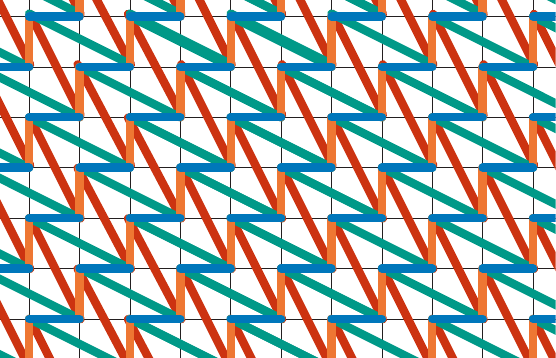}
\hskip 0.5cm     \includegraphics[width=1.1\columnwidth]{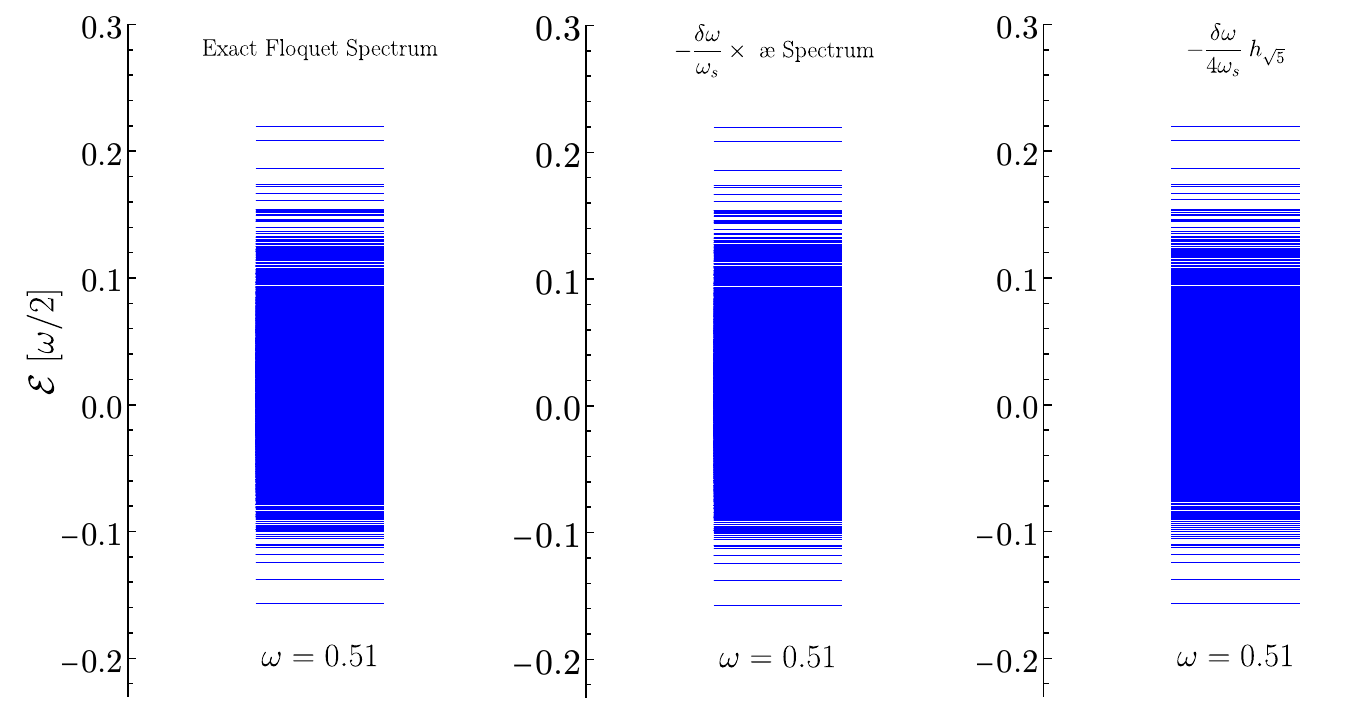}
    \caption{Left: snapshot of all interaction terms appearing in $U_F (\omega)$ at first order in $\delta \omega$ on a closed system (e.g. torus).  
     The bond sets $a$, $d$, $b'$ and $c'$ are respectively represented in blue, orange, green and red. Right: comparison between the exact Floquet spectrum, the exact average-energy spectrum given by Eq.~(\ref{eq:Aver2}) (rescaled by a factor $-\frac{\delta\omega}{\omega_s}$, with $\omega_s=1/2$) and the spectrum of the 1st-order Floquet Hamiltonian $-\frac{\delta\omega}{4\omega_s}{\hat h}_{\sqrt{5}}$ given by Eq.~(\ref{eq:HFloquet}), computed on a $4\times4$ torus for $\omega=0.51$ and $\Delta=1$. Energy is in units of $\omega/2$. Note the ferromagnetic state corresponds here (i.e. $\delta\omega>0$ case) to the lowest level at energy $-\frac{N}{4}\delta\omega$. }
     \label{fig:longrange1}
\end{figure*}

The previous empirical findings are based on numerical simulations on small systems. Next, we provide additional understanding  using some analytical calculations.
We consider a small deviation $\delta \omega$ around a Swap model characterized by a pair $(\omega_s,\Delta_s)$. We shall first consider a system with no boundary, like a torus geometry. After expanding $u_l$ of Eq.~(\ref{eq:UnitaryGateMatrix}) in first order in $\delta \omega$, one can extract the Floquet hamiltonian to lowest order (details are left to Appendix \ref{app:aeSpectrumAndHfExpansion}),
\begin{equation} 
H_F (\omega)  = - \frac{\delta \omega}{4\omega_s} {\hat h}_{\sqrt{5}} + \mathcal{O}\left(\delta \omega ^2\right)\, , 
\label{eq:HFloquet} 
\end{equation}
where 
\begin{equation}
    {\hat h}_{\sqrt{5}} = \sum_{\substack{(i, j) \in a,d\\(i, j) \in b',c'}}  S_i^x S_j^x +S_i^y S_j^y
+ \Delta_s S_i^z S_j^z \, . \nonumber
\end{equation} 
Note that $H_F^{\Delta} (\omega)$ is (in lowest order)  invariant under time reversal symmetry (i.e it is not chiral) but is odd in $\delta\omega$.
The expansion at first order contains interactions at length $1$ (bonds $a$ and $d$) and $\sqrt{5}$ (bonds $b'$ and $c'$) as shown in figure~\ref{fig:longrange1}.  This pattern has the same spatial symmetries as the drive Hamiltonian: translational symmetries of the square lattice are partially broken, the system being invariant only under $T_x^2$, $T_y^2$ and $T_x\, T_y=T_y\, T_x$. Also the $\pi/2$-rotation is broken.
Note that the Floquet hamiltonian $H_F[t_0]$ depends on the initial time $t_0$. E.g. $H_F[T/4]$ is the same as $H_F\equiv H_F[0]$ shown in figure~\ref{fig:longrange1} (left panel) but rotated by $\pi/2$ around a lattice site.

We have compared the exact Floquet spectrum and the spectrum of the 1st-order Floquet (bulk) Hamiltonian on a $4\times 4$ torus.
As shown in the right panel of figure~\ref{fig:longrange1}, on the torus the agreement between the two spectra is excellent for $\delta\omega=0.01 J$. As seen in figure~\ref{fig:FloquetSpectraT} (top panel), we also find that changing $\delta\omega$ into $-\delta\omega$ reverses the Floquet spectrum, i.e. $\{{\cal E}_n\}\rightarrow \{-{\cal E}_n\}$, as expected from Eq.~(\ref{eq:HFloquet}).

We now consider the case where boundaries are present: At fine-tuning $\delta\omega=0$, we observe massively degenerate (edge) states located at ${\cal L}_{\rm geo}$ discrete energy levels equally spaced by $\omega/{\cal L}_{\rm geo}$. 
On a (e.g. horizontal) cylinder this is easy to understand: in that case the Floquet unitary takes the simple form,
\begin{equation}
    U_F (\omega_s)= T_{-2a}^{\rm A, left}\otimes\mathbb{1}_{\rm bulk} \otimes  T_{2a}^{\rm B, right} \,
    \label{eq:UF_omegas}
\end{equation} 
where $T_{2a}^{\rm A, right}$ and $T_{-2a}^{\rm B, left}$ are translations by 2 lattice spacing along the edge (e.g. the $x$ axis) in opposite directions, acting only on the $L/2$ sites of opposite sublattices on the right and left boundaries, respectively. The identity operators acts on the complementary set of sites (i.e. on $L_y-2$ rows of $L_x$ sites and the remaining $L_x$ sites on the edges). The eigenstates are therefore simple tensor products of Bloch states defined on (half of) the left and right boundaries with momenta $p=-L_x/4+1,\ldots,L_x/4$, by arbitrary spin configurations leaving on the bulk $(L_y-2)L_x$ sites, and the complementary set of boundary sites (or linear combination of those). The quasi-energies are therefore given by discrete levels $e_P=\frac{k_P}{T}=P\frac{\omega}{(L_x/2)}$ where $P=p_L+p_R$. A similar  derivation is also possible for an open $L_x\times L_y$ system of levels,
\begin{equation}
    e_P=P \frac{\omega}{{\cal L}_{\rm geo}}, \hskip 0.5cm  P=-{\cal L}_{\rm geo}/2+1,\ldots,{\cal L}_{\rm geo}/2 \, .
    \label{eq:eP}
\end{equation}
On the cylinder, the full system is invariant under a global translation $T_{2a}^{\rm global}$ and the Floquet eigenstates can be labeled according to their momentum $k=N_k \frac{2\pi}{(L_x/2)}$. Note that $N_k\ne P$.
\begin{figure*}
	\begin{center}
\includegraphics[width=1.95\columnwidth]{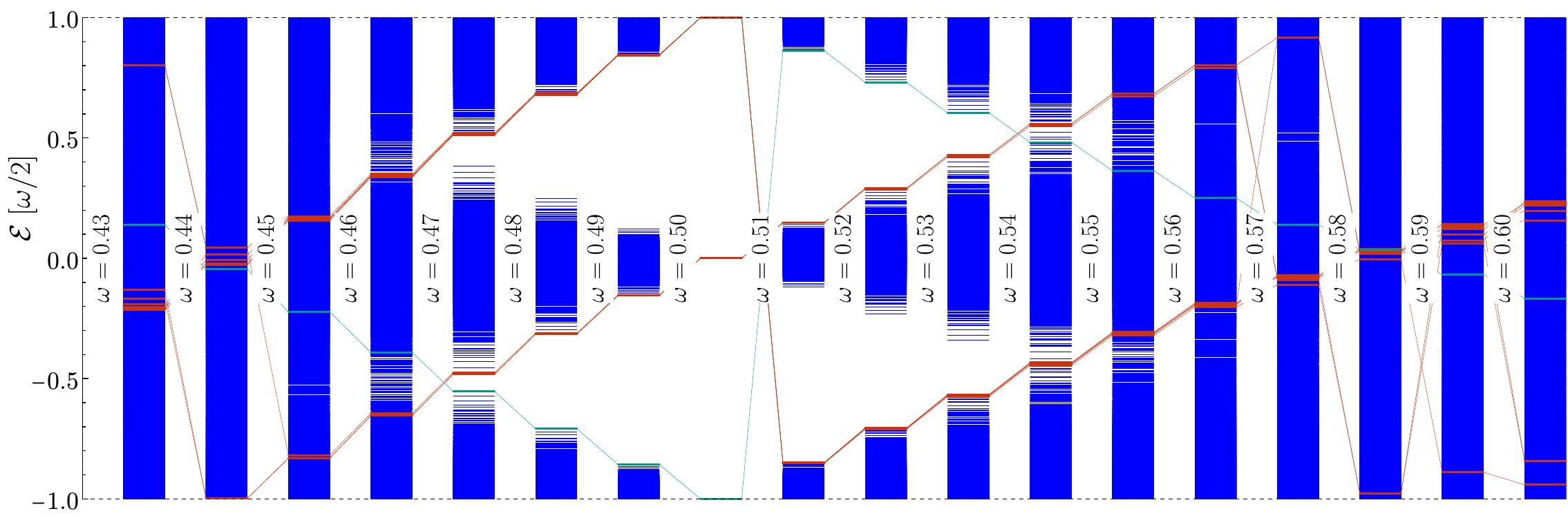}
\includegraphics[width=1.95\columnwidth]{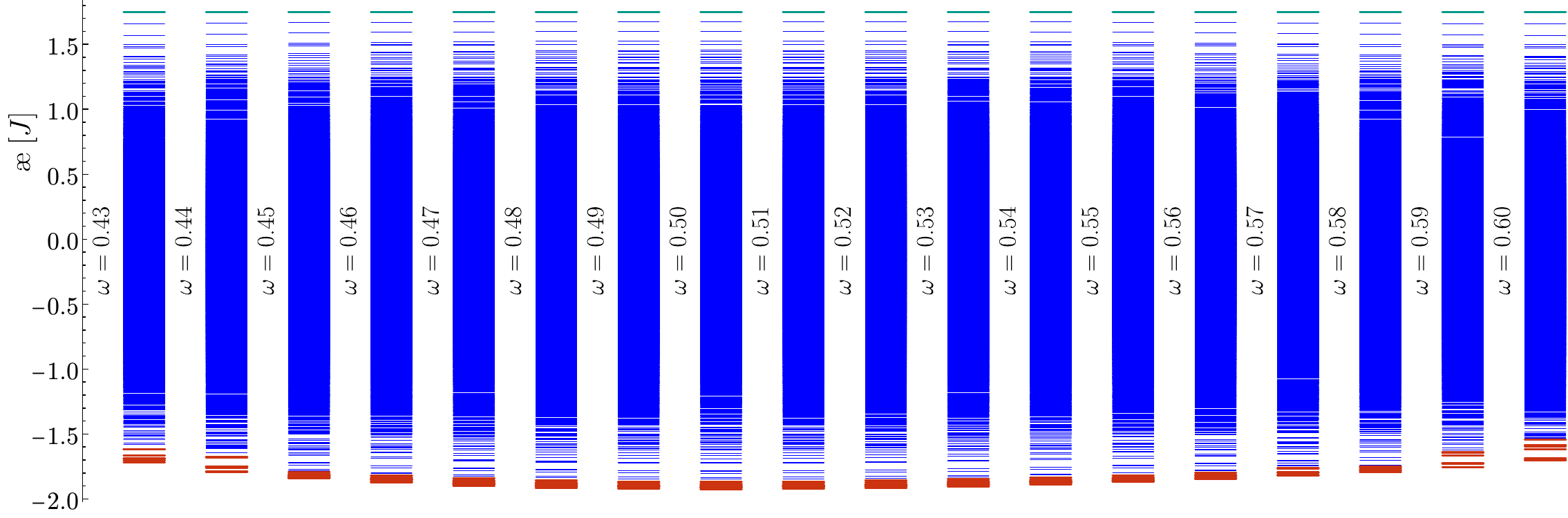}
\includegraphics[width=1.95\columnwidth]{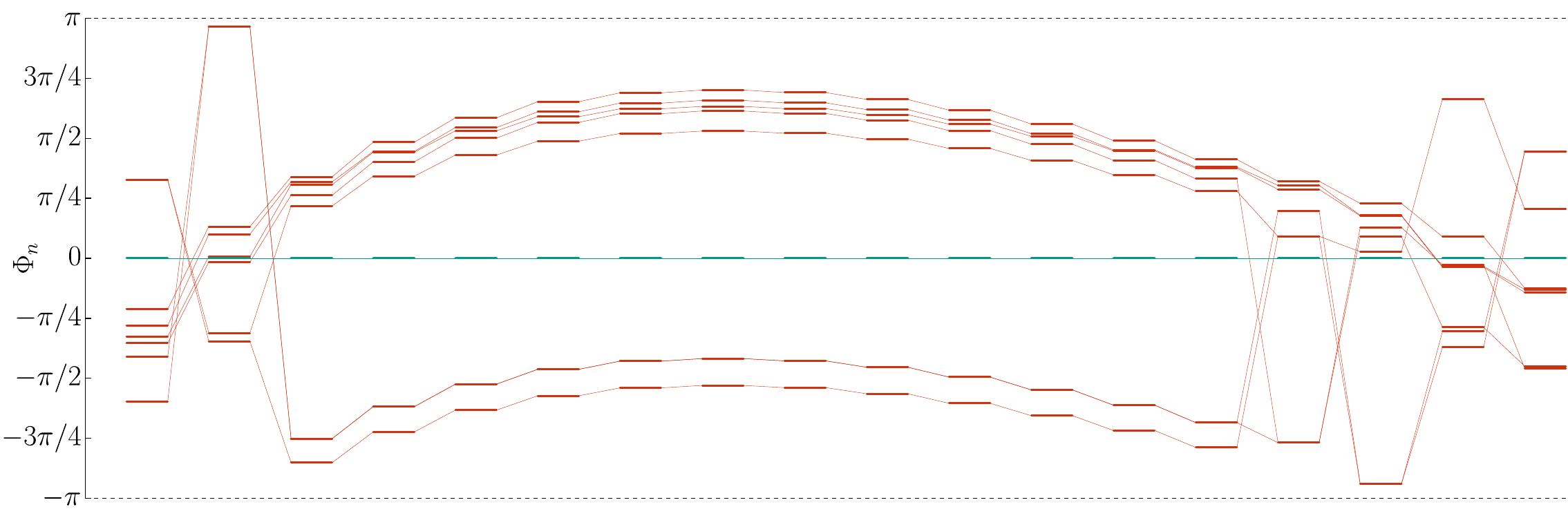}
\end{center}
  \caption{Same as figure~\ref{fig:FloquetSpectraT} for a $4\times 4$ cylinder geometry.}
	\label{fig:FloquetSpectraC}
\end{figure*}

\begin{figure*}
	\begin{center}
	\includegraphics[width=1.93\columnwidth]{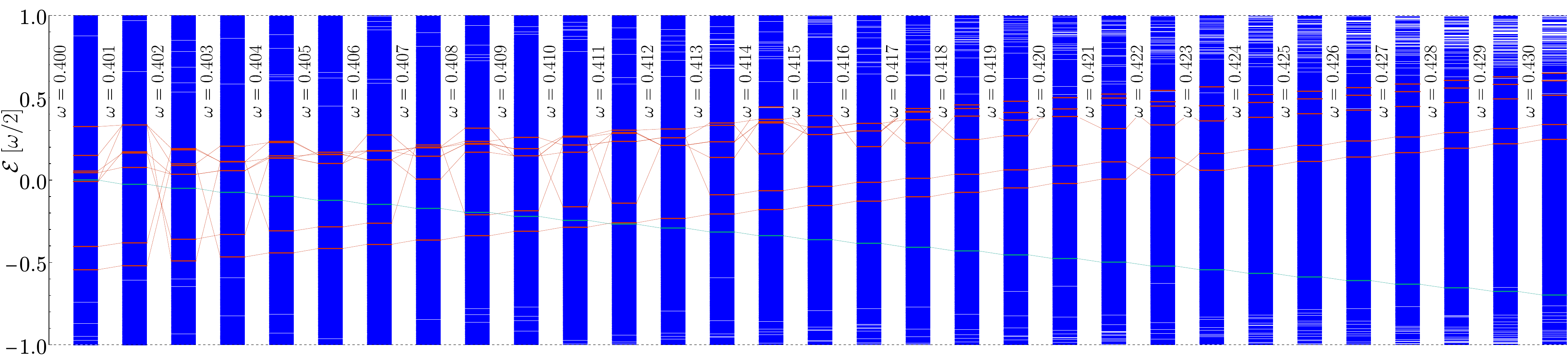}
 	\includegraphics[width=1.93\columnwidth]{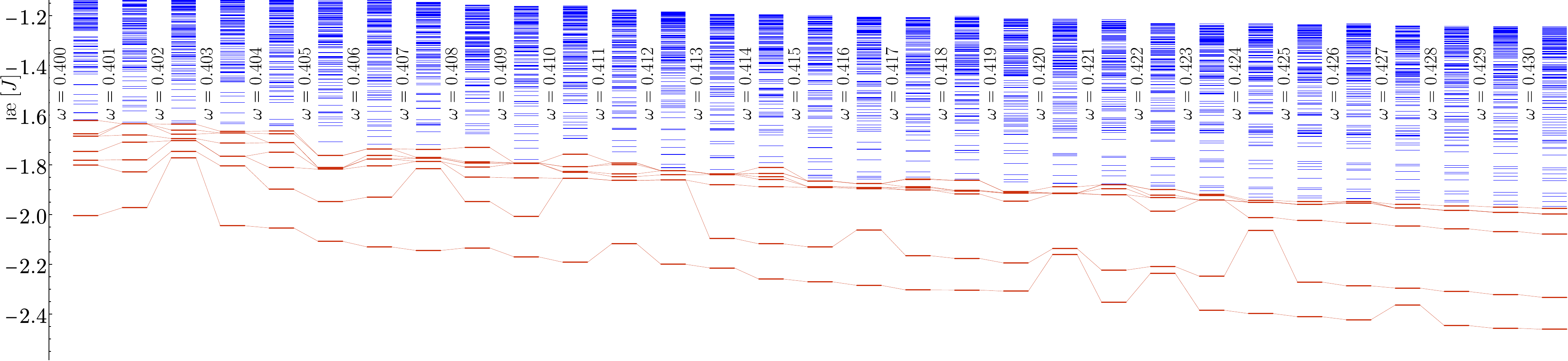}
\hspace*{-0.03\columnwidth}\includegraphics[width=1.97\columnwidth]{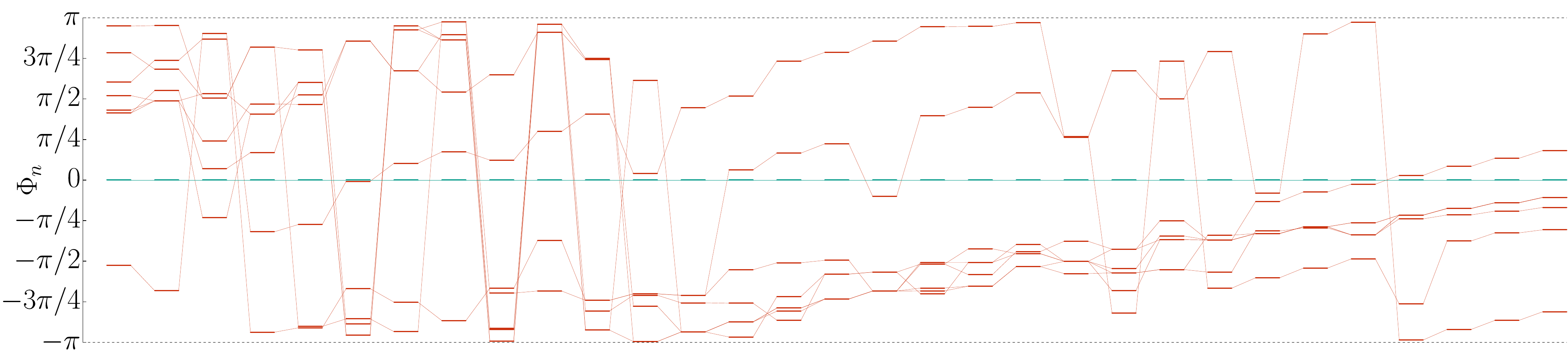}
	\end{center}
  \caption{Same as figure~\ref{fig:FloquetSpectraT} but for larger detuning in the range $\omega\in [0.40,0.43]$, with completely folded Floquet spectra, zooming (400 first levels) on the low-energy region of the ${\text{\AE}}$ spectra. The eight lowest average-energy states are indicated by red lines and the ferromagnetic state by a green line in the Floquet and average-energy ${\text{\AE}}$ spectra. Resonances are responsible for the erratic structure of the data, especially at the largest detuning (left side of the plots). 
 }
	\label{fig:FloquetSpectraT_zoom}
\end{figure*}

These ${\cal L}_{\rm geo}$ levels are broadened into bands by the detuning, in the same way as the zero-energy states on the torus geometry. 
An example is shown in figure~\ref{fig:FloquetSpectraC} (top panel) for the case of a $4\times 4$ cylinder, with ${\cal L}_{\rm cyl}=2$ sub-bands centered around $P\, \omega/2$, $P=0,1$. 
The case of a long $2\times 8$ ribbon, with ${\cal L}_{\rm cyl}=4$ sub-bands centered around $P\, \omega/4$, $P=-1,0,1,2$ is also shown in Appendix~\ref{app:ladder}. 
As observed numerically, the broadening of each edge mode (i.e. the width of the sub-bands) is very similar to that of the bulk band on the torus (and linear with the detuning) so that, empirically, we find that the overall bandwidth $W(\delta\omega)$ (defined as the sum of all the many-body bandwidths) behaves as:
\begin{equation}
    W(\delta\omega)\simeq w_{p,q}\, {\cal L}_{\rm geo}\, N\,\delta\omega\, ,
    \label{eq:bandwidth}
\end{equation}
where $N$ is the number of sites and $w_{p,q}$ is a non-universal prefactor depending on the pair $(p,q)$ defining the model, which can be estimated numerically. Note that the effect of the boundary(ies) is (approximately) taken care of by the geometrical factor ${\cal L}_{\rm geo}$. A crossover occurs when the bandwidth reaches the size $\omega$ of the energy Floquet-Brillouin zone i.e. above a finite-size (relative) detuning
\begin{equation}
\frac{\delta\omega_{\rm fold}}{\omega} (N) \simeq 1 / ( w_{p,q}\,{\cal L}_{\rm geo}\, N  )\, .    
\label{eq:detune_crit}
\end{equation}
When increasing detuning, the crossover to a folded spectrum should appear sooner for a system with a boundary (like a $L\times L$ cylinder) where ${\cal L}_{\rm geo}\sim L=N^{1/2}$  and, hence,
$\frac{\delta\omega_{\rm fold}}{\omega} (N) \sim 1/{N}^{3/2}$, in contrast to the $1/N$ behavior expected on a torus.

Note that, on the cylinder, the 1st-order {\it bulk} (non-chiral) local Floquet Hamiltonian is not able to reproduce the chiral Floquet states. This shows that the Floquet spectrum on a system with boundaries cannot be obtained from an expansion of the bulk Floquet Hamiltonian alone, but an additional non-local edge Hamiltonian is necessary. This is to be contrasted with static and local chiral Heisenberg models whose many-body spectra show edge modes on geometries with edges, like a cylinder~\cite{Poilblanc2017b,Chen2018b}, a disk~\cite{Chen2021} or a half-infinite plane~\cite{Hasik2022}. In this case the chiral modes appear within the many-body bulk energy gap and no sub-band structure like in figure~\ref{fig:FloquetSpectraC} is present.  Overall, this reflects the ‘Floquet anomalous’ nature of the edge modes depicted in Figure~\ref{fig:FloquetSpectraC}.

\subsection{Average-energy spectrum}
\label{subsec:average}

The previous study reveals a (finite size) frequency range around the $\omega_s=0.5 J$ anomalous CSL where a finite $\delta\omega$ expansion seems valid on a torus. However, the effective Floquet Hamiltonian does not give insight on how the micromotions of the Floquet states are affected by detuning neither does it take into account the role of resonances associated to folding of the Floquet spectrum when $|\delta\omega| > \delta\omega_{\rm fold}(N)$ i.e. beyond the finite-size frequency interval $\propto 1/N$.
The procedure described in Ref.~\cite{Schindler2025} enables a ``spectral sorting'' and permits an ordering of the Floquet eigenstates which uniquely defines a Floquet ground state. 
For that, one defines the average energy spectrum as
\begin{equation}
{\text{\ae}}_n=\frac{1}{T} \int_0^T  \langle\phi_n(t)|H(t)|\phi_n(t)\rangle\, dt \, ,
\label{eq:AverEner}
\end{equation} 
where $|\phi_n(t)\rangle$ are the micromotion Floquet eigenstates $U(t,0) |\phi_n\rangle$. Here the average-energy spectrum is expected to give insights into the effect of detuning on the micromotion of the Floquet eigenstates. In the case of the simple periodic drive considered here, the calculation of (\ref{eq:AverEner}) can be greatly simplified as shown in Appendix~\ref{app:aeSpectrumAndHfExpansion}. One gets:
\begin{equation}
 {\text{\ae}}_n =  \frac{1}{4} \langle\phi_n|H_{abcd}|\phi_n\rangle \, ,
 \label{eq:Aver2}
 \end{equation}
 with
 \begin{equation}\label{eq:Habcd}
 H_{abcd}=  H_a +  U_a^\dagger H_b U_a +
U_d H_c U_d^\dagger + H_d  \, ,
 \end{equation}
 which can easily be computed numerically. Note however that $H_{abcd}$ may in general be different from the average-energy operator $\hat{\text{\AE}}=\sum_n {\text{\ae}}_n |\phi_n\rangle\langle\phi_n|$ defined in Ref.~\cite{Schindler2025}, although the eigenvalue spectrum of $\hat{\text{\AE}}$ is the diagonal of $H_{abcd}$ in the Floquet basis.

We shall first concentrate on small detuning, typically within the finite-size frequency interval $|\delta\omega| < \delta\omega_{\rm fold}(N)$ or just slightly beyond. In that regime, the average-energy spectrum computed on a $4\times 4$ torus is shown on the middle panel of figure~\ref{fig:FloquetSpectraT} and compared to the Floquet quasi-energy spectrum on the same system. Interestingly we find that the average-energy bandwidth remains constant (of order $4.5J\sim 9\omega$ for the parameters of Fig.~\ref{fig:FloquetSpectraT}) even in the limit $\delta\omega\rightarrow 0$, in stark contrast to the Floquet spectrum's bandwidth which scales as $\delta\omega$. In fact, at small $\delta\omega$, we observe a close similarity between the average-energy spectrum $\{{\text{\ae}}_n\}$ and the (rescaled) Floquet spectrum $\{{\cal E}_n\}$. This suggests a simple relation between these spectra at leading order in $\delta\omega$,
\begin{equation}
    {\cal E}_n \simeq -\frac{\delta\omega}{\omega_s}\, {\text{\ae}}_n  + {\cal O}(\delta\omega^{2})\, .
\end{equation}
Such a relation is indeed correct, holding from the operator equality at leading order in $\delta\omega$ (see Appendix \ref{app:aeSpectrumAndHfExpansion}),
\begin{equation}
    H_{abcd}=J {\hat h}_{\sqrt{5}} + {\cal O}(\delta\omega)\, ,
    \label{eq:Aver3}
\end{equation}
and from equations (\ref{eq:HFloquet}) and (\ref{eq:Aver2}).
Note also that, since $|\phi_n\rangle$ are eigenstates of $\hat h_{\sqrt{5}}$ in lowest order, we also have
\begin{equation}
    {\hat{\text{\AE}}}\vert_{\delta\omega\rightarrow 0}=\frac{1}{4} H_{abcd}\vert_{\delta\omega\rightarrow 0} =\frac{J}{4} {\hat h}_{\sqrt{5}}
    \end{equation}
 in the $\delta\omega\rightarrow 0$ limit.
Since $h_{\sqrt{5}}$ is a short-range antiferromagnetic Heisenberg model with {\it no magnetic frustration} we expect it exhibits a magnetic N\'eel phase in the thermodynamic limit, with a gapless spectrum. The existence of both even and odd spin states in the low-energy average-energy spectrum and the absence of a clear gap confirms this prediction.

Nevertheless, such simple facts must of course break down beyond a limited frequency range $|\delta\omega| \lesssim \delta\omega_{\rm fold}(N)$, whenever the expansion in power of $\delta\omega$ breaks down.
However we find that the average-energy spectrum still varies very smoothly beyond $\delta\omega_{\rm fold}$ in an intermediate frequency range $\delta\omega_{\rm fold} <  |\delta\omega| \lesssim \delta\omega_\times$ characterized by a folded Floquet spectrum. 
We continue to observe the same (reverse) ordering of the energy levels in the Floquet and average-energy spectra at small $\delta\omega<0$ ($\delta\omega>0$) when the folding of the Floquet spectrum occurs for $|\delta\omega| \gtrsim \delta\omega_{\rm fold}$. As an illustration we have marked the ferromagnetic state (at maximum energy $N J/8$) and some low-energy states of the ${\text{\AE}}$ spectrum with the same colors in both spectra. 

\begin{figure*}
\begin{center}
\hspace*{0.03\columnwidth}\includegraphics[width=1.9\columnwidth]{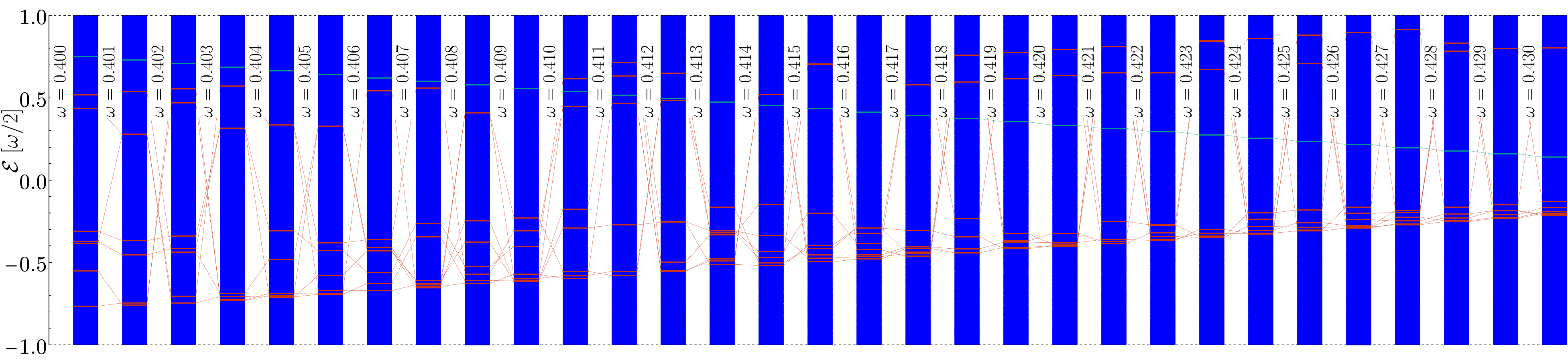}
\hspace*{0.03\columnwidth}\includegraphics[width=1.9\columnwidth]{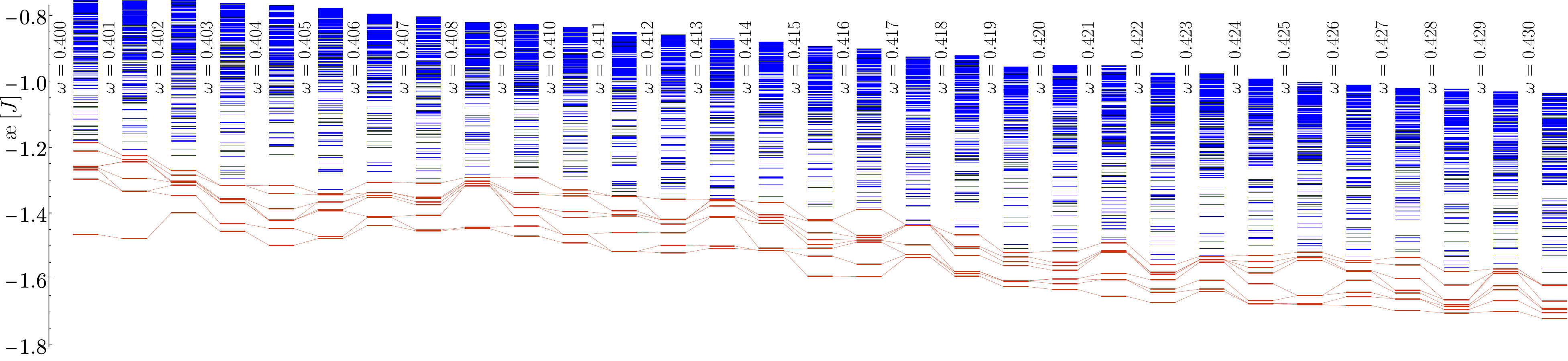}
\includegraphics[width=1.95\columnwidth]{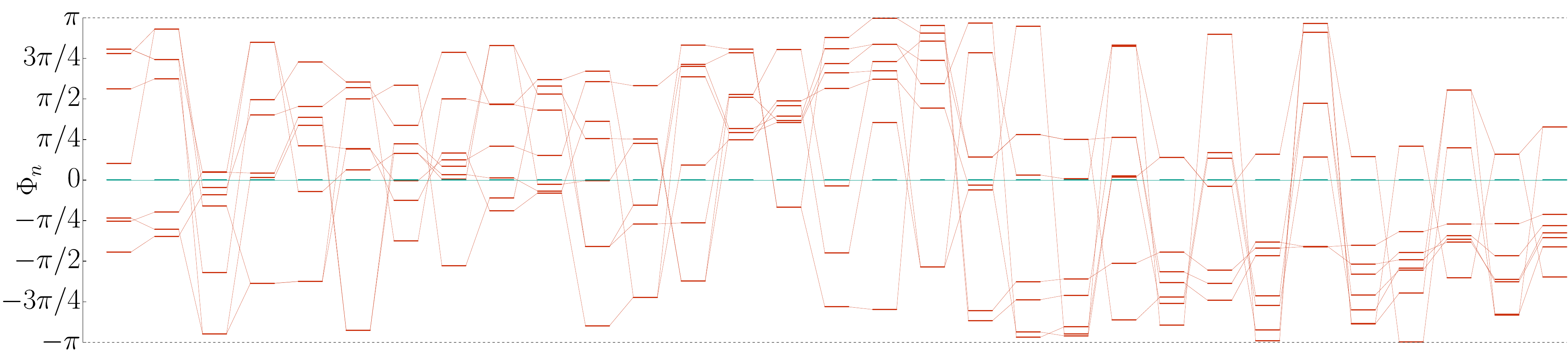}
\end{center}
  \caption{Same as figure~\ref{fig:FloquetSpectraT_zoom} for a 4x4 cylinder geometry. 
 }
	\label{fig:FloquetSpectraC_zoom}
\end{figure*}

The average-energy spectrum on a $4\times 4$ cylinder shown on the middle panel of figure~\ref{fig:FloquetSpectraC} bears similar features as the torus case. However, the low-energy part of the spectrum shows many more closely packed levels which may be associated to edge states, as suggested by edge mode spectroscopy (see next in \ref{subsec:spectro}). 
    Also, the lowest average-energy states (the ferromagnetic state) are (is) located at the bottom (top) or at the top (bottom) of the Floquet subbands in the finite-size regime for $\delta\omega<0$ or $\delta\omega>0$, respectively. All levels evolve smoothly and continuously with increasing detuning and, surprisingly, even beyond the finite-size regime. 
We have checked numerically that, in contrast to the torus geometry, the average-energy spectrum cannot be described by the Hamiltonian $\frac{J}{4} {\hat h}_{\sqrt{5}}$ in the small $\delta\omega$ limit. In particular, it is easy to see that the exact average-energy $(2N-L)J/16$ of the fully polarized ferromagnetic state (an exact Floquet eigenstate) is different from   
 its energy $(2N-\frac{3}{2}L)J/16$ in the spectrum of $\frac{J}{4} {\hat h}_{\sqrt{5}}$ (upper bounds in both spectra). However a simple relation between the average-energy spectrum and the Floquet spectrum holds in lowest order in $\delta\omega$ (see Appendix~\ref{app:FloquetAE}):
 \begin{equation}
    {\cal E}_n 
              - \epsilon_{P_n} =  -\frac{\delta\omega}{\omega_s} {\text{\ae}}_n \, 
 \end{equation}
 where, in contrast the torus case, there is an energy shift $-{\epsilon_P}=-P \frac{\omega}{2}$ to move the $P$ Floquet sub-bands (here the $P=1$ sub-band) down to zero (average) energy, overlapping with the $P=0$ sub-band. Note also that the $H_{abcd}$ Hamiltonian entering in the expression of the average-energy spectrum and in $H_F(\omega)$ (see Appendix~\ref{app:aeSpectrumAndHfExpansion}) now involves, in addition to the bulk part $\frac{J}{4} {\hat h}_{\sqrt{5}}$,
long-range terms acting on the boundaries. 
 
We now examine the case of larger detuning, significantly beyond the finite-size frequency interval. Results are shown in figure~\ref{fig:FloquetSpectraT_zoom} and \ref{fig:FloquetSpectraC_zoom} in the frequency interval $[0.40,0.43]$ which, although narrow, shows a rapid change of behavior of the low average-energy spectrum. On the torus (figure~\ref{fig:FloquetSpectraT_zoom}), decreasing the frequency from $\omega=0.43$ we see the progressive occurrence of more and more ``anomalous'' points at which both the associated Floquet energy (top panel) and average-energy (middle panel) deviates/jumps from the expected continuous lines. On the cylinder geometry (figure~\ref{fig:FloquetSpectraC_zoom}), the density of anomalous points is even higher, eventually blurring any continuous behavior when reaching the lowest frequency shown $\omega=0.40$. These anomalies are direct signatures of resonances which occur due the folding of the quasi-energy Floquet spectrum. Indeed, sharp (anti-)crossings occur for quasi-energy levels separated by an integer multiple of the frequency $\omega$ in the extended FBZ scheme (provided the corresponding states belong to the same symmetry sector). 
Simultaneously, these resonances lead to sudden (but still continuous) changes in the average-energy spectrum. The latter provides then a fundamental tool to identify resonances in the Floquet spectrum.

Based on these findings we tentatively identify three frequency domains: i) a finite-size regime where $|\delta\omega|\lesssim \delta\omega_{\rm fold}$ with a many-body Floquet bandwidth smaller than the frequency, ii) an intermediate (narrow) frequency interval $\delta\omega_{\rm fold}(N) <  |\delta\omega| \lesssim \delta\omega_\times$ where folding of the Floquet spectrum occurs but does not significantly induce resonances and iii) above $\delta\omega_\times$ a regime with an increasing density of resonances in the Floquet spectrum, possibly characterizing heating. On our $4\times 4$ systems we roughly estimate $\delta\omega_{\rm fold}\simeq 0.05 J$ and $\delta\omega_\times\simeq 0.075 J$. When $N\rightarrow\infty$, region i) shrinks like $1/N$ or $1/N^{3/2}$, depending on the geometry, and disappears in the thermodynamic limit. It is not clear whether regime ii) will survive in that limit and more advanced numerical techniques enabling to attack larger systems are needed to resolve such an issue. 

Geometrical (nonadiabatic) Berry phases have been introduced for any cyclic evolution of a quantum system~\cite{Aharonov1987,Moore1990,Moore1991b}. In our set-up these geometric phases take the simple form~\cite{Schindler2025},
\begin{equation}
    \Phi_n= T({\cal E}_n - {\text{\ae}}_n)\hspace{0.2cm} {\rm mod}[2\pi]\, ,
   \label{eq:GeoPhases}
\end{equation}
whose behavior consistently supports the existence of three regimes. The geometric phases are shown in the bottom panels of figure~\ref{fig:FloquetSpectraT} and \ref{fig:FloquetSpectraT_zoom} on the same $4\times 4$ torus and figures~\ref{fig:FloquetSpectraC} and \ref{fig:FloquetSpectraC_zoom} on the cylinder and for the same selection of low average-energy states. While the (highly excited) ferromagnetic state has zero Berry phase (since $T{\cal E}_{\rm ferro}=T{\text{\ae}}_{\rm ferro}\hspace{0.1cm}{\rm mod}[2\pi]$) the geometrical phases of the low average-energy states are finite, even in the $\delta\omega\rightarrow 0$ limit. The latter tend to wind 
continuously around the circle
when moving away from the fine tuned frequency $\omega=\omega_s$. We interpret these features as being signatures of the anomalous character of the CSL around $\omega=\omega_s$. However, the occurrence of more frequent resonances when $|\delta\omega|$ increases above $\delta\omega_\times$ leads to an increasingly erratic behavior of the geometric phases. 

\begin{figure*}
	\centering
\includegraphics[width=0.33\textwidth]{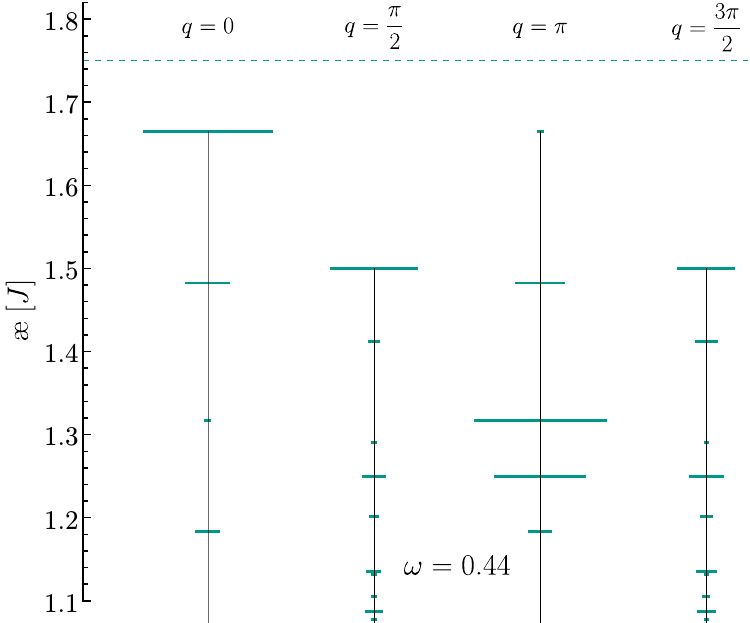}\hfill
\includegraphics[width=0.33\textwidth]{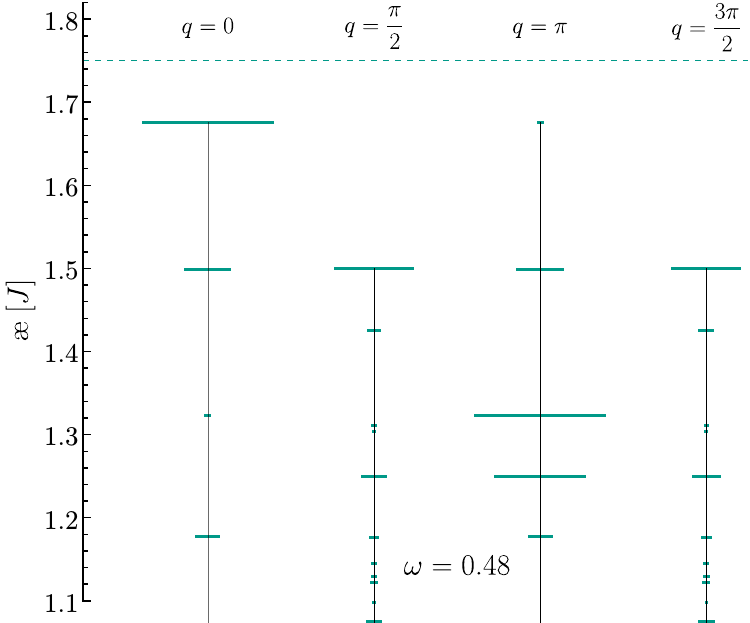}\hfill
\includegraphics[width=0.33\textwidth]{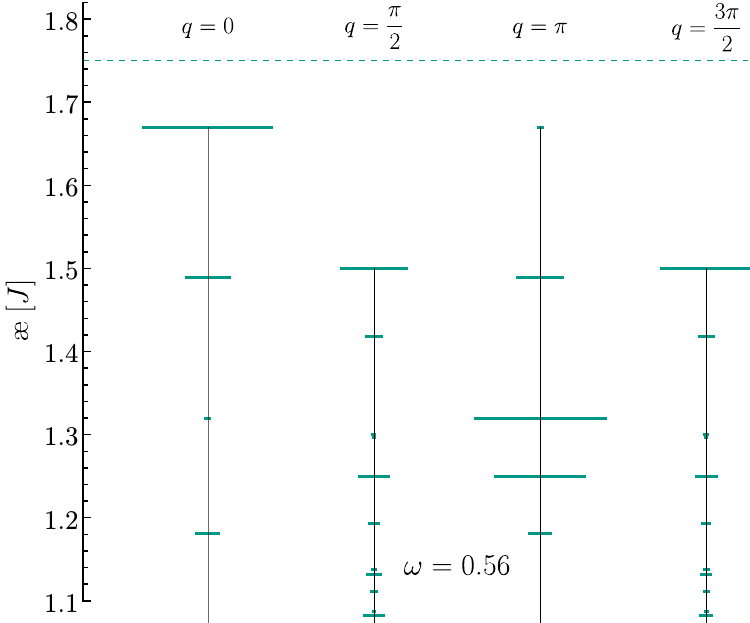}

\vspace{5mm}

\includegraphics[width=0.33\textwidth]{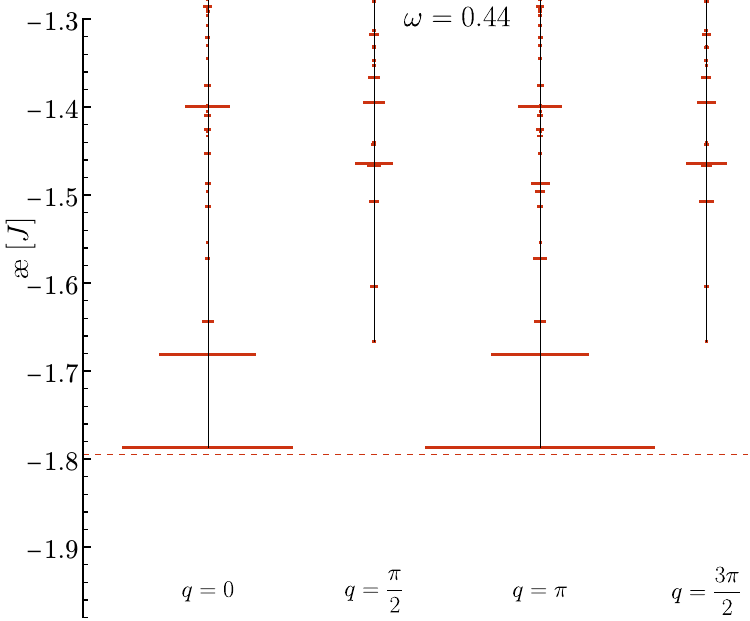}\hfill
\includegraphics[width=0.33\textwidth]{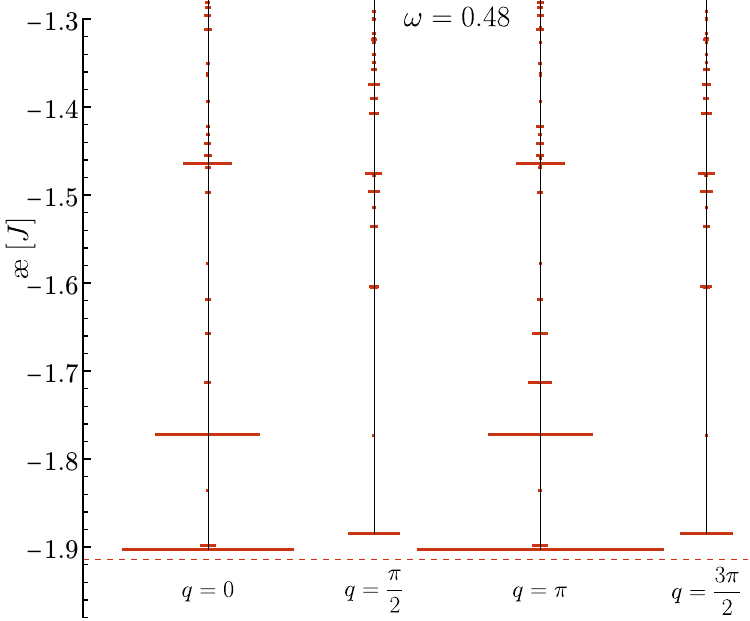}\hfill
\includegraphics[width=0.33\textwidth]{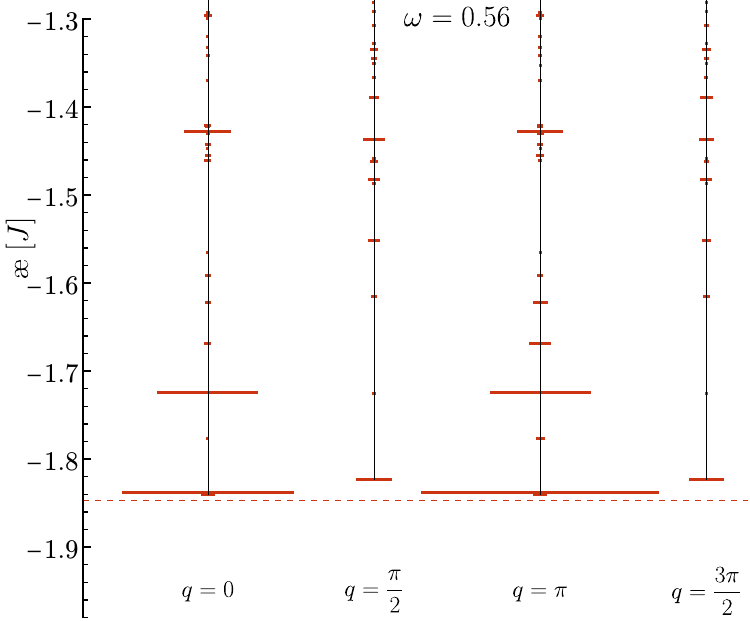}
 \caption{
 Edge dynamical structure factor $S_q({{\text{\AE}}})$ vs the average-energy ${{\text{\AE}}}$ (in units of $J$) for the allowed momenta $q$ of a $4\times 4$ cylinder at frequencies $0.44 J, 0.48 J$ and $0.56 J$. Horizontal segments of length proportional to $|{\cal W}_q^\alpha|^2$ are displayed at each energy pole ${{\text{\AE}}}={{\text{\ae}}}_\alpha$. Only low excitation energy (carrying most of the spectral weight) is displayed. Bottom: The reference state is taken as the $k_0=0$ (modulus $\pi$) singlet GS of the average-energy spectrum. Top: The reference state is taken as the $S_z=0$ component of the fully polarized state located at the top of the average-energy spectrum. The energy ${\text{\AE}}_0$ of the reference state is indicated by a dashed line on each panel.
 }
	\label{fig:EdgeSpectra_4x4_detuned}
\end{figure*}

\subsection{Spectroscopy of the edge modes}
\label{subsec:spectro}

To identify the edge states of the anomalous chiral phase on a system  with boundaries (typically a cylinder) we introduce a spectral operator ${\hat S}_q^{\rm edge}$ acting on one of the edges. We assume the edge is periodic of length $L$ so that $q=p\frac{2\pi}{L}$, $p\in{\mathbb N}$, is the edge momentum. More precisely we define,
\begin{equation}
    {\hat S}_q^{\rm edge}= \left (\sum_{j\in {E}} \exp{(iq x_j)} S_j^z \right ) \otimes {\mathbb I}_{\bar E}\, ,
\end{equation}
where $E$ is the set of edge sites and $x_j$ the linear coordinate of $j$ along the edge. ${\mathbb I}_{\bar E}$ is the identity operator acting on the complementary set of sites $\bar E$. We note that this approach is inspired from spectroscopic methods introduced to probe chiral edge modes in ultracold atoms~\cite{Goldman2012,Binanti2024}.

To perform the spectroscopy of the edge modes we first choose a reference state $|\Psi_0\rangle$ -- typically a singlet Floquet eigenstate with zero momentum -- and compute the edge dynamical structure factor defined as, 
\begin{eqnarray}
    \label{eq:matrixelements}
    S_q({{\text{\AE}}})&=& \sum_\alpha |{\cal W}_q^\alpha|^2 \delta({{\text{\AE}}} -{{\text{\ae}}}_\alpha)\, ,\\
  \text{with}\hspace{1cm} {\cal W}_q^\alpha &=& \langle \phi_{\alpha} | {\hat S}_q^{\rm edge} |\Psi_0\rangle \, ,
  \end{eqnarray}
where the sum is performed over all many-body Floquet eigenstates labeled by $\alpha$, of average-energies ${{\text{\ae}}}_\alpha$. Note that, the spectral function is computed as a function of the average-energy ${\text{\AE}}$ and not the quasi-energy $\cal E$ since the former naturally i) enables to define a 'ground-state' $|\Psi_0\rangle$ and ii) provides an energy axis to conveniently order the edge excitations. 
When $|\Psi_0\rangle$ is chosen to
have zero momentum and to be even (under spin inversion), selection rules apply, namely the sum is restricted to odd-spin states with momentum $k=q$ mod$[\pi]$. The corresponding static structure factor 
\begin{equation}
    S(q)=\langle\Psi_0|{\hat S}_{-q}^{\rm edge} {\hat S}_{q}^{\rm edge} |\Psi_0\rangle
\end{equation} 
gives the energy-integrated weight as a function of $q$.

Here we shall consider a periodic $4\times 4$ cylinder for which momenta $q=0,\pi/2,\pi$ and $3\pi/2$ are allowed. 
For the fine-tuned Swap models the eigenstates are exact tensor products of edge states $|\phi(p_L)\rangle\otimes |\phi(p_R)\rangle$ where the edge (pseudo-)momenta $p_L$ and $p_R$ are taken in opposite directions along the two edges.
Under detuning, the edge states start to couple to each other but, if the detuning is small and the transverse correlation length remains short the two physical $R$ and $L$ edges should remain weakly entangled.

Here, to investigate the ${\text{\AE}}$-spectroscopy of the edge states, the reference state is chosen as the lowest-energy $k=0$ (non-degenerate singlet) state of the average energy spectrum. For comparison we have also considered the ferromagnetic state as a reference state since, in this trivial case, we expect to see exact 1-magnon edge modes not subject to any resonance.
The dynamical edge spectral function $S_q({{\text{\AE}}})$ is shown in figure~\ref{fig:EdgeSpectra_4x4_detuned} both i) at a small detuning $\omega=0.48$ i.e. in the finite size region $|\delta\omega|<\delta\omega_{\rm fold} (N)$ and ii) at larger detuning $\omega=0.44$ and $\omega=0.56$ i.e. for completely folded quasi-energy spectra.
Note that, in the latter regime, the unfolding of the quasi-energy spectrum is essential to define a ``groundstate'' to be used as a reference state. 
As expected, at high-energy, edge magnon excitations are seen below the ferromagnetic reference state. The existence of edge modes is also clearly revealed at low energy. A significant fraction of the spectral weight is in fact concentrated on some of the low-energy Floquet states marked in red on the right panels of figure~\ref{fig:FloquetSpectraT} and \ref{fig:FloquetSpectraC}. Note that, due to total spin selection rules, our spectral function does not probe the even-integer spin sector and e.g. the singlet states also present in these tagged levels.

We also remark that, in principle, the chiral nature of the edge modes should be revealed by a dissymmetry between $S_{q}$ and $S_{-q}$.
In our case, the only available pair of opposite momenta are $\pm\pi/2$ which are in fact equal mod($\pi$) so that they sit at the reduced zone boundary. This renders the identification of the chiral nature impossible.

\subsection{Anomalous winding number}

We now turn to the investigation of the anomalous winding number. According to Ref.~\cite{gavensky2025}, the anomalous Floquet spectral flow $W_A$ can be obtained by computing the (linear) response to a uniformly distributed (magnetic) flux piercing the system surface,
\begin{equation}
    \label{eq:awn}
    W_A=  \frac{\partial}{\partial\phi}N_1(\omega,\phi)\, ,
\end{equation}
where $\phi$ is the reduced flux in units of the flux quantum $\Phi_0=2\pi$. $N_1$ is the first-order winding
number associated to the micromotion over one period
of the driving cycle,
\begin{equation}
     N_1 = \frac{1}{\omega} \{\, {\rm Tr} [H_F] - \frac{1}{T} \int_0^T {\rm Tr} [H_{\rm drive}(t)] dt \,\}
          \end{equation}
          which can be expressed in terms of the geometrical Berry phases defined in (\ref{eq:GeoPhases}),
\begin{equation}
     N_1= \frac{1}{2\pi}\sum_n \Phi_n \, ,
          \end{equation}         
which is always quantized, i.e. $N_1\in {\mathbb Z}$.
In our set-up, ${\rm Tr}  [H_{\rm drive}(t)]=0$ at all times and the winding number $N_1$ reads:
\begin{equation}
    \label{eq:N1}
 N_1 (\omega,\phi) =   \frac{1}{\omega} {\text{Tr}}\Bigl[ H_F(\omega,\phi)\Bigr] \, ,
\end{equation}
and, as can be seen explicitly in Appendix~\ref{app:winding}, 
\begin{equation}
    N_1 (\omega,\phi) \in{\mathbb Z}\, . 
        \label{eq:integer}
        \end{equation}
Note that it is crucial to perform the summation over quasi-energies prior to taking the flux derivative. Indeed, the restricted sum over a given Floquet zone and the magnetic-field derivative are not commuting operations. When changing the flux we expect discontinuities of (\ref{eq:N1}) generated by edge and bulk states that enter or leave the Floquet zone. $N_1$ versus $\phi$ is therefore a steplike function with integer jumps. The derivative (\ref{eq:awn}) can then be obtained from coarse-graining, i.e. by estimating the average slope at $\phi=0$~\cite{gavensky2025}. 

Let us also mention that, in the following, we shall consider particular $S_z$ sectors corresponding to given hard-core boson fillings. Therefore the trace in the formula above will be a partial trace over the corresponding reduced Hilbert space and, in that case, ${\rm Tr}  [H_{\rm drive}(t)]$ no longer vanishes. However, the magnetic flux appears only in the off-diagonal elements of $H_{\rm drive}(t)$ and ${\rm Tr}  [H_{\rm drive}(t)]$ is just a flux-independent constant that does not contribute to $W_A$.

\begin{figure}
	\centering
\includegraphics[width=0.7\columnwidth]{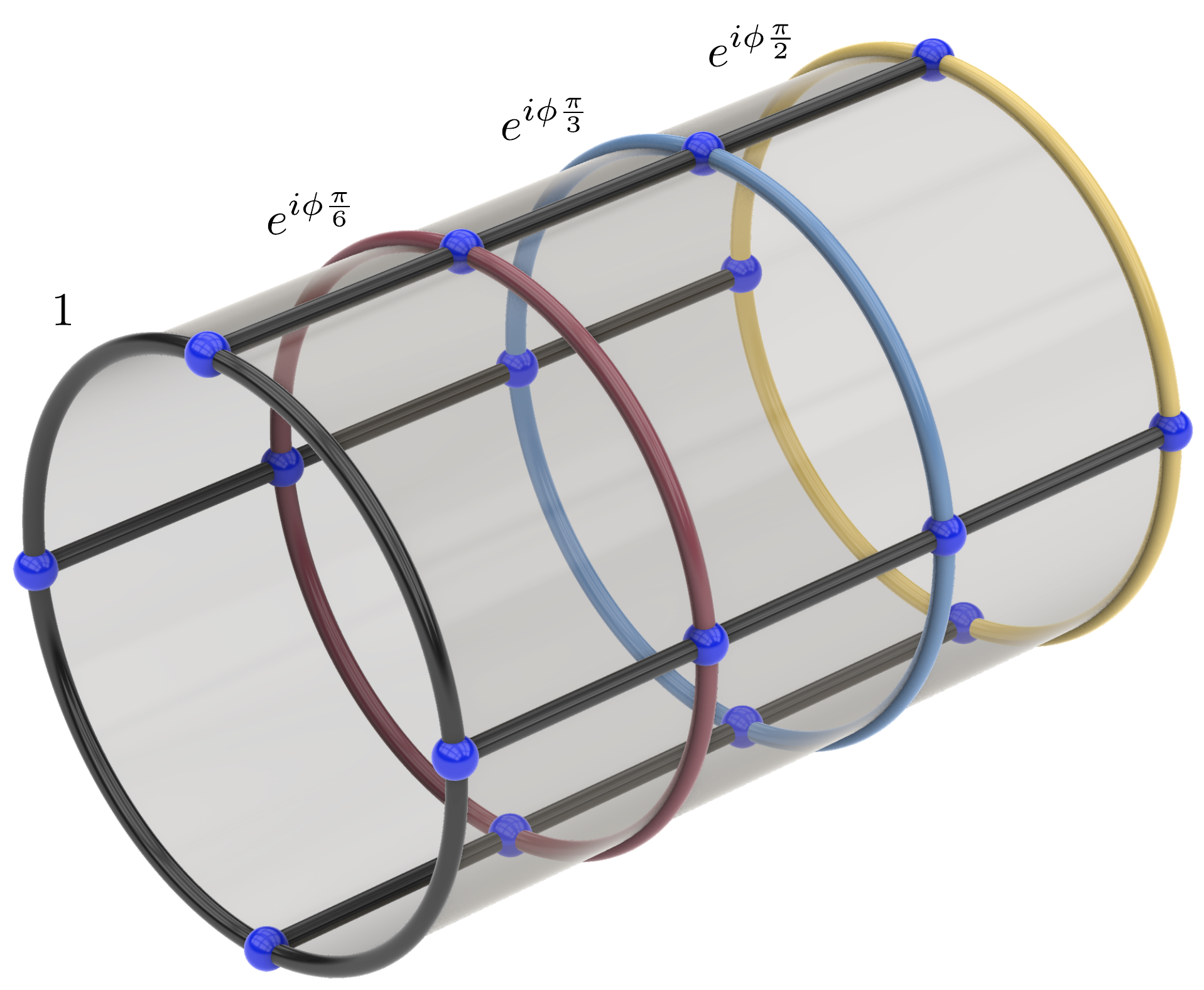}
 \caption{Peierls phases on the nearest-neighbor bonds of a $4\times4$ cylinder pierced by a uniform flux $\phi/12$ through its $N_{\rm pl}=12$ plaquettes. No flux (A flux $\phi$) enters through the left (right) side of the cylinder. }
	\label{fig:flux}
\end{figure}

To avoid issues related to total flux quantization (i.e. \hbox{$\phi\in{\mathbb N}$}) inherent to the torus geometry, we shall here consider a  cylinder geometry since, in an open-boundary sample, the flux can take continuous values. The flux threading the system is implemented by adding a Peierls phase $\exp{(i\theta_{ij})}$ to some of the bonds, i.e. replacing some of the ${\bf S}_i\cdot {\bf S}_j$ nearest-neighbor couplings of $H_{\rm drive}(t)$ by  $S_i^z S_j^z + \frac{1}{2} (S_i^+ S_j^- \exp(i\theta_{ij}) + h.c.)$. 
More precisely, assuming we take a $L_h\times L_v$ (e.g. horizontal) cylinder of $N=L_v L_h$ sites we can include a phase $\theta_{ij}=(m-1)\theta$, where $\theta=2\pi\phi/N_{\rm pl}$, uniformly in all the Heisenberg couplings on the vertical bonds of the $m$th column (Landau gauge) so that a flux $\phi/N_{\rm pl}=\theta/2\pi$  (in unit of $2\pi$) is piercing each of the $N_{\rm pl}=L_v (L_h-1)$ plaquettes ($L_v=L_h=4$, $N_{\rm pl}=12$ sites in our case). For the $4\times4$ cylinder considered here, this setup is depicted on figure~\ref{fig:flux}.
Note that there are infinitely many ways to introduce a flux $\phi$ piercing (uniformly) the cylinder surface. We can thread a flux $\phi_L$ through the left side of the cylinder and $\phi_R$ through the right side such that $\phi_R+\phi_L=\phi$. In our setup we have chosen $\phi_L=0$ and $\phi_R=\phi$. Hence, the edge states on the left boundary do not feel any Aharonov-Bohm flux while the ones on the right boundary are subject to the total flux $\phi$ by circulating around the sample.

\begin{figure*}
	\centering
\includegraphics[width=0.25\textwidth]{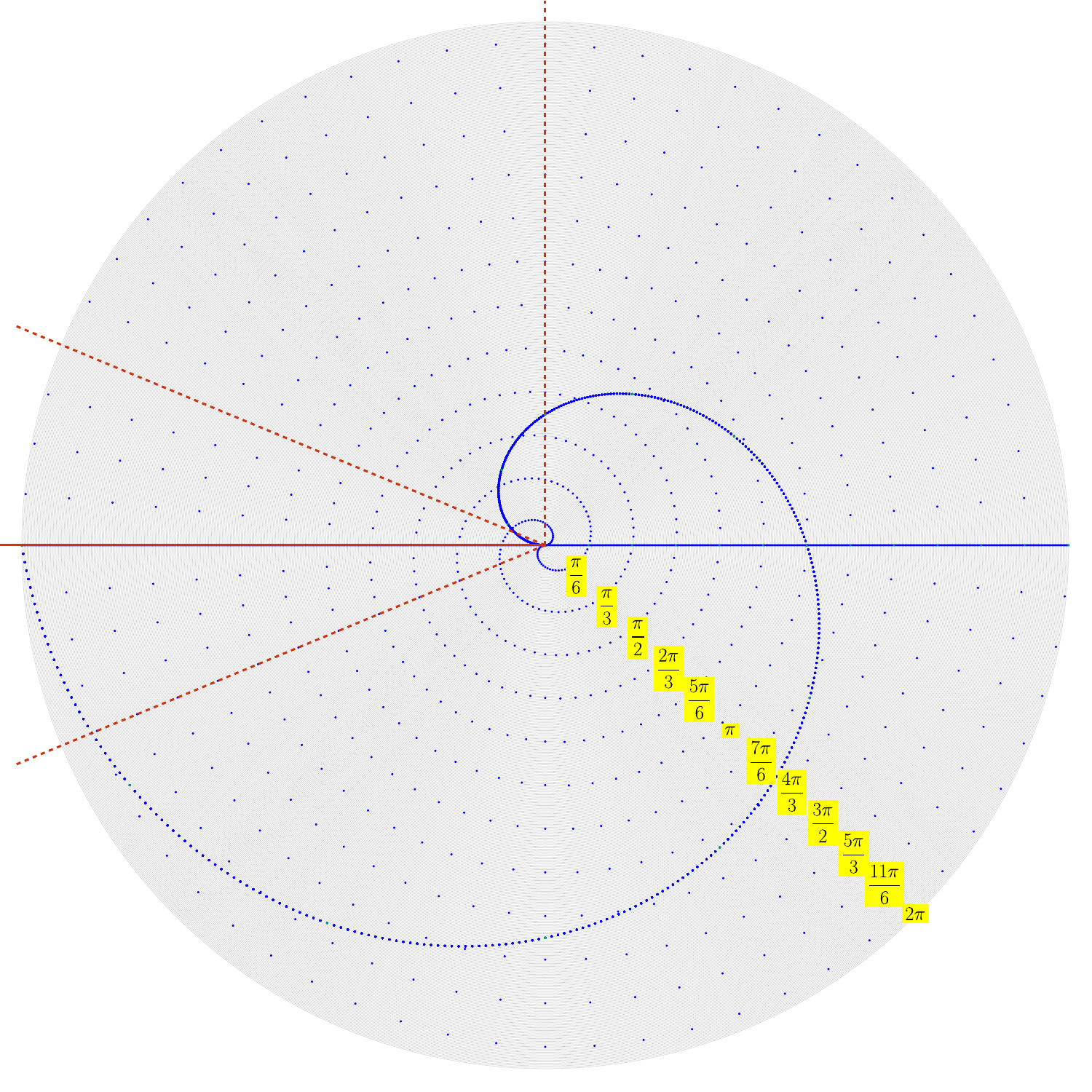}\hfill
\includegraphics[width=0.25\textwidth]{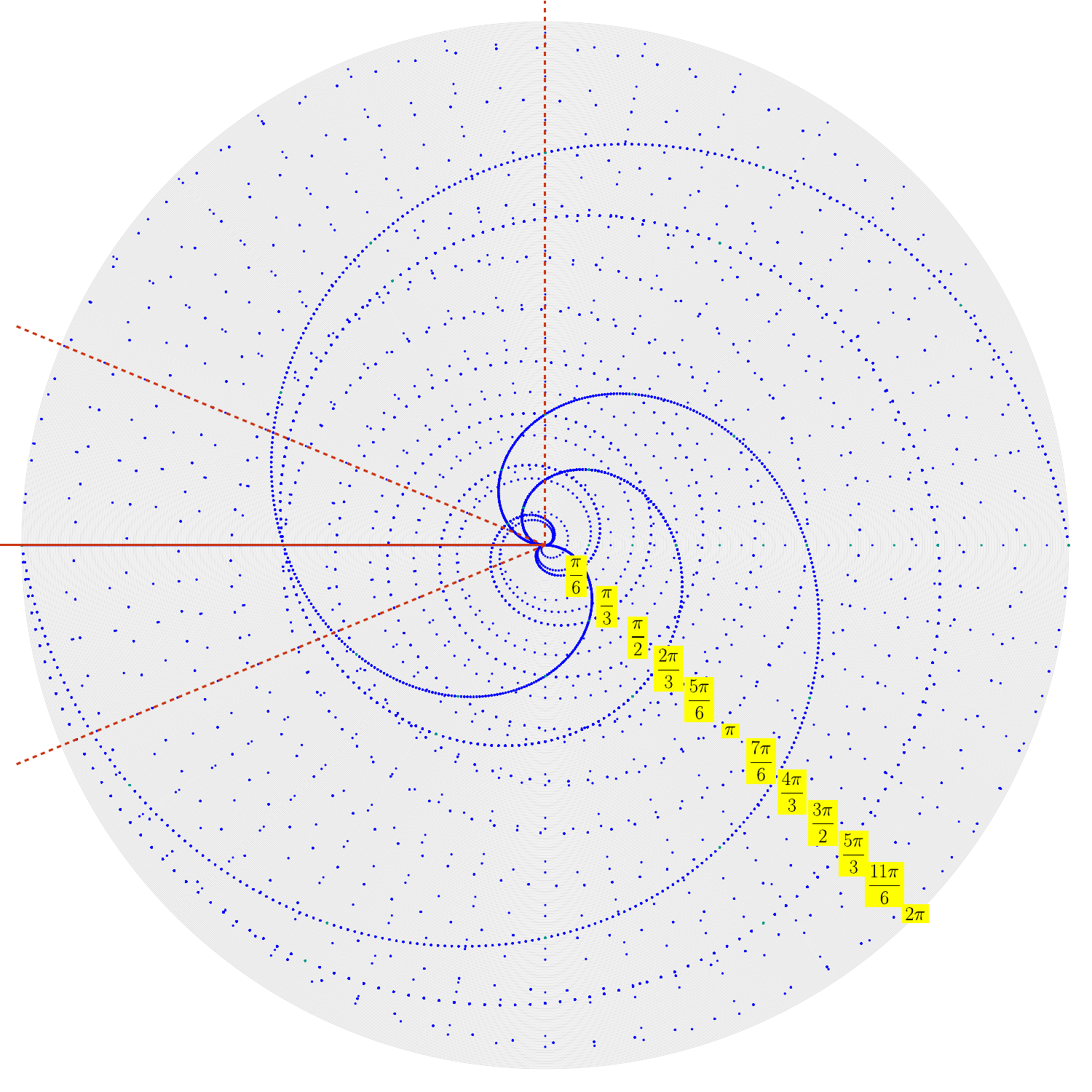}\hfill
\includegraphics[width=0.25\textwidth]{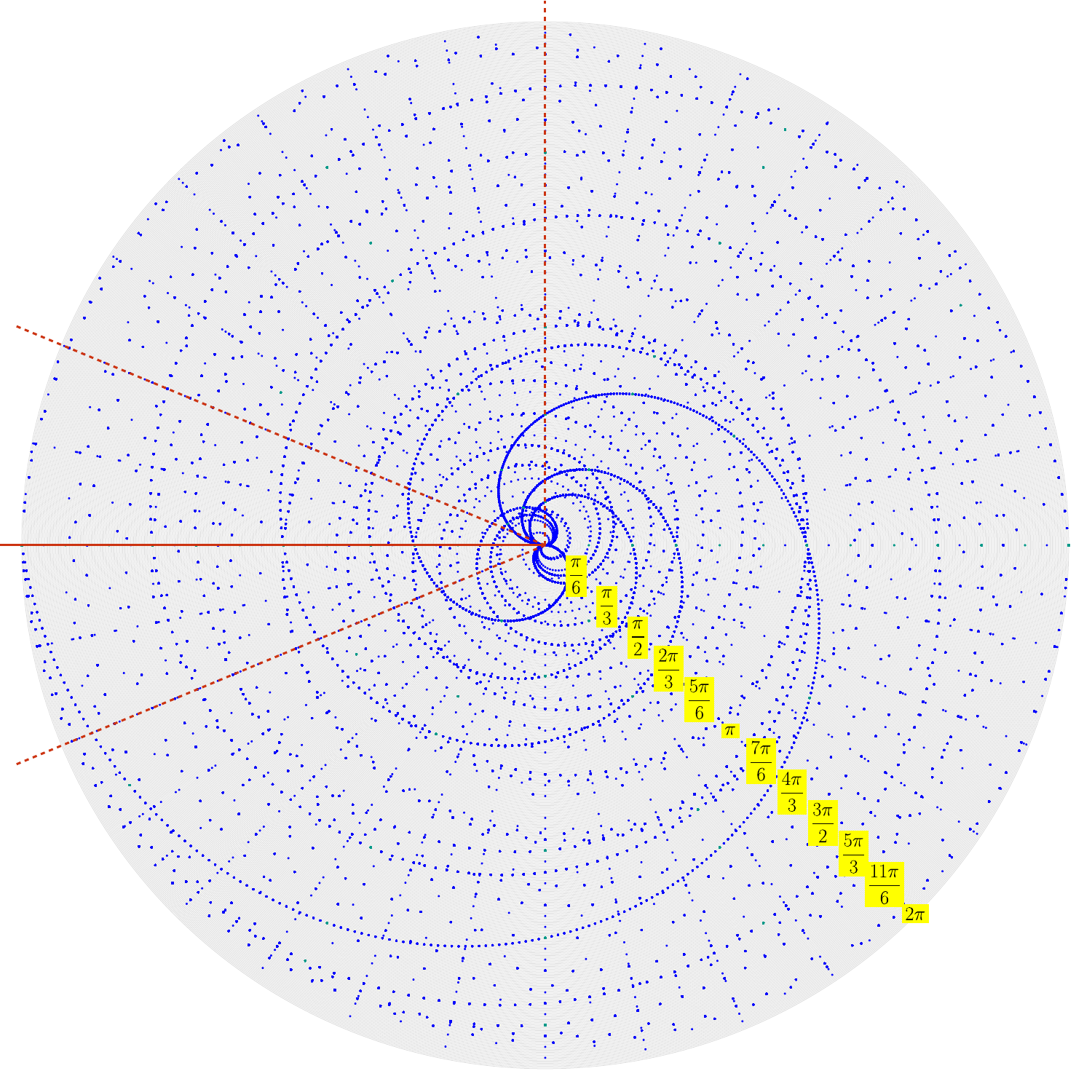}\hfill
\includegraphics[width=0.25\textwidth]{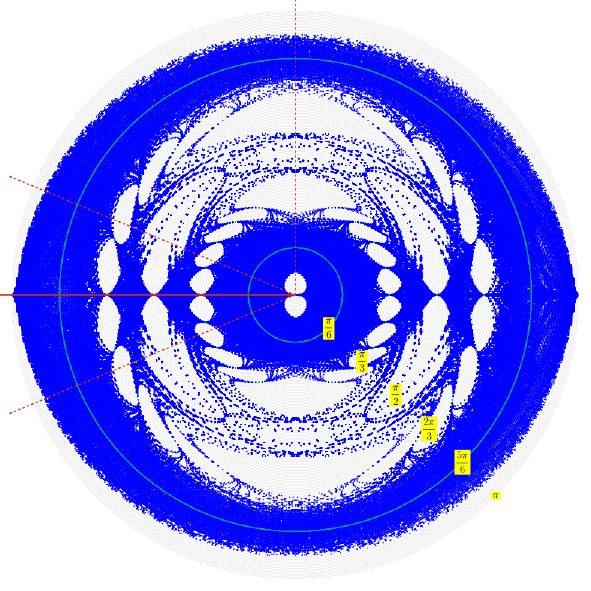}\hfill
 \caption{
 $U_F$ spectra of a $4\times 4$ cylinder in the complex plane for $\omega=0.5$ as a function of $\theta=2\pi\phi/N_{\text{pl}}$, for a total flux $\phi$ varying from $0$ (central plots) to $N_{\rm pl}=12$ (left) or $N_{\rm pl}/2=6$ (right) flux quanta (perimeters of plots). The four panels, from left to right, correspond to a single spin-flip ($S_z=N/2-1=7$) described by the tight-binding model of~\cite{Rudner2013}, two ($S_z=N/2-2=6$) and three ($S_z=N/2-3=5$) spin-flips and a half-filled ($S_z=0$) interacting hard-core boson system, respectively. Although all eigenvalues have a modulus 1, we represent them on radius-$\phi$ concentric circles to give a visual representation of the states flowing through the various chosen branch-cuts (represented as red lines at angles $\Phi_{\text{cut}} = \pi, \pi/2, 7\pi/8, -7\pi/8$). Values of $\theta$ multiple of $\pi/6$ corresponding to integer values of flux quanta threading the whole sample are indicated. Note that, for $N_p=1$ and $N_p=2$, there is a (flux independent) L-edge state at $U_F=+1$ (hidden on the plot by the red line representing the horizontal branch-cut). }
\label{fig:ComplexUfSpectra}
\end{figure*}

\begin{figure*}
	\centering
\includegraphics[width=0.25\textwidth]{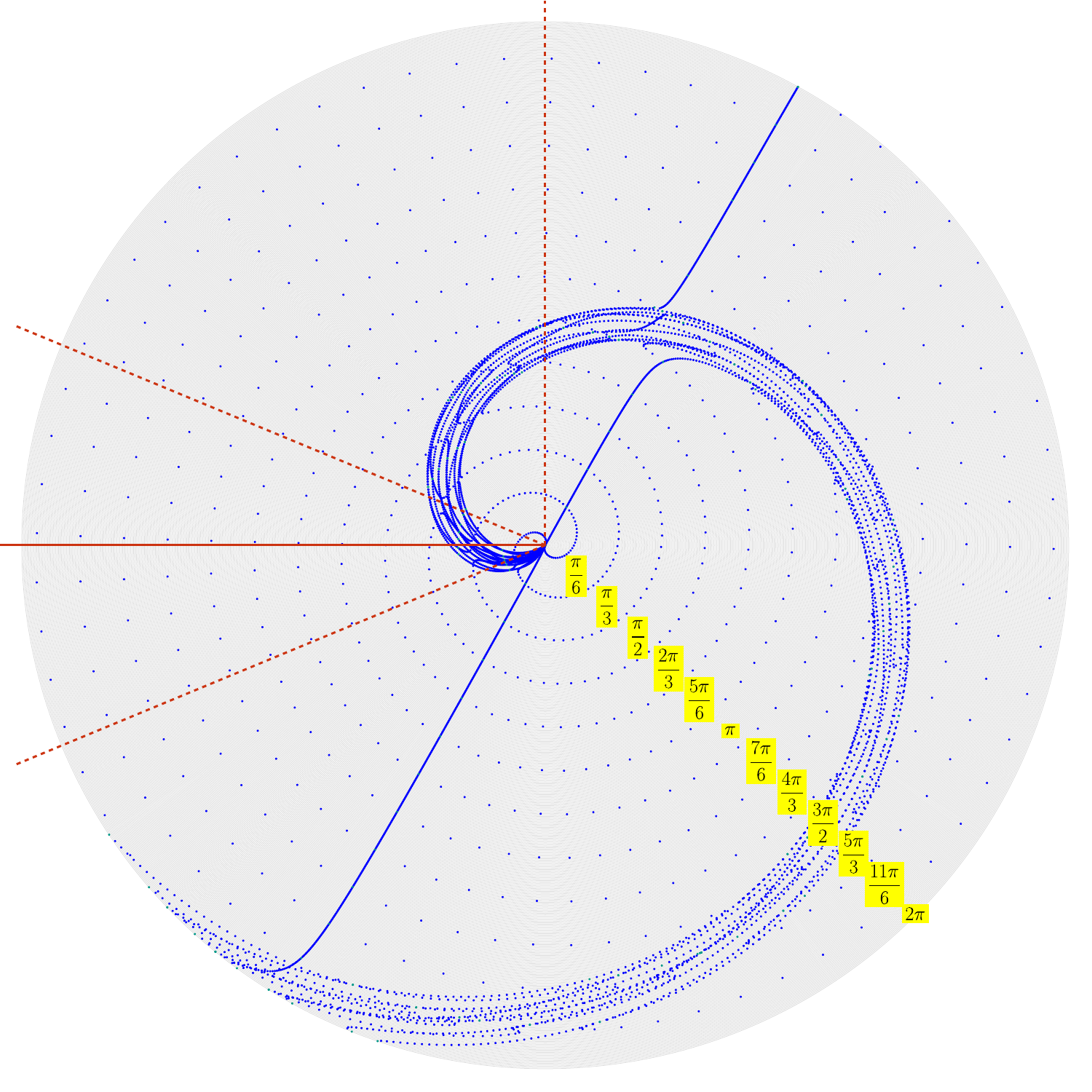}\hfill
\includegraphics[width=0.25\textwidth]{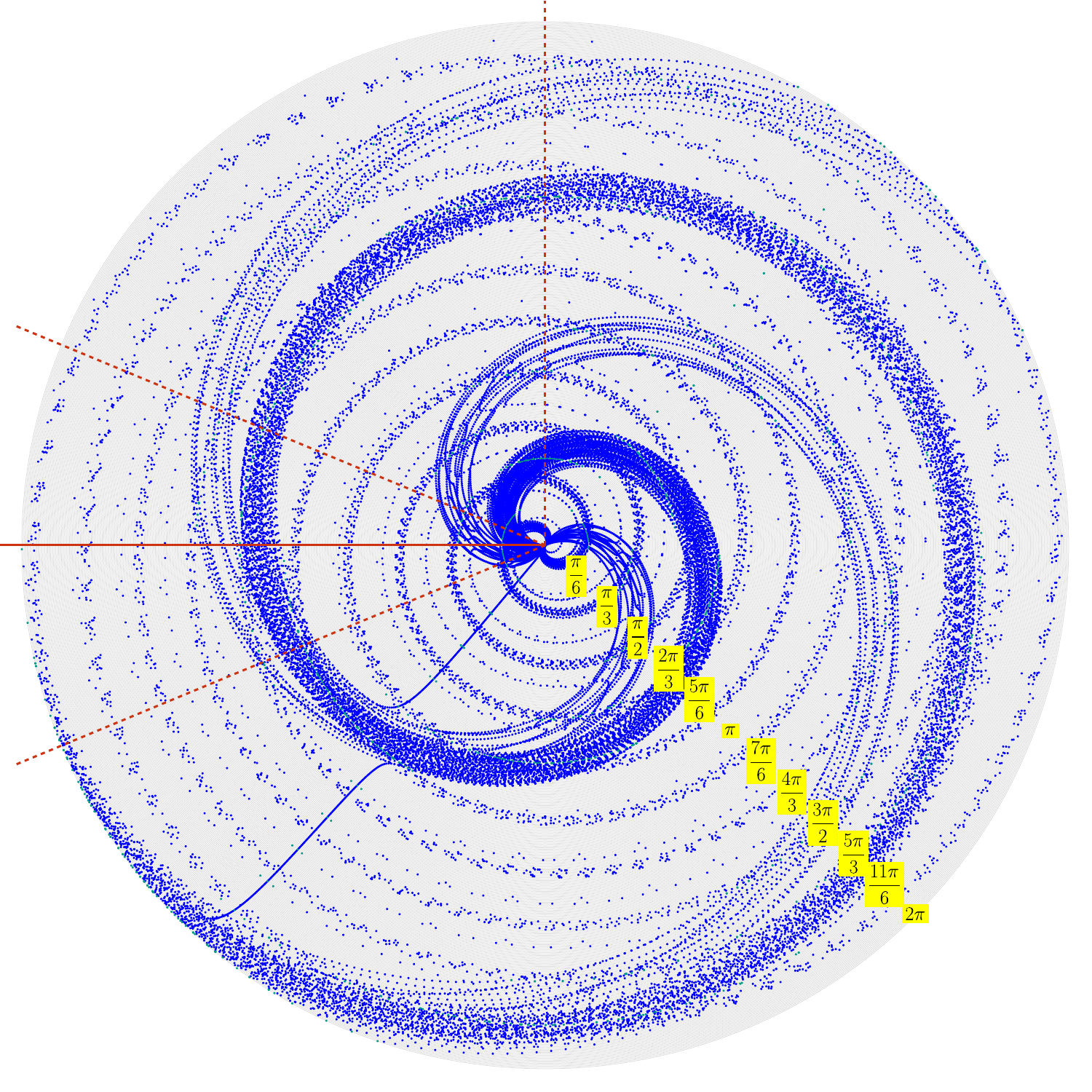}\hfill
\includegraphics[width=0.25\textwidth]{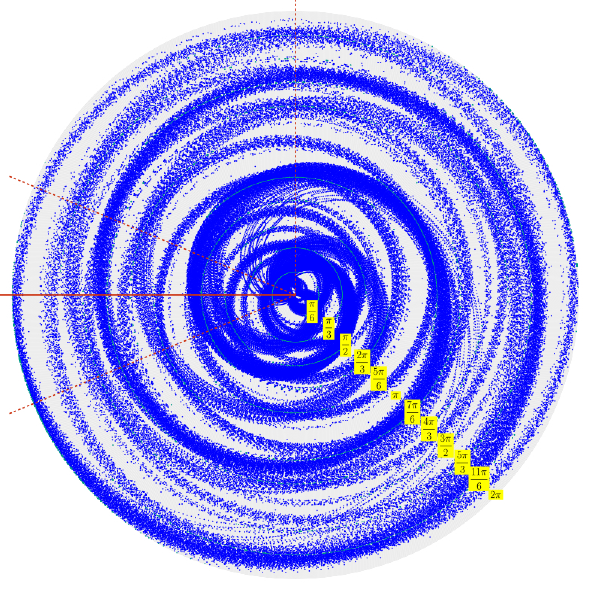}\hfill
\includegraphics[width=0.25\textwidth]{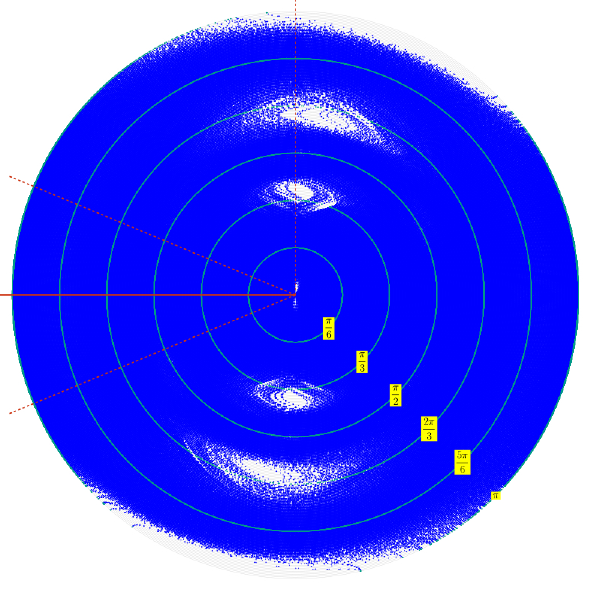}\hfill
 \caption{
 Same as Fig.~\ref{fig:ComplexUfSpectra} for a small detuning $\omega=0.53$.}
\label{fig:ComplexUfSpectra2}
\end{figure*}

As a simple test, we shall first consider the case of a single spin flip ($S_z=N/2 -1$) whose dynamics is described by the (non-interacting) single particle tight-binding model introduced in~\cite{Rudner2013}. The diamagnetic response of this model has been studied in~\cite{gavensky2025} for an open geometry in the plane. We show here that similar results are obtained in our setting on the cylinder. Two typical spectra of $U_F$ are shown in the complex plane in the left panel of figure~\ref{fig:ComplexUfSpectra} at the fine-tuned frequency and in the left panel of figure~\ref{fig:ComplexUfSpectra2} for a small detuning of $6\%$. At fine-tuning, we observe $N_{\rm pl}=12$ degenerate bulk states (localized on plaquettes) of eigenvalue $-\exp{(-i\theta)}$ and $L_x/2=2$ edge modes on both left (L) and right (R) edges. As expected the L edge modes (of eigenvalues $\pm 1$) do not show any flux dependence while the two R edge modes have eigenvalues $\pm\exp{(6i\theta)}$. In our representation of figure~\ref{fig:ComplexUfSpectra}
the bulk states and the R edge modes give two sets of spirals winding around the center of the plots in opposite directions (and with different pitch angles).
Note that the R edge states switch their eigenvalues upon increasing the flux $\phi$ by one unit (i.e. $\theta$  by $\pi/6$). The edge state contribution to $N_1$ in Eq.~(\ref{eq:N1}) 
is therefore a periodic function of the flux of period $\Delta\phi_{\rm edge}=1$, as also found in~\cite{gavensky2025} for a planar geometry. In contrast, the contribution from the bulk states is periodic of period $\Delta\phi_{\rm bulk}=N_{\rm pl}=12$. When a small detuning is introduced we find that the exact degeneracy of the bulk states is lifted and that bulk and edge modes get coupled.

\begin{figure*}
	\centering
\includegraphics[width=0.9\columnwidth]{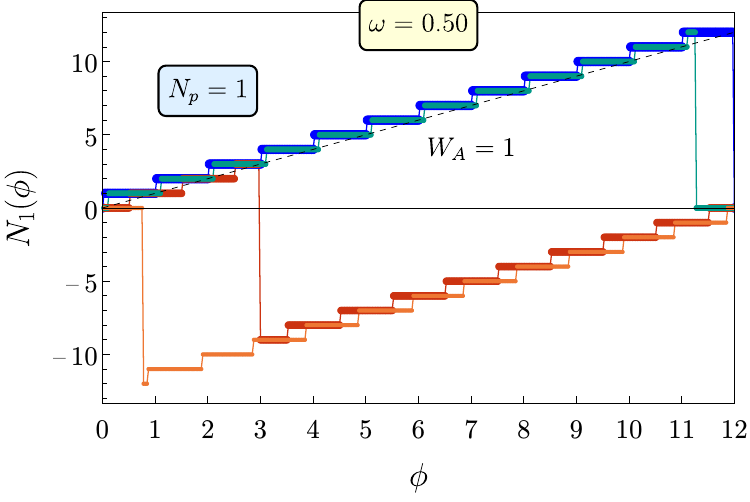} 
\includegraphics[width=0.9\columnwidth]{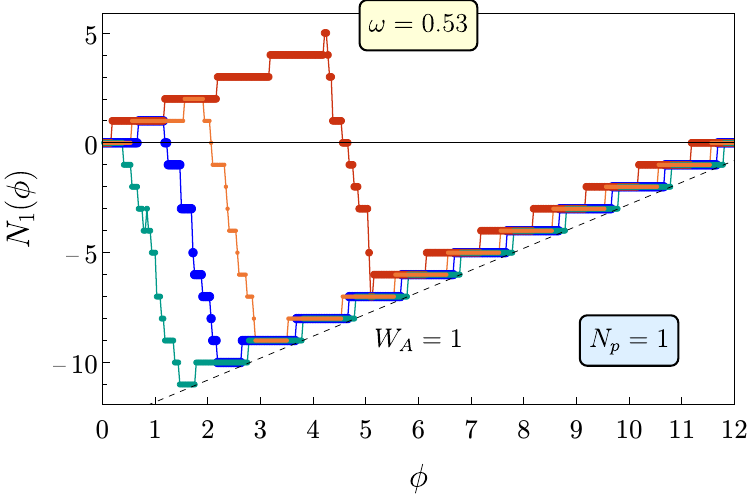} 
\includegraphics[width=0.9\columnwidth]{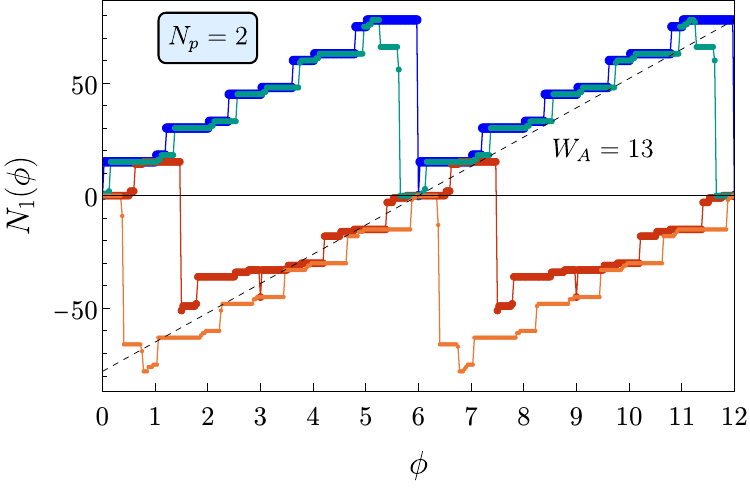}
\includegraphics[width=0.9\columnwidth]{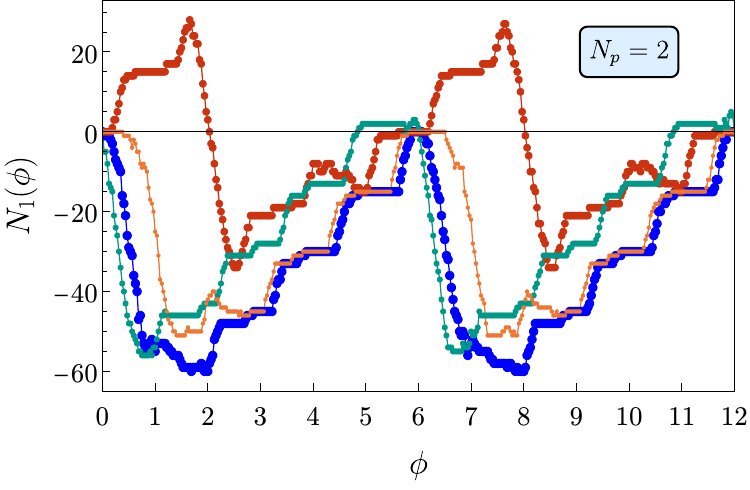}
\includegraphics[width=0.9\columnwidth]{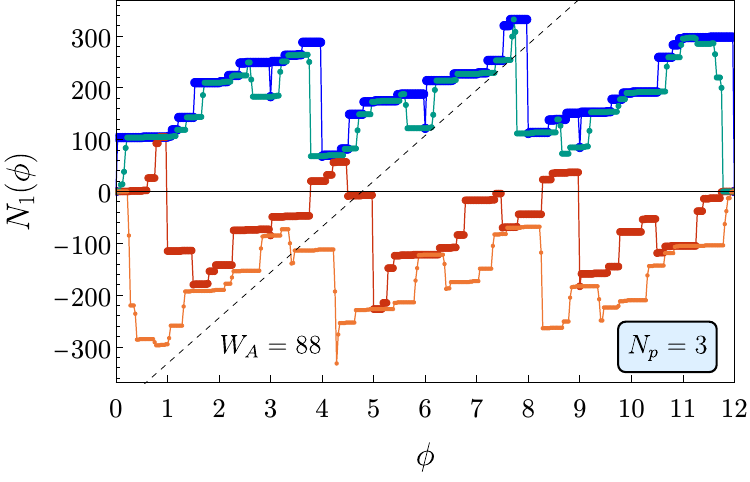} 
\includegraphics[width=0.9\columnwidth]{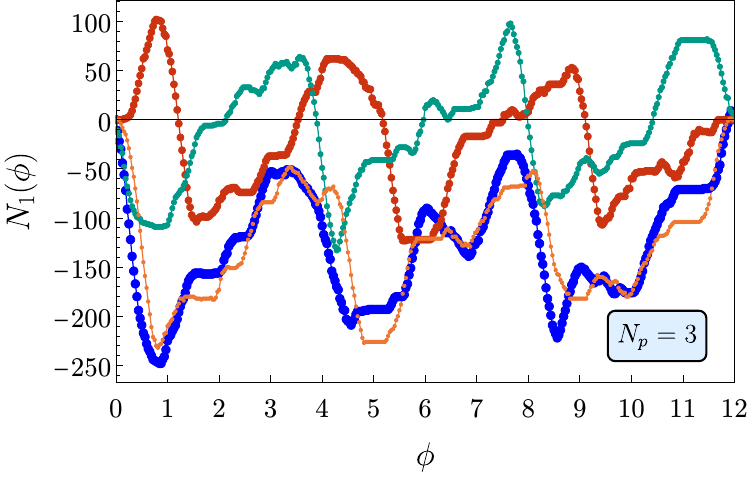} 
\includegraphics[width=0.9\columnwidth]{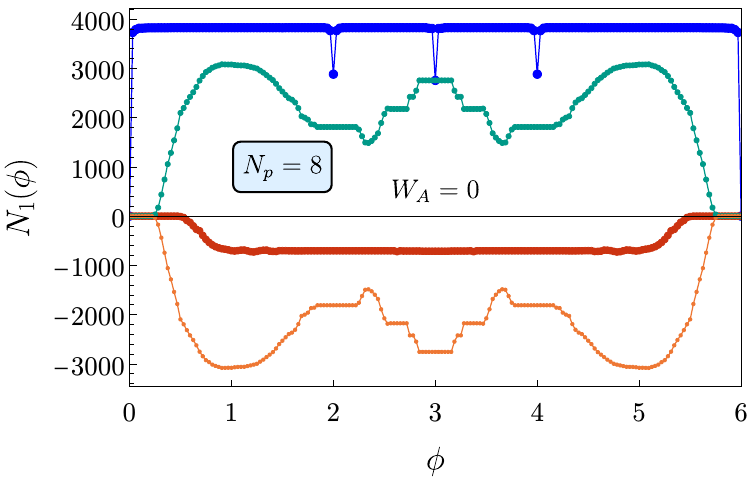}
\includegraphics[width=0.9\columnwidth]{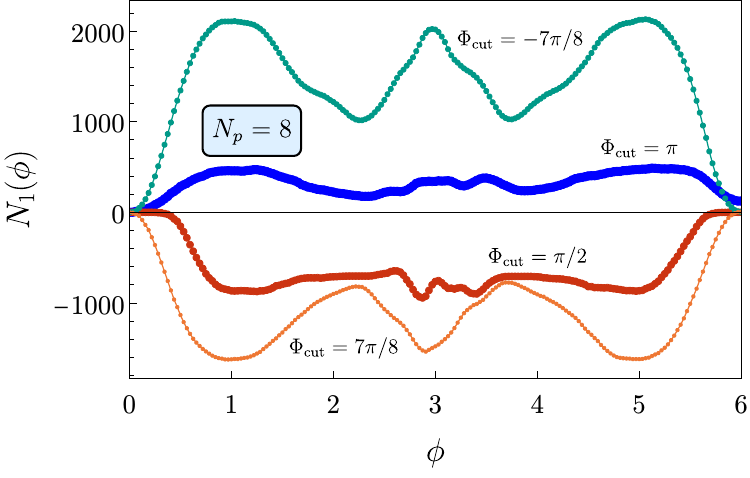}
 \caption{Winding number $N_1(\phi)$ versus $\phi$ for $N_p=1$ (top panels), $N_p=2$, $N_p=3$ (middle panels) and $N_p=N/2=8$ (bottom panels) hard-core bosons, in the flux intervals $[0,12]$ (top and middle panels) or $[0,6]$ (bottom panels), for $\omega=0.5$ (left panels) and $\omega=0.53$ (right panels).  Data obtained for the four different branch-cuts of figure~\ref{fig:ComplexUfSpectra} are shown using different colors as indicated on the bottom-right plot. To eliminate a trivial dependence of $N_1(0)$ with the choice of the branch-cut, all data have been shifted by $N_1(0)$ for clarity. Then, all data points displayed have integer values, even away from fine-tuning.
Estimations of the slope $W_A$ are shown on the plots as dashed lines. For $N_p=3$ the dashed line is only a guide to the eye. }
	\label{fig:N1_vsPhi}
\end{figure*}

In order to obtain the eigenvalues of $H_F$ from those of $U_F$ on the unit circle (taking the logarithm) one has to choose a branch cut (defining a particular quasi-energy first FBZ). Four different branch-cuts have been considered in figure~\ref{fig:ComplexUfSpectra}. The negative horizontal axis (red line) represents the branch-cut in the complex plane defining the familiar first FBZ $]-\omega/2,\omega/2]$ used throughout this work. Three other branch cuts at $\Phi_{\rm cut}=\pi/2$ and $\Phi_{\rm cut}=\pm 7\pi/8$  angles from the horizontal axis (in the anticlockwise direction) are also shown by dashed red lines. We have computed $N_1(\phi)$ from the trace formula (\ref{eq:N1}) for different choices of the branch-cuts, for 
the fine-tuned frequency and for a small detuning $\omega=0.53$. A  close look of the results shown in figure~\ref{fig:N1_vsPhi} reveals that $N_1$ versus $\phi$ has a staircase behavior. Note that $N_1(\omega,\phi)$ is quantized (i.e it is an integer) at $\omega=0.5$ for all $\phi$, while it is not away from fine-tuning. However, the variation of $N_1(\omega,\phi)$ versus $\phi$ always occurs by integer jumps, physically representing the number of states crossing the branch-cut at a particular value of the flux.
We also see that, strictly speaking, the choice of the branch-cut matters, with jumps in $N_1(\phi)$ at different locations. However, the slope (\ref{eq:awn}) is independent of this choice. We find $W_A=+1$ in agreement with~\cite{gavensky2025} for a planar geometry. The same (coarse grained) slope is also found at small detuning but the jumps associated to the bulk states crossing the branch-cuts are no longer abrupt. We note that choosing $S_z=-(N/2-1)$, also corresponding to a unique spin-flip, gives, for the same orientation of the physical flux, a reverse winding orientation of the spirals in figure~\ref{fig:ComplexUfSpectra}. Therefore, we obtain $W_A=-1$ in that case. 

Next, we turn to the case of two and three spin flips, i.e. $S_z=N/2-2=6$ and $S_z=N/2-3=5$ respectively, which can be viewed as $N_p=2$ and $N_p=3$ interacting hard-core bosons in our finite cylinder. For completeness we also consider the $S_z=0$ sector, as in the previous parts of this work, which corresponds to an interacting half-filled hard-core bosonic system ($N_p=8$ bosons on $N=16$ sites). 
In contrast to the single particle case, 
the nearest-neighbor Ising coupling $S_i^z S_j^z$ (corresponding to a repulsion between two neighboring bosons) becomes relevant and plays a key role, even at fine-tuned frequency $\omega_s=0.5$
for which the tensor product structure (\ref{eq:UF_omegas}) of $U_F(\omega_s)$ breaks down.

Results for the spectra of the many-body unitary operator $U_F(\omega,\phi)$ at 
the fine-tuned frequency $\omega=0.5$ are shown versus the flux $\phi$ in the three panels on the right of figure~\ref{fig:ComplexUfSpectra}. For $S_z=0$ we restrict here to $\phi\in[0,6]$ since the spectra are symmetric under $\phi\rightarrow N_{\rm pl}-\phi$ ($N_{\rm pl}=12$). Since we have now many-body systems whose Hilbert space size scales exponentially with system size, the density of points becomes larger than in the one-particle case, especially for $S_z=0$ for which it is now impossible to follow by eye the individual state trajectories. 
However, for the case of two and three particles, one can still identify (5 and 9) degenerate edge modes forming spirals winding counterclockwise  and (3 and 4) degenerate ''bulk modes" winding in the opposite direction. The degeneracy of these modes indicates some degree of localization in the bulk and/or on the boundary. Note that, for $N_p=2$, there is an additional sixth (non-degenerate) edge mode of (flux-independent) eigenvalue $U_F=+1$ corresponding to the two particles localized on the two $A$ sites of the L edge.
For $S_z=0$ (half-filled bosonic system), the density of points is very high (forming a quasi-continuum) and numerous gaps (the white regions) are observed in the spectrum (right panel). The two white blobs at the center of the figure corresponds to the $\phi=0$ gap (separating the two degenerate sub-bands at quasi-energies $0$ and $\pm\omega/2$) which survives up to a small flux value $\phi\sim 0.45$. The gaps appearing at $\phi>1$ are due to lattice commensurability effects, like in the Hofstadter tight-binding model, made here relevant by the fact that (most) bulk states become spatially extended, in contrast to the single particle case.

Similar spectra are shown in the three right panels of figure~\ref{fig:ComplexUfSpectra2} for the same 
detuned frequency $\omega=0.53$ as for the single particle case (left panel). 
The detuning, although quite small, has a strong effect on the spectrum: the degeneracies of the edge and bulk modes are immediately lifted and edge and bulk states couple to each other. 
At $S_z=0$, gaps are rapidly reduced/washed out, as seen on the far-right panel. Note that, in this case, we observe that a small number ($=346$) of states are flux independent with $U_F$ eigenvalues $\pm 1$ (i.e. on the horizontal axis of figure~\ref{fig:ComplexUfSpectra2}). We believe that they correspond to a (small) subset of the edge states strictly localized on the left boundary. 

As for the single particle case, we have computed $N_1(\phi)$ from the trace formula (\ref{eq:N1}) for the same choices of the branch-cuts and for the same frequencies $\omega=0.5$ and $\omega=0.53$.  
A close look of the results shown in figure~\ref{fig:N1_vsPhi} reveals that $N_1$ versus $\phi$ has always a staircase behavior with integer jumps as expected ($N_1$ itself is quantized for $\omega=0.5$ for all particle number $N_p$).  
The behavior of $N_1(\phi)$ for $N_p=2$ is qualitatively similar to the single particle case with well-defined linear portions from which it is easy to extract the anomalous spectral flow $W_A=13$. The case $N_p=3$ gives a larger value around $W_A=88$ but a precise determination is no longer possible. 

The $S_z=0$ ($N_p=N/2$) case bears some special features and is qualitatively different from the previous cases. First, we see that the density of integer jumps is so high that $N_1(\phi)$ looks now like a continuous function. We also note that, when the branch-cut is chosen as the negative horizontal axis (continuous red line), we have now to pay attention to the fact that some states are located exactly on this axis (for $\omega=0.5$). Our ``traditional'' choice of the first FBZ $]-\omega/2,\omega/2]$ corresponds in fact to the branch-cut lying just {\it below} the negative horizontal axis. We observe that $N_1(\phi)$ has generically some flat portions in the center of the plots, especially for the two branch cuts corresponding to the horizontal and vertical semi-infinite axis. This suggests that $W_A=0$ and we argue that this is expected for a half-filled hardcore bosonic systems: A particle-hole transformation in the bosonic language corresponds to a $S_z\rightarrow -S_z$ transformation in the spin language. As explicitly seen above in the case of a single spin-flip, a particle-hole transformation leads to $W_A\rightarrow -W_A$. Since, the $S_z=0$ sector is particle-hole symmetric, we then expect $W_A=0$. 

Finally, we stress that the effect of detuning is much more severe in the interacting case than for a single particle. For $N_p\ge 2$ 
the linear behaviors of $N_1(\phi)$ are quickly altered by detuning, even very small, as seen in the right panels of figure~\ref{fig:N1_vsPhi}. This shows that the Streda response is a very sensitive tool to probe the stability of anomalous interacting topological phases. 

\section{Interpolation to the high-frequency regime}
\label{sec:interpol}

\subsection{High-frequency limit}

We shall here investigate the connection between the anomalous CSL realized in the Swap models  and the (dynamical) CSL realized in some simple setup at intermediate and high frequencies. We shall restrict ourselves here to SU(2)-invariant Heisenberg couplings ($\Delta=1$) for which the properties of the high-frequency (static) Floquet Hamiltonian have been investigated previously. It was shown that the dynamical CSL is smoothly connected (upon increasing the frequency) to a (Abelian) SU(2)-symmetric spin-1/2 CSL~\cite{Poilblanc2015} hosted in some parameter range of a static chiral Heisenberg model~\cite{Poilblanc2017b}.

We assume that the time-dependent Hamiltonian $H(t)$ may now include an additional static homogeneous Heisenberg term acting on all bonds, in addition to the periodic drive. Namely,  $H(t)=H_0 + H_{\rm drive}(t)$ where $H_0=J_\omega \sum_{\langle ij\rangle} {\bf S}_i\cdot {\bf S}_{j}$ and the amplitude $J_\omega$ will be adjusted, depending on the drive frequency. 

In the following we perform exact numerical calculations of the Floquet unitary $\exp{(-iH_F(\omega)T)}={\cal T}_{t}\exp{(-i\int_{0}^{T} H(t) dt)}$ on a $4\times 4$ torus/cylinder and on a $2\times 8$ ribbon. As in (\ref{eq:UnitaryGateProduct}) the 
Floquet unitary can be written as a product of four (non-commuting) unitaries ${\tilde U}_\alpha$, $\alpha=a,b,c,d$, but each ${\tilde U}_\alpha=\exp{(-i(H_0+H_\alpha)\frac{T}{4})}$ can no longer be decomposed in terms of 2-site gates. It should be noticed that, to diagonalize the Floquet unitary, in addition to translation symmetry we have also used the $S_z\rightarrow -S_z$ symmetry, hence separating between integer even ($S=0,2,\cdots$) and odd ($S=1,3,\cdots$) total spin sectors. Finally, note that, in the case of a sinusoidal modulation to which we compare at high frequency, a Trotter-Suzuki decomposition of the unitary operator becomes necessary~\cite{Mambrini2024,Poilblanc2024}.

We show now that the Floquet Hamiltonian in the high-frequency limit is in fact identical in first-order in $1/\omega$ to the one obtained for a monochromatic drive studied previously~\cite{Mambrini2024,Poilblanc2024}.
The effective Hamiltonian of a generic periodic sequence involving $M$ discrete steps, 
\begin{equation}
U (T) = e^{-i (T/M)  H_M} \dots e^{-i (T/M)  H_2} e^{-i (T/M)  H_1}
\end{equation}
has been derived in the high-frequency limit in Ref.~\cite{Goldman_Dalibard}. The lowest-order effective Hamiltonian reads
$H_{\rm eff}= H_{\rm eff}^{(0)} + H_{\rm eff}^{(1)} + O(1/\omega^2)$, where
the zeroth order is the time-average of the time-dependant Hamiltonian over one period,
\begin{equation}
    H_{\rm eff}^{(0)}= \langle H(t)\rangle_T =  J_1 \sum_{\langle ij\rangle} {\bf S}_i\cdot {\bf S}_{j}\, ,
\label{eq:JF0}
\end{equation}
where $J_1=J_\omega + J/4$. For a sequence involving $M\!=\!4$ steps, as in the Swap model, the 1st order term reads
\begin{equation}
 H_{\rm eff}^{(1)} = \frac{i \pi}{32 \omega} \left ( [ H_1,  H_2] + [ H_2,  H_3] + [ H_3,  H_4] + [ H_4,  H_1]  \right )\, .
\end{equation}
Introducing the operators
\begin{equation}
 H_x =  H_1 -  H_3 , \quad  H_y =  H_2 -  H_4 ,
\end{equation}
one obtains a simple expression
\begin{align}
 H_{{\rm eff}}^{(1)} &= \frac{i \pi}{32 \omega} [ H_x ,  H_y]
 \label{eq:H1}
\end{align}
Considering the operators of the swap sequence, $ H_{1,2,3,4}\!=\! H_0 + H_{a,b,c,d}$, we obtain static (staggered) Heisenberg couplings 
   $H_x=H_a-H_c$ and $H_y=H_b-H_d$ acting on the vertical and horizontal nearest-neighbor bonds, respectively. One can then relate the commutator $[ H_x ,  H_y]$ to the chiral operators~\cite{Mambrini2024},
\begin{equation}
[ H_x ,  H_y] = \frac{J^2}{2} \sum_{\square} ( P_{i,j,k,l} -  P_{i,j,k,l}^{-1}),
\end{equation}
where $P_{ijkl}$ is the 4-site cyclic permutation which applies to all plaquettes $\square$.
We then readily obtain the expression for the effective Hamiltonian in the high-frequency limit,
  \begin{equation}
   H_{\rm eff}^{(1)}= J_F \, i\sum_\square (P_{ijkl} - P_{ijkl}^{-1}),  
   \label{eq:JF1}
   \end{equation}
   where 
   \begin{equation}
         J_F=\frac{\pi J^2}{64\omega}   
         \end{equation}
 is the emerging Floquet energy scale.
We note that the next-order correction can also be calculated explicitly using the expressions provided in Ref.~\cite{Goldman_Dalibard}.

In order to stabilize a CSL in the high-frequency regime, the two terms (\ref{eq:JF0}) and (\ref{eq:JF1}) of $H_{\rm eff}$ should be of the same order~\cite{Poilblanc2017b}, i.e. of order $J_F\ll J$. Therefore we choose a static ferromagnetic $J_\omega<0$ coupling to partially cancel the average antiferromagnetic coupling $J/4$ originating from the drive i.e. $J_\omega=-J/4 + \lambda_\omega J_F(\omega)$, where $\lambda_\omega$ is of order $1$ for large frequency. In fact, from previous work we know that the Floquet Hamiltonian in the $\omega\rightarrow\infty$ limit, 
\begin{equation}
      H_{\rm eff}=\lambda_\omega J_F \sum_{\langle ij\rangle} {\bf S}_i\cdot {\bf S}_{j}+ J_F \, i\sum_\square (P_{ijkl} - P_{ijkl}^{-1})\, ,
      \label{eq:fullFloquet}
      \end{equation}
hosts a CSL phase for $\lambda_\omega$ around $2$~\cite{Hasik2022}. Note that, in this limit, $J_F$ is just an overall (vanishing) energy scale. 

Note that a Magnus expansion of $H_F[t_0]$ can also be performed (see Appendix~\ref{app:high}). 
Limiting ourselves to a particular regime where the $1/\omega$ contribution, of order $J J_1 / \omega$, actually becomes a $1/\omega^2$ contribution due to the scaling $J_1 \sim 1/\omega$, we obtain the same expression as Eq.~(\ref{eq:fullFloquet}), with no $t_0$-dependence at leading order. 

We have compared the square drive used in this study to the sinusoidal one used in previous studies~\cite{Mambrini2024,Poilblanc2024}. In the limit of large frequency, both lead to the same effective Hamiltonian (\ref{eq:fullFloquet}) (but with different prefactors $J_F$) provided the homogeneous Heisenberg coupling of amplitudes $-J/4+\lambda_\omega J_F$ and $\lambda_\omega J_F$ added to the square and sinusoidal drives, respectively, have the same value of $\lambda_\omega$. 
Comparisons shown in Appendix~\ref{app:high} reveal a perfect agreement between the high-frequency Floquet spectra and the spectrum of the effective Hamiltonian (\ref{eq:fullFloquet}).

\begin{figure}
	\centering
    \includegraphics[width=0.9\columnwidth]{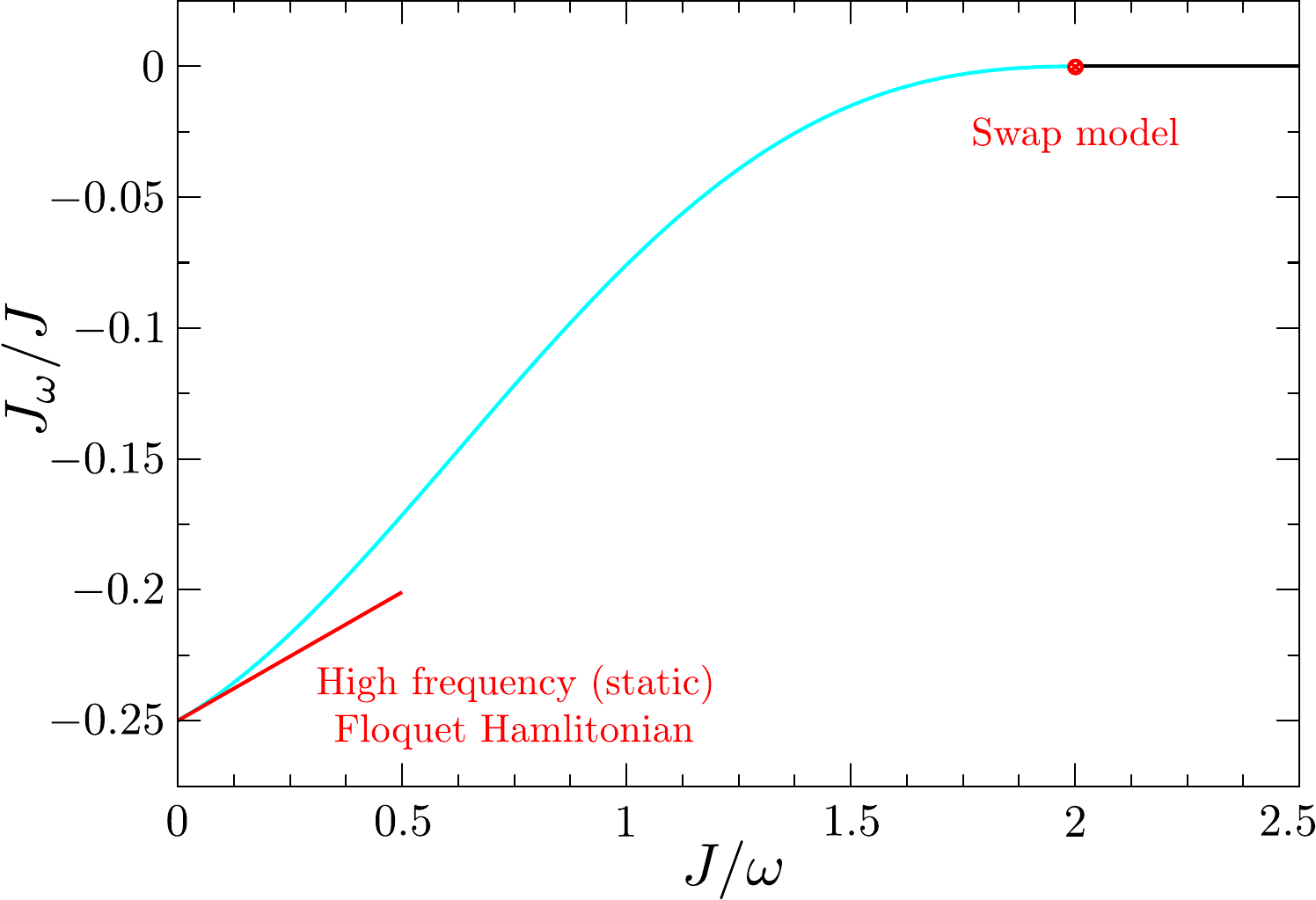}      
           \caption{Relative amplitude $J_\omega/J$ of the time-independent nearest-neighbor Heisenberg coupling, added to the periodic drive, as a function of $J/\omega$. A smooth interpolation between the $\omega\rightarrow\infty$ limit ($J_\omega/J=-\frac{1}{4}+\frac{\pi J}{32\omega}$, red line) and the Swap model at $J/\omega=2$ ($J_\omega=0$, red point) is realized.}
	\label{fig:Interpol}
\end{figure} 

It is interesting to compute the average-energy spectrum (\ref{eq:AverEner}) in this limit, showing a very different behavior than in the vicinity of the tuned Swap model. Since the leading order of the Floquet hamiltonian $H_F[t_0]$ given by (\ref{eq:fullFloquet}) does not depend on $t_0$, the corresponding Floquet eigenstates $|\psi_n(t)\rangle$ do not depend on $t$ (up to small $1/\omega$ time-dependent corrections). Therefore, at high-frequency, the average-energy levels are given by,
\begin{equation}
      {\text{\ae}}_{n}=\lambda_\omega J_F(\omega) \langle h_{\rm AF} \rangle_n  + O(1/\omega^2)\, ,
      \label{eq:AverEner2}
\end{equation}
where $h_{\rm AF}$ is the sum of nearest-neighbor Heisenberg couplings appearing in the first term of (\ref{eq:fullFloquet}) and $\langle\cdots\rangle_n$ is the average in the $|\psi_n(0)\rangle$ Floquet states. This is consistent with the fact that $H_{abcd}\rightarrow\lambda_\omega J_F(\omega) h_{\rm AF}$ in the large-$\omega$ limit.
The variation of each energy level with decreasing $\omega$ is smooth as checked numerically.  In that limit it is also easy to compute the leading term of the (many-body) geometric Berry phases~\cite{Schindler2024},
\begin{eqnarray}
    \Phi_n &=& T({\cal E}_n - {\text{\ae}}_n)\hspace{0.2cm} {\rm mod}(2\pi) \\
    &=&  \pi\,  \frac{\pi J^2}{16 \omega^2}  \langle h_{\rm chiral} \rangle_n + O(1/\omega^3) \, .
   \label{eq:GeoPhases2}
\end{eqnarray}
where $h_{\rm chiral}$ is the sum over the $N$ (pure-imaginary) chiral plaquette couplings appearing in the second term of (\ref{eq:fullFloquet}), taking $\lambda_\omega=\lambda_\infty=2$. Note that $\Phi_n$ is of order $1/\omega^2$ and not $1/\omega$ since the leading Floquet Hamiltonian is of order $1/\omega$ and does not have a constant term. 
For system size $N$ small compared to $\sim\frac{16}{\pi}\omega^2/J^2$, the Berry phases (in absolute value) remain smaller than $\pi$.

\subsection{Interpolation}

We would like now to interpolate smoothly our Floquet drive model between large and low frequencies, using the fact that $J_\omega$ is a free parameter of the model. At high-frequency, we enforce that $\lambda_\omega= 2 +{\cal O}(1/\omega)$ i.e. $J_\omega=-J/4 + \pi J^2/32\omega + {\cal O}(1/\omega^2)$. Decreasing the frequency, we gradually turn off the (ferromagnetic) $|J_\omega|$ amplitude to smoothly reach $J_\omega=0$ for the Swap model at $\omega/J=1/2$, i.e. as $J_\omega\propto -(\omega-J/2)^2$. 
A simple cosine interpolation fulfilling these requirements
\begin{equation}
    J_\omega/J= \left (-1/4 + \frac{\pi J}{32\omega} \right ) \left (1+\cos{\frac{\pi J}{2\omega}} \right )/2 \, ,
    \end{equation}
is shown in figure~\ref{fig:Interpol}. For lower frequency $\omega\le J/2$ we keep $J_{\omega}=0$. 

\begin{figure}
	\centering
    \includegraphics[width=0.98\columnwidth]{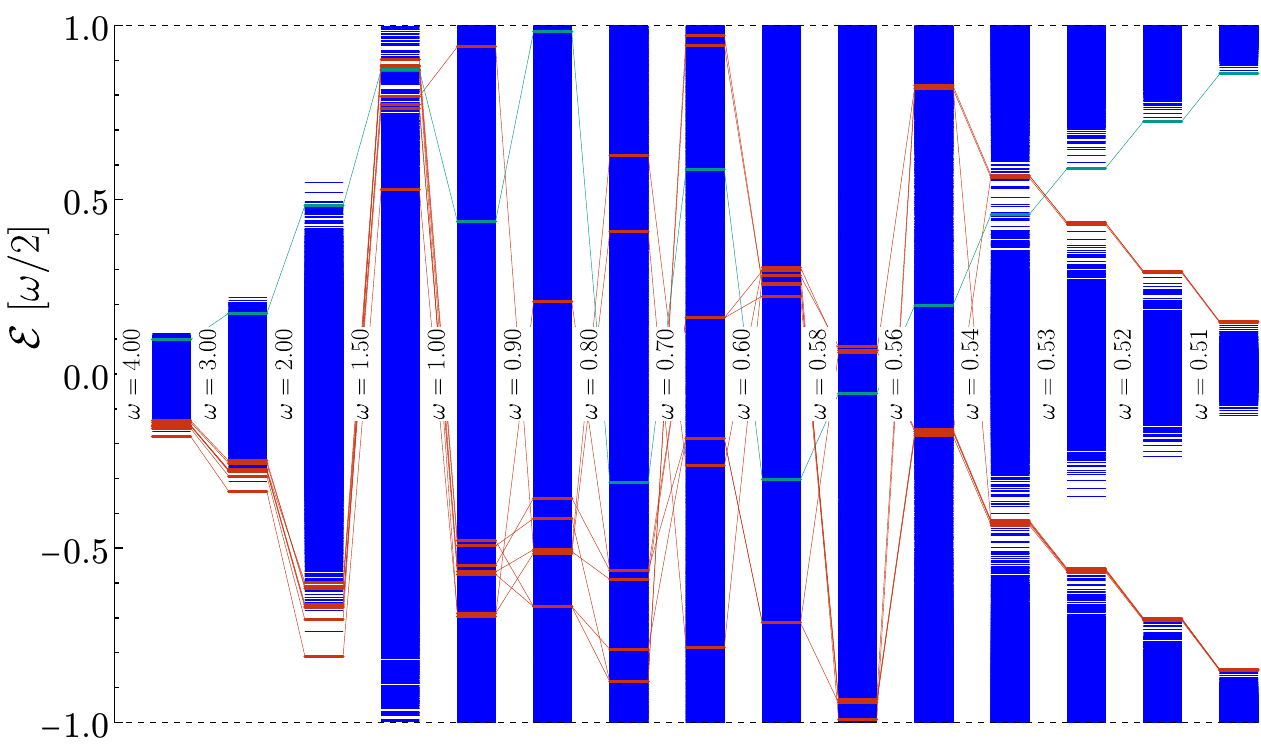}
    \hfill \includegraphics[width=0.98\columnwidth]{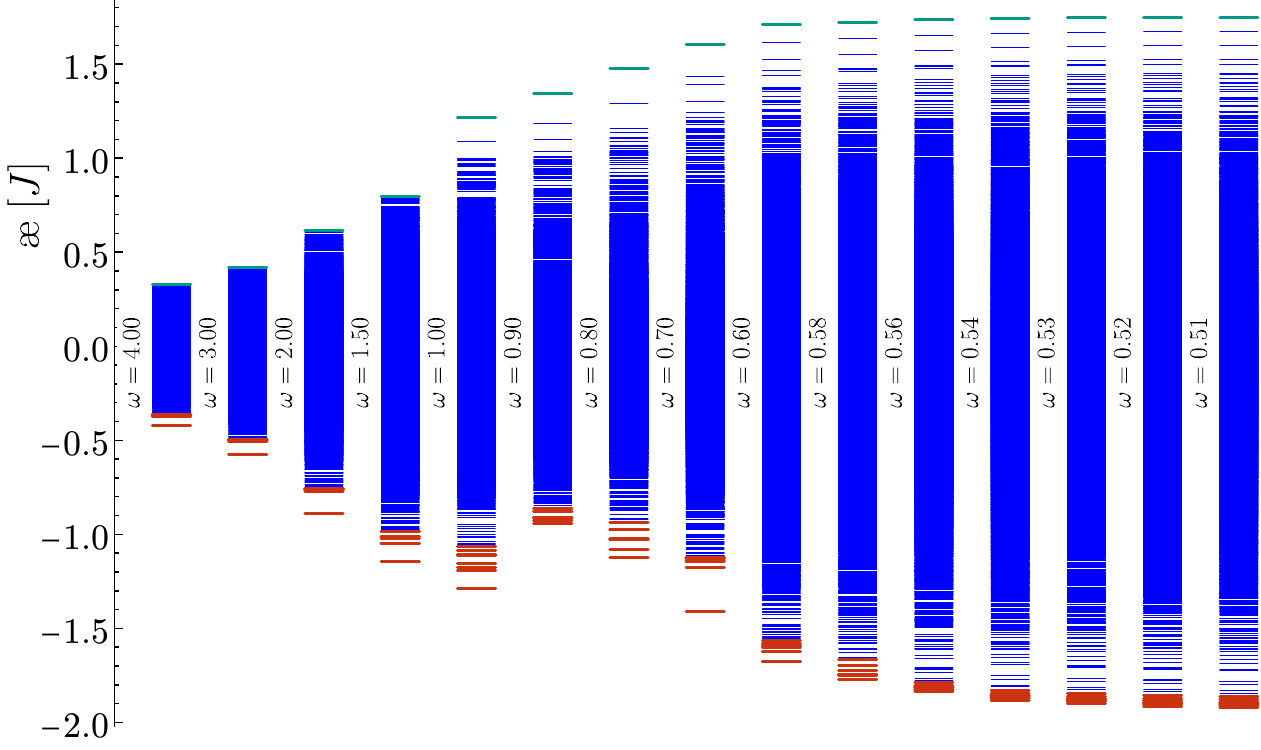}
 	\hspace*{-0.01\columnwidth}\includegraphics[width=0.98\columnwidth]{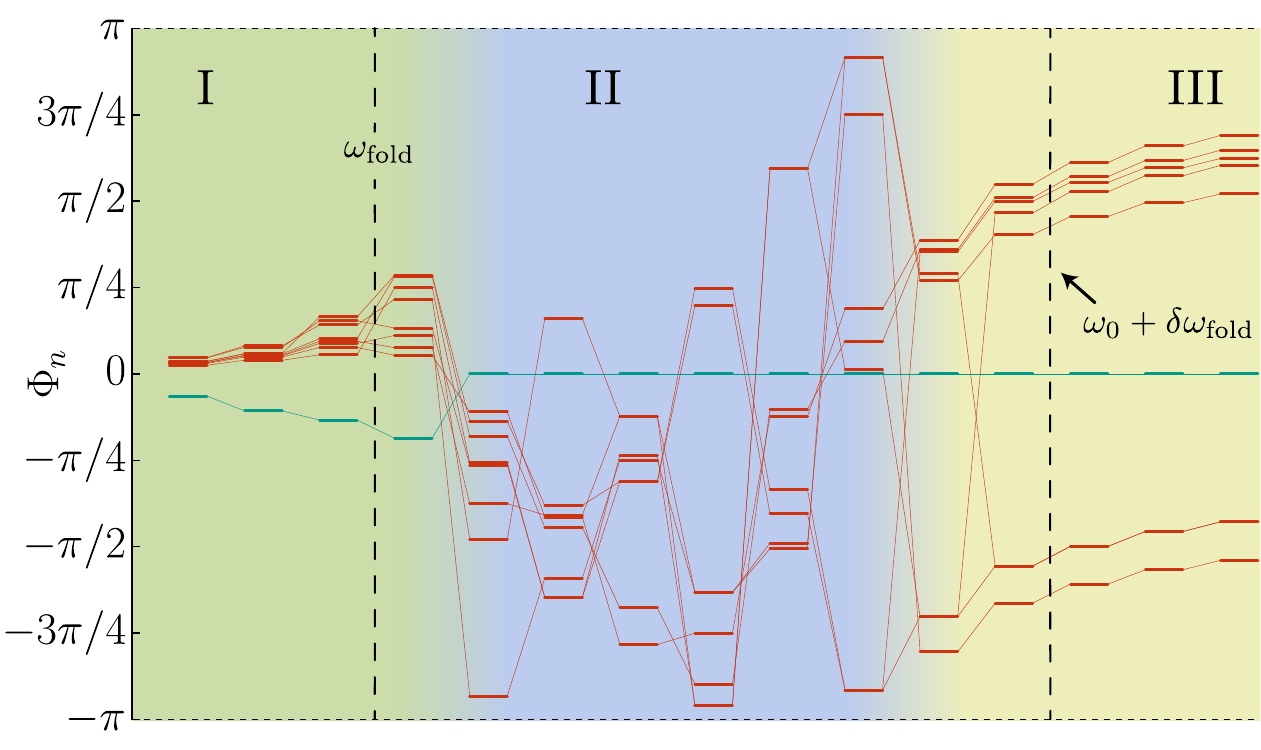}
           \caption{Floquet quasi-energy spectra in units of $\omega/2$ (top), average-energy spectra (middle) and geometrical Berry phase (bottom) on a $4\times 4$ cylinder. 
           Frequency is decreased from the finite size high-frequency regime I (left side) to the ``folded'' regime II (delimited by dashed vertical lines) and, finally, to the (detuned) Swap model III (right side). 
           Note that the top of the average-energy spectrum in the high-frequency regime $\omega > 1$ (left side) no longer corresponds to the ferromagnetic state. The significance of the yellow and blue regions is given in the text.
           }
	\label{fig:FloquetInter}
\end{figure} 

We have considered several frequencies interpolating between the high-frequency limit and the Swap model ($\omega/J=1/2$) and computed the corresponding many-body Floquet spectra on a $4\times 4$ cylinder
as shown in figure~\ref{fig:FloquetInter}. Let us first focus on the large frequency regime for e.g. $\omega\gtrsim 2$. As discussed above the large-$\omega$ spectrum agrees very well with the one of the sinusoidal driving considered in~\cite{Mambrini2024,Poilblanc2024}, once each of the spectra is normalized by its associated  Floquet energy scale. In fact, for both types of driving sequence, the effective {\it local} Floquet Hamiltonian (\ref{eq:fullFloquet}) can well describe the physics of the dynamical CSL phase, both on torus or cylinder geometries. Note that, in this dynamical CSL regime, a chiral branch should be observable in the edge dynamical structure factor at low energy within the bulk gap (with a non-universal velocity). However, although the correlation length should be relatively small below one lattice spacing (see Refs.~\cite{Poilblanc2017b,Hasik2022} for the static effective $\omega=\infty$ model), the two edges of the cylinder are still significantly coupled, preventing any clear signature of a chiral branch. In this high-frequency limit, the average-energy spectrum is given by (\ref{eq:AverEner2}) and its (many-body) bandwidth scales as $J_F\sim J^2/\omega$ like the Floquet spectrum. Nevertheless, the two spectra show some clear differences, e.g. some reordering of levels.

When the frequency is decreased from the high-frequency regime, the many-body bandwidth broadens as $N/\omega$ and eventually occupies the $\omega$-wide full quasi-energy FBZ at a characteristic cross-over frequency $\omega_{\rm fold}(N)\propto \sqrt{N}J $, typically $\omega\sim 1.5 J$ in our case for $N=16$. This behavior is exactly the same as the one encountered previously for a purely sinusoidal drive. Interestingly, $\omega_{\rm fold}$ corresponds also to the characteristic frequency for which some of the Berry phases $\Phi_n$ will reach $\pm\pi$. For $\omega\lesssim\omega_{\rm fold}(N)$, folding of the Floquet spectrum induces more and more resonances. 
Upon further reducing the frequency, when one gets closer to the fine-tuned Swap model, many-body sub-bands characteristic of the anomalous CSL suddenly appear in the Floquet spectrum as shown in figure~\ref{fig:FloquetInter} (top panel).

Our data suggest that a direct transition between the two phases does not exist. Indeed, the average-energy spectrum and the geometrical Berry phase in the two lowest plots of figure~\ref{fig:FloquetInter} clearly show an erratic behaviour in an intermediate frequency interval (represented by the blue color in figure~\ref{fig:FloquetInter})
attributed to the proliferation of resonances. In this regime the system may heat up (to infinite temperature). Nonetheless, prethermal regimes may still exist in the vicinity of the (finite-size) high-frequency CSL and anomalous CSL, as marked by the yellow regions extending within the ``folded" region II of Fig.~\ref{fig:FloquetInter}.

\section{Conclusion and outlook}

The models studied here involve a periodic driving sequence of four discrete steps acting on the quantum spin-1/2 of a square lattice.  When tuned at a precise frequency such models realize a swap circuit and can emulate anomalous CSLs. Constructing an (infinite discrete) family of Swap models with specific anisotropic XXZ couplings on nearest-neighbor bonds, we have studied the effect of frequency detuning on the Floquet quasi-energy spectrum. Folding of the Floquet spectrum occurs beyond a small typical detuning $\delta\omega_{\rm fold}(N) \sim J / N$ (which vanishes in the thermodynamic limit). Using the average-energy procedure to unfold the Floquet spectrum, we have found an intermediate (narrow) frequency regime $\delta\omega_{\rm fold} (N) <  |\delta\omega| \lesssim \delta\omega_\times$ where the folding of the Floquet spectrum does not induce visible resonances in the average-energy spectrum and edge modes are revealed by spectroscopic means. 
The investigation of the Floquet topological invariants (winding numbers) obtained via quantized response functions~\cite{gavensky2025} seems also to indicate a relative stability of the anomalous CSL under detuning. 
We speculate that this intermediate regime may correspond to a prethermalisation regime~\cite{Abanin2015,Luitz2020} which could survive for large systems. 
In such a scenario, crossings of Floquet quasi-energy levels (due to folding) involve very small matrix elements so that resonances remain very weak. This would lead to very long time-scales for thermalization to set up. In contrast, for detuning above $\delta\omega_\times$, an increasing density of resonances in both the Floquet and the average-energy spectra suggests the onset of heating (to infinite temperature). 

In the high-frequency limit we recover the same physical behavior as seen previously in the case of a sinusoidal driving, emulating in that limit a different type of CSL, associated to the low-energy physics of an effective {\it static and local} Floquet Hamiltonian. This is contrary to the anomalous CSL regime where the edge modes cannot be described by any local static Hamiltonian. 
In all, the above investigations consistently point towards the absence of a direct transition between the two CSL phases, likely separated by an intermediate frequency regime where heating takes place. This is to be contrasted to the case of the exactly solvable honeycomb Kitaev model where a continuous transition is expected between the high-frequency (non-Abelian) CSL and the anomalous CSL upon decreasing the frequency~\cite{Fulga2019}. This model can be mapped to a non-interacting problem equivalent to the Kitagawa model~\cite{Kitagawa2010}, where this transition is well established.
In other words, numerous extra conservation laws lead in this case to vanishing matrix elements at the crossings between Floquet (many-body) eigenvalues and, hence, no resonances. In the more generic case of our driven system, the two phases are believed to be separated by a region where the system absorbs energy until it reaches “infinite temperature”.  However, the time scale of heating may strongly depend on the detuning, with the possibility of a prethermal regime~\cite{Luitz2020} in the vicinity of the fine-tuned point, as it has been predicted for high-frequency driven systems~\cite{Abanin2015}. 

Interestingly, the model, the probes and physical regimes explored in this work can be accessed on existing experimental quantum-engineered platforms. 
As already suggested in \cite{Mambrini2024}, Heisenberg interactions can be realized with cold atoms loaded on an optical
lattice, for which a microscopic Hubbard-like description applies. The excellent
isolation of the optical lattice from its environment enables to realize ideal unitary dynamics.
Assuming one atom per site (on average) and a large on-site repulsion to be in the Mott
insulating phase, hopping terms can be turned on sequentially 
on the four types of NN bonds a, b, c, d of
Fig.~\ref{fig:model} by (abruptly) changing the tunneling barriers of the optical lattice appropriately. Note that (strongly correlated) two-component bosonic atoms~\cite{Altman2003} are more appropriate than fermionic atoms to reach low temperatures and 
realize both antiferromagnetic and ferromagnetic couplings. Finally, we note that Rydberg atom arrays can also emulate
XXZ quantum spin systems~\cite{Browaeys2020}. On such setups edge spectroscopy can be implemented~\cite{Goldman2012,Binanti2024} to investigate edge physics and detect chiral edge modes. The Streda response which gives access to (Floquet) topological invariants~\cite{gavensky2025,gavensky2025b} could be measured experimentally~\cite{leonard2023realization}.

On a more theoretical side it will also be interesting to investigate whether the low average-energy states are weakly entangled states, as in the case of driven spin chains~\cite{Schindler2025}, which would open the door to 2D tensor network studies using e.g. Projected Entangled Pair States ansatze~\cite{Poilblanc2024}, even in the thermodynamic limit. 

Finally, it has been claimed that disorder is a key ingredient to stabilize intrinsically dynamical anomalous CSL phases via many-body localisation of the bulk~\cite{Po2016,Fulga2019}. Further investigation on the role of disorder could take advantage of the average-energy concept to address low-entangled states in the presence of disorder.

{\it Acknowledgements --}
This work
was granted access to the HPC resources of CALMIP
center (Toulouse, France) under the allocation 2017-P1231. D.P. thanks the Kavli Institute for Theoretical Physics (KITP) at the University of California Santa Barbara (USA) for hospitality, where discussions with Dima Abanin, Leon Balents, Vedika Khemani, Roderich Moessner, Zlatko Papic, and Maksym Serbyn are acknowledged.  D.P. is particular indebted to Paul Schindler and  Mar{\' i}n Bukov for insightful comments and suggestions during the final stage of the work at KITP, as well as for a careful reading of the manuscript. N.G. acknowledges support from the ERC project LATIS.

\appendix

\section{Detuning the Ising parameter $\Delta$}
\label{app:Delta_detuning}

In this appendix we address the question of detuning the model by varying $\Delta$ for a fixed value of $\omega=\omega_s$.
Starting from a fined-tuned point $(\omega_s,\Delta_s)$ (see Eq.~(\ref{eq:Deltas}) and (\ref{eq:Omegas})) and detuning it to $(\omega_s,\Delta_s+\delta)$ we note that the 2-site gate given by equation (\ref{eq:UnitaryGateMatrix}) takes a remarkably simple form:
\begin{equation}
u_l (\omega_s,\Delta_s+\delta) =  \kappa
\begin{pmatrix}
  \mu^* & 0 & 0 & 0 \\
 0 & 0 & \mu & 0 \\
 0 & \mu & 0 & 0 \\
 0 & 0 & 0 & \mu^* \\
\end{pmatrix}_{\!\!\!l},
\label{eq:UnitaryGateMatrixDelta}
\end{equation}
with $\kappa=(-1)^q e^{-i\frac{\pi}{4}} e^{\frac{i \pi  p}{2}}$ and $\mu=e^{\frac{1}{4} i \pi  \delta  (2 p+1)}$. 

This operator shares an important property with Eq.~(\ref{eq:Swap0}): its only effect is to swap the two sites of bond $l$. But, contrary to the pure swap gate case (\ref{eq:Swap0}), it introduces a phase $\kappa \mu^*$ on parallel spin configurations and $\kappa \mu$,on anti-parallel spin configurations. As a direct consequence, any spin configuration is an eigenstate of $U_F$ in the bulk. As in the case of a swap gate, every spin returns to its initial position after 4 steps. Noting that $\kappa^4=-1$, we remark that $U_F$ is a diagonal operator only depending on $\mu$.

\begin{figure}
	\centering
    \includegraphics[width=0.95\columnwidth]{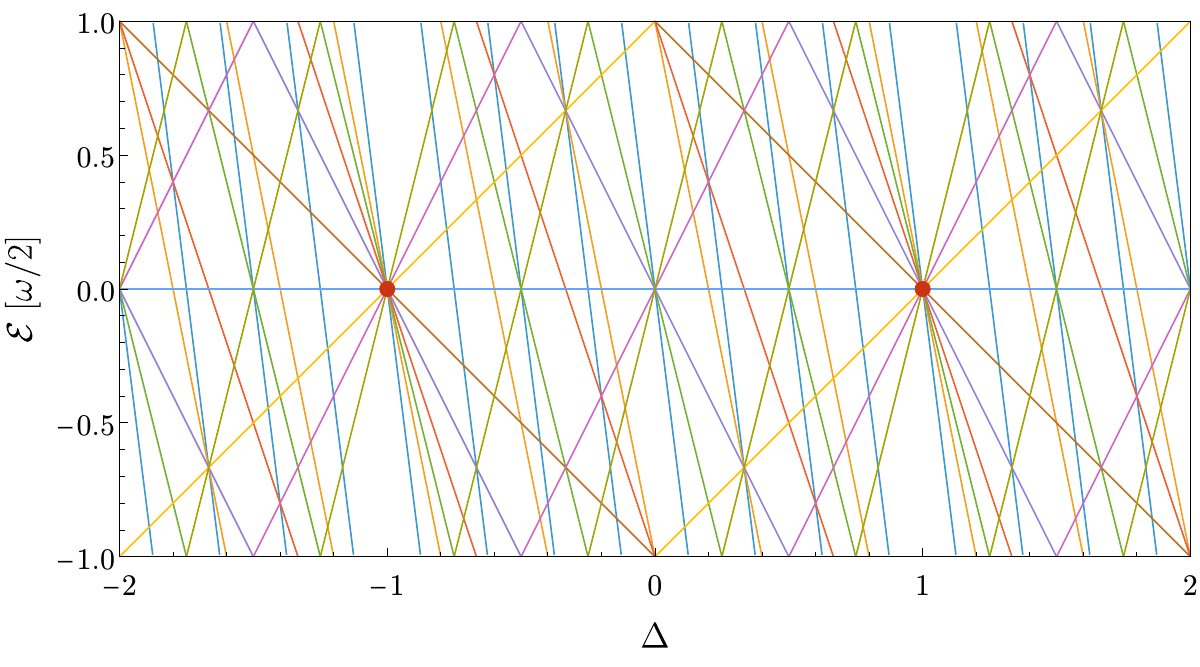}
     \caption{Quasi-energy spectrum as a function of $\Delta$ for $\omega=1/2$ on a $4\times4$ torus. Fined-tuned points (corresponding to odd integer values of $\Delta$) are represented as red circles. The $\Delta$-periodicity of the spectrum is 2.}
	\label{fig:deltadetuning}
\end{figure}

This shows that the quasi energy spectrum $\cal E$ consists in (degenerate) branches linear in $\Delta$ as numerically confirmed (see figure \ref{fig:deltadetuning}) on a $4\times4$ torus for $\omega=1/2$. As expected all branches collapse at fined-tuned points (odd integer values of $\Delta$).

For open geometries, like a cylinder, we expect a diagonal contribution from the bulk (linear in $\Delta$) and a boundary dynamics similar to Eq.~(\ref{eq:UF_omegas}) but with different velocities for each branch of states.

\section{Symmetry properties of the Floquet spectra}
\label{app:sym}

Each many-body Floquet spectrum on the cylinder (or on the ribbon) can be labeled by a momentum $k$ (defined mod $\pi$ because of the doubling of the unit cell). We show here that the many-body spectra at momenta $k$ and $-k$ are identical. We first note that the combination $R_x\Theta$ of time reversal with the reflection symmetry $R_x$ exchanging the two edges (oriented e.g. along x) is a symmetry of $H_F$ i.e. $[R_x\Theta, H_F]=0$. Then, if $|\phi_k\rangle$ is an eigenstate of energy $E_k$, $|\tilde{\phi}_{-k}\rangle=R_x\Theta|\phi_k\rangle$ is also an eigenstate with the same energy $E_k$ but with momentum $-k$, hence establishing the equivalence of the two spectra. 

On the torus the momentum $\bf k$ has two components and is defined mod $(\pi,\pi)$. If $\bf k$ is invariant with respect to one of the lattice reflection symmetry, the same conclusion applies i.e. the spectra at $\bf k$ and $-\bf k$ are degenerate. In that case, this property is not a generic property of the spectra at arbitrary $\bf k$ and $-\bf k$ points.

\section{Calculation of the average-energy spectrum and detuning expansion of the Floquet Hamiltonian}
\label{app:aeSpectrumAndHfExpansion}

In this appendix, we consider two seemingly unrelated calculations : the average-energy spectrum and the expansion of the Floquet Hamiltonian obtained by detuning the frequency around a swap point as defined in Eqs~(\ref{eq:Deltas}) and (\ref{eq:Omegas}). Indeed, it appears that the particular choice of piecewise constant $H_{\text{drive}} (t)$ (cf. figure\ref{fig:model}) leads to similar expressions involving the operator $H_{abcd}$ introduced in Eq.~(\ref{eq:Habcd}).

\subsection{Average-energy spectrum}
The average-energies are defined as
\begin{equation}
{\text{\ae}}_n=\frac{1}{T} \int_0^T  \langle\phi_n(t)|H(t)|\phi_n(t)\rangle\, dt \, . 
\label{eq:AverEnerApp}
\end{equation} 
Here $|\phi_n(t)\rangle$ are the micromotion Floquet eigenstates given by $U(t,0) |\phi_n\rangle$. 
Note that $|\phi_n(t)\rangle$ are also, up to a phase irrelevant in Eq.~(\ref{eq:AverEnerApp}), the eigenstates of the unitary evolution operator $U(t+T,t)$.

In the general case of an arbitrary drive Hamiltonian, evaluating (\ref{eq:AverEnerApp}) would require first to diagonalize $U(T,0)$ to obtain all eigenvectors $\{|\phi_n\rangle\}$ at $t=0$ and then compute the actual time evolution of the Floquet states $|\phi_n(t)\rangle$. Numerically the method consists in splitting the time interval $[0,T]$ into $N_\tau$ steps $[t_p,t_{p+1}]$, where $t_p=p\tau$, $p=0,\cdots,N_\tau -1$, with $\tau=\frac{T}{N_\tau}$ where $N_\tau$ is a multiple of $4$ in order to have an {\it integer} number $N/4$ of $\tau$ steps in each of the four $T/4$ time intervals. Next, elementary unitaries are applied recursively  on all the eigenvectors, starting from $t_0=0$,
\begin{equation}
    |\phi_n(t_{p+1})\rangle= \delta U_\alpha |\phi_n(t_p)\rangle\, ,
\end{equation}
where, within each of the four time intervals, the same elementary unitary $\delta U_\alpha=\exp{(-i(H_\alpha + H_0) \tau)}$, $\alpha=a,b,c,d$, is used at every step. Note that when $H_0=0$ the elementary unitaries $\delta U_\alpha$ can be split in the tensor product of two-site gates (see main text).

In our specific case another method, much simpler and more efficient numerically, can be used to compute Eq.~(\ref{eq:AverEnerApp}). It fully takes advantage of the fact that the drive $H(t)$ is {\it piecewise constant}. Hence $U(t,0)$ has only four expressions :
 \begin{equation*}
U(t,0) = \left\{
\renewcommand\arraystretch{1.3}
        \begin{array}{ll}
            e^{-i t H_a } & \quad {\text{if }} t \in [0,\frac{T}{4}[ \\
            e^{-i (t-\frac{T}{4}) H_b} U_a & \quad {\text{if }} t \in [\frac{T}{4},\frac{T}{2}[ \\
            e^{-i (t-T/2) H_c} U_b U_a & \quad {\text{if }} t \in [\frac{T}{2},\frac{3T}{4}[ \\
            e^{-i (t-3T/4) H_d} U_c U_b U_a & \quad {\text{if }} t \in [\frac{3T}{4},T[
        \end{array}
    \right. 
\end{equation*}
Note that if $H_0$ is finite it is then implicitly added to all $H_\alpha$ and to the expressions of all $U_\alpha$. 
Generically, we can then use the fact that $e^{-i t H_\alpha}$ commutes with $H_\alpha$ to get:
\begin{equation*}
\begin{split}
\langle\phi_n(t)&|H(t)|\phi_n(t)\rangle  \\ 
&=\left\{
\renewcommand\arraystretch{1.3}
        \begin{array}{ll}
            \langle\phi_n|H_a|\phi_n\rangle &  {\text{if }} t \in [0,\frac{T}{4}[ \\
            \langle\phi_n|U_a^\dagger H_b U_a |\phi_n\rangle &  {\text{if }} t \in [\frac{T}{4},\frac{T}{2}[ \\
            \langle\phi_n|U_a^\dagger U_b^\dagger H_c  U_b U_a|\phi_n\rangle &  {\text{if }} t \in [\frac{T}{2},\frac{3T}{4}[ \\
            \langle\phi_n|U_a^\dagger U_b^\dagger U_c^\dagger H_d U_c U_b U_a |\phi_n\rangle &  {\text{if }} t \in [\frac{3T}{4},T[
        \end{array}
    \right.
    \end{split}
\end{equation*}
Then ${\text{\ae}}_n$ takes the simple form
\begin{equation}
 \begin{split}
 {\text{\ae}}_n = \frac{1}{4}  \langle\phi_n|H_a
 &+ U_a^\dagger H_b U_a + U_a^\dagger U_b^\dagger H_c  U_b U_a  \\
 &+U_a^\dagger U_b^\dagger U_c^\dagger H_d U_c U_b U_a |\phi_n\rangle 
 \end{split}
 \end{equation}

 But since $U_F = U_d U_c U_b U_a$, this equation can be rewritten as
 
\begin{align}
\label{eq:aeH}
\nonumber
 {\text{\ae}}_n &= \frac{1}{4} \langle\phi_n|H_a + U_a^\dagger H_b U_a + U_F^\dagger U_d U_c H_c  U_c^\dagger U_d^\dagger U_F \\ \nonumber &+U_F^\dagger U_d H_d U_d^\dagger U_F |\phi_n\rangle \\
 &=\frac{1}{4}\langle\phi_n| H_a +  U_a^\dagger H_b U_a +
U_d H_c U_d^\dagger + H_d |\phi_n\rangle \nonumber \\ 
&=  \frac{1}{4}\langle  \phi_n| H_{abcd} |\phi_n\rangle
 \end{align}
 with
 \begin{equation}\label{eq:HabcdDef}
 H_{abcd}=  H_a +  U_a^\dagger H_b U_a +
U_d H_c U_d^\dagger + H_d .
 \end{equation}

 Note that expressions (\ref{eq:aeH}) and (\ref{eq:HabcdDef}) are independent of the geometry of the system (torus, cylinder or open boundary conditions, cf. figure\ref{fig:geometries}).

\subsection{Expansion near a swap point}
{\it Floquet Hamiltonian --} We now turn to the problem of computing the Floquet Hamiltonian in the vicinity $\omega=\omega_s+\delta \omega$ of a Swap model characterized by a pair $(\omega_s,\Delta_s)$ (see Eqs.~({\ref{eq:Deltas}) and (\ref{eq:Omegas})}). Up to the first order in $\delta \omega$, each term $U_{\alpha=a,\ldots,d}^{\Delta} (\frac{\pi}{2\omega})$ of Eq.~(\ref{eq:UnitaryGateProduct}) takes the form
\begin{align}
\label{eq:UfirstOrder}
U_\alpha &= U_\alpha^s \; e^{\frac{i \pi \delta \omega}{2 \omega_s^2} H_\alpha} \nonumber \\
&= U_\alpha^s \left ( \mathbb{1} -\delta \omega \frac{i \pi}{2 \omega_s^2} H_\alpha \right ),
\end{align}
where we use the shortcut notation $U_\alpha$ for $U_\alpha^{\Delta} (\frac{\pi}{2\omega})$ and $U_\alpha^s = U_\alpha^{\Delta_s} (\frac{\pi}{2\omega_s})$.

Using this expression, we can write the Floquet unitary $U_F = U_d U_c U_b U_a$ up to first order in $\delta \omega$ as
\begin{align}
\label{eq:UFfirstOrder}
U_F &= U_d^s U_c^s U_b^s U_a^s \nonumber \\
&-i \delta \omega \frac{\pi}{2 \omega_s^2}  \Bigl(
U_d^s H_d U_c^s U_b^s U_a^s + U_d^s U_c^s H_c U_b^s U_a^s \nonumber \\
&+U_d^s U_c^s U_b^s H_b U_a^s+U_d^s U_c^s U_b^s U_a^s H_a \Bigr) + {\cal O}(\delta \omega^2)
\end{align}

Using the identity $U^s_F = U_d^s U_c^s U_b^s U_a^s$, we obtain
\begin{align}
\label{eq:UFfirstOrder2}
U_F &= U^s_F -i \delta \omega \frac{\pi}{2 \omega_s^2}  \Bigl( U^s_F H_a+ H_d U^s_F \nonumber \\
&+ U^s_F {U_a^s}^\dagger H_b U_a^s + U_d^s H_c {U_d^s}^\dagger U^s_F \Bigr) + {\cal O}(\delta \omega^2)
\end{align}

Note that the above expressions are independent of the geometry of the system. To go further, as shown in the next sections, 
we have to treat separately the cases of the torus and the cylinder to obtain more explicit expressions for
 $H_F$.

{\it Average energy spectrum --} In contrast, on all geometries, the calculation of the ${\text{\AE}}$ spectrum defined by equation (\ref{eq:Aver2}) can be performed in first order using: 
 \begin{equation}
     H_{abcd}=H_{abcd}^s + {\cal O}(\delta\omega)\, ,
 \end{equation}
 with
 \begin{equation}
 H_{abcd}^s=  H_a +  {U_a^s}^\dagger H_b U_a^s +
U_d^s H_c {U_d^s}^\dagger + H_d.
\label{eq:Habcd_s}
 \end{equation}
 
 \subsection{Floquet Hamiltonian on the torus}

\begin{figure}
	\centering
    \includegraphics[width=0.9\columnwidth]{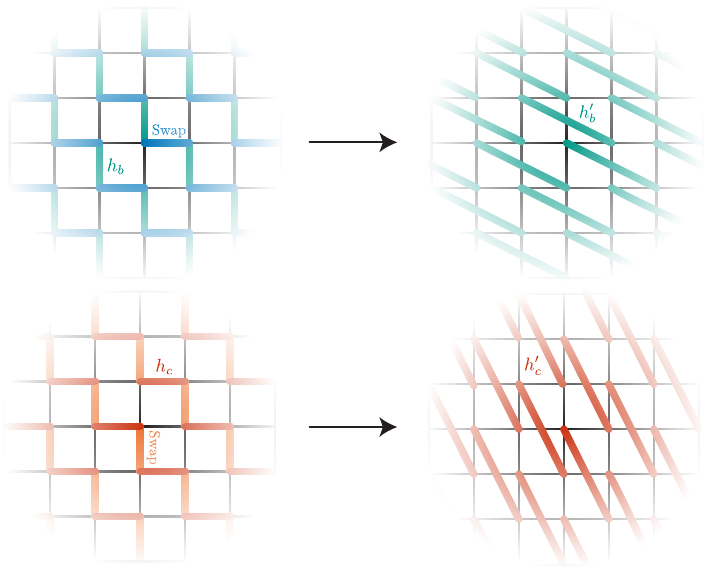}
     \caption{Range $\sqrt{5}$ interactions generated under unitary transforms of $H_b$ by ${U_a^s}$ (top row) and $H_c$ by ${U_d^s}$ (bottom row) on a torus or an infinite system.}
	\label{fig:longrange2}
\end{figure}

Let us consider a system with no boundary (either infinite or a finite cluster on a torus). 
In this case, using the identity $U_F^s = \mathbb{1}$, equation (\ref{eq:UFfirstOrder2}) takes the form
\begin{align}
\label{eq:UFfirstOrder3}
U_F &= \mathbb{1} -i \delta \omega \frac{\pi}{2 \omega_s^2}  \Bigl( H_a+ H_d \nonumber \\
&+ {U_a^s}^\dagger H_b U_a^s + U_d^s H_c {U_d^s}^\dagger  \Bigr) + {\cal O}(\delta \omega^2) \nonumber \\
&=e^{\frac{i \pi \delta \omega}{2 \omega_s^2} H_{abcd}^s} + {\cal O}(\delta \omega^2)\, .
\end{align}

A simple identification of $U_F = \exp(-2i\pi H_F / \omega)$ and (\ref{eq:UFfirstOrder3}) leads to
 \begin{equation}
 \label{eq:HFfirstOrder}
 H_F(\omega) = -\frac{\delta \omega}{4\omega_s} H_{abcd}^s + {\cal O}(\delta \omega^2)
 \end{equation}

Note that this identification is well defined on the torus since, for a small detuning, the spectrum of $U_F$ consists in a single narrow band around eigenvalue 1, which makes the correspondence between $H_F$ and $U_F$ unambiguous.

The action of $U_a^s$ on $H_b$ involved in the ${U_a^s}^\dagger H_b U_a$ term of (\ref{eq:UFfirstOrder3}) is nothing but reconfiguring the nearest neighbor $b$
bonds into length-$\sqrt{5}$ $b'$ bonds, as illustrated in the top row of figure~\ref{fig:longrange2}. The same mechanism occurs considering $U_d H_c {U_d^s}^\dagger$ terms (see figure~\ref{fig:longrange2}, bottom row) which reconfigure into .length-$\sqrt{5}$ $c'$ bonds.

As a result $H_{abcd}^s$ takes the simple explicit form on a torus
\begin{equation}
    H_{abcd}^s = J {\hat h}_{\sqrt{5}},
    \label{eq:sqrt5}
\end{equation}
where ${\hat h}_{\sqrt{5}}$ is the dimensionless XXZ Hamiltonian 
\begin{equation}
\label{eq:hsqrt5}
    {\hat h}_{\sqrt{5}} = \sum_{\substack{(i, j) \in a,d\\(i, j) \in b',c'}}  [ (S_i^x S_j^x +S_i^y S_j^y)
+ \Delta_s S_i^z S_j^z ],
\end{equation}
represented on figure~\ref{fig:longrange1} (left panel).

The numerical agreement between the exact Floquet spectrum and its first order approximation using $-\delta \omega\, (J / 4 \omega_s) {\hat h}_{\sqrt{5}} $ on a $4\times4$ torus for $\omega=0.51$ is shown in figure~\ref{fig:longrange1} (right panel).

\subsection{Floquet and average-energy spectra in the presence of a boundary}
\label{app:FloquetAE}

\begin{figure}
	\centering
    \includegraphics[width=\columnwidth]{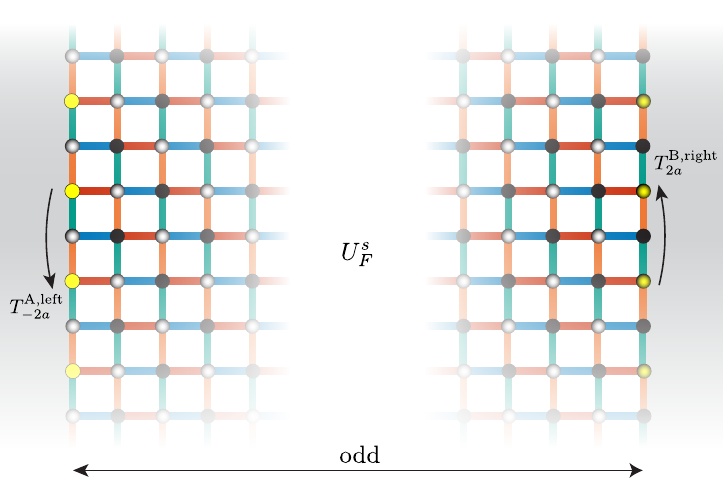}
     \caption{Floquet unitary $U_F^s$ action at fine tuning for an even-length horizontal cylinder (respectively an  infinite vertical strip) along its axis direction. Open boundaries are represented as grey regions. The system is periodic (resp. infinite) in the other direction. ``Static'' sites are represented as (solid or shaded) black dots and ``active'' sites by (solid or shaded) yellow dots. Solid dots (black or yellow) are used for A-sublattice while shaded dots (black or yellow) are used for B-sublattice).}
	\label{fig:Ufs}
\end{figure}

On a cylinder, $U_F^s$ no longer reduces to $\mathbb{1}$ but is given by (see Section \ref{sec:QuasiEnergies}) :
\begin{equation}
\label{eq:NNUFloquetCylinderFineTuned}
 U_F^s= T_{-2a}^{\rm A, left}\otimes\mathbb{1}_{\rm bulk} \otimes  T_{2a}^{\rm B, right}
 \end{equation}
which now involves, as shown if figure~\ref{fig:Ufs},  the translations by two lattice spacings of half the sites of each boundaries (belonging to opposite sublattices).

\begin{figure}
 \includegraphics[width=0.95\columnwidth]{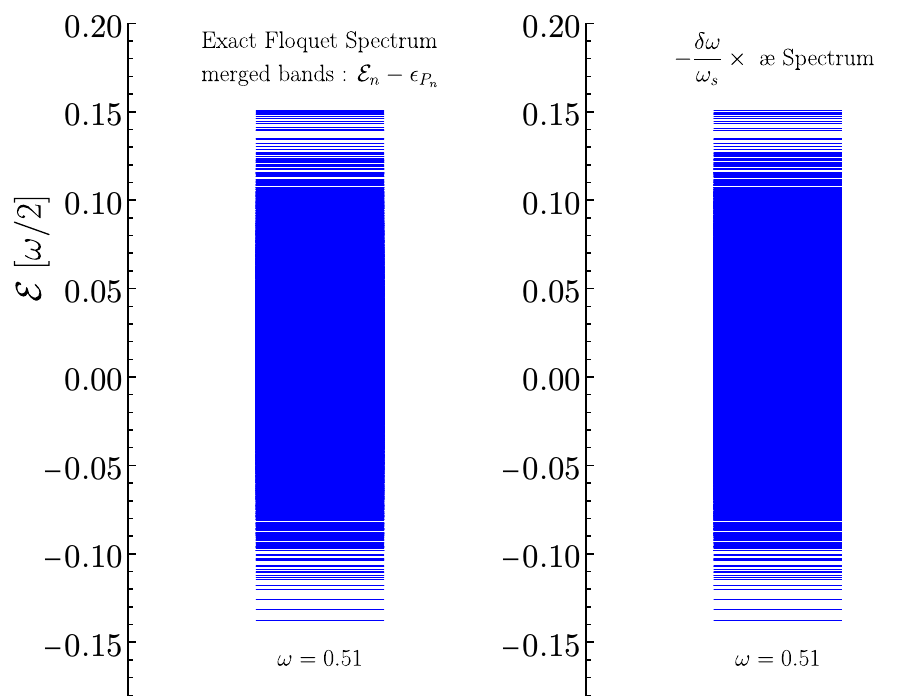}
 \caption{
 Comparison between the ``compressed'' Floquet spectrum (left) and the rescaled average-energy spectrum (right) on a $4\times 4$ cylinder at $\omega=0.51$.
 The $P=1$ sub-band of the Floquet spectrum has been shifted down by $-\omega/2$, overlapping with the $P=0$ sub-band.
Energy is displayed in units of $\omega/2$.  }
	\label{fig:CompCyl}
\end{figure}

Eigenstates $| \phi^s_{P}\rangle$ of $U_F^s$ are exactly known (see Section~\ref{sec:QuasiEnergies}):
 \begin{equation}
     | \phi^s_{P}\rangle =  |p_L\rangle\otimes\mathbb{1}_{\rm bulk} \otimes |p_R\rangle,   
     \end{equation}
with $P=\{ p_L, p_R \}$ and
 \begin{equation}
     U_F^s | \phi^s_{P}\rangle =  e^{-i \frac{2\pi}{\omega_s} {\cal E}^s_P}| \phi^s_{P}\rangle,   
     \end{equation}
where ${\cal E}^s_P=p_L+p_R$ mod(${\cal L}_{\rm geo}$).

We denote $| \phi^\omega_{P}\rangle$ the eigenstates of $U_F (\omega)$ such as
\begin{equation}
U_F (\omega) | \phi^\omega_{P}\rangle =  e^{-i \frac{2\pi}{\omega} {\cal E}^\omega_P}| \phi^\omega_{P}\rangle\, ,
\end{equation}
where, for clarity, we have omitted the extra index in $| \phi^\omega_{P}\rangle$ labeling the different states in the $P$ sub-band.
In the following we establish an interesting general relation between ${\cal E}^s_P$, ${\cal E}^\omega_P$ and ${\text{\ae}}_P$ up to the first order in $\delta \omega = \omega-\omega_s$.

Let us consider the expansion in $\delta\omega$ of Floquet eigenstates around $\omega_s$ :
\begin{equation}
\label{eq:phiexpand}
| \phi^\omega_{P} \rangle = | \phi^s_{P} \rangle  + \delta\omega | \phi^{(1)}_{P} \rangle + {\cal O} (\delta \omega^2).
\end{equation}

By convention both $| \phi^s_{P} \rangle$ and $| \phi^\omega_{P} \rangle$ are normalized, hence $\langle \phi^s_{P} \vert \phi^{(1)}_{P}\rangle=0$ and $\langle \phi^\omega_{P} \vert \phi^{(1)}_{P}\rangle={\cal O}(\delta \omega)$. We now rewrite Eq.~({\ref{eq:UFfirstOrder2}}) as
\begin{align}
\label{eq:UFfirstOrder4}
{U^s_F}^\dagger  U_F &= \mathbb{1} -i \delta \omega \frac{\pi}{2 \omega_s^2}  \Bigl( H_a+ {U^s_F}^\dagger  H_d U^s_F + {U_a^s}^\dagger H_b U_a^s \nonumber \\
& + {U^s_F}^\dagger U_d^s H_c {U_d^s}^\dagger U^s_F \Bigr) + {\cal O}(\delta \omega^2)
\end{align}
and take its expectation value in $| \phi^s_{P} \rangle$. We obtain
\begin{equation*}
    e^{i \frac{2\pi}{\omega_s} {\cal E}^s_P} \langle \phi^s_{P}  | U_F | \phi^s_{P} \rangle= 1-i \delta \omega \frac{\pi}{2 \omega_s^2} \langle \phi^s_{P}  | H_{abcd}^s | \phi^s_{P} \rangle  +{\cal O}(\delta \omega^2).
\end{equation*}
But (\ref{eq:phiexpand}) implies
\begin{align}
\langle \phi^s_{P}  | H_{abcd}^s | \phi^s_{P} \rangle&=\langle \phi^\omega_{P}  | H_{abcd}^s | \phi^\omega_{P} \rangle + {\cal O} (\delta \omega), \nonumber \\
\langle \phi^s_{P}  | U_F | \phi^s_{P} \rangle&=\langle \phi^\omega_{P}  | U_F | \phi^\omega_{P} \rangle + {\cal O} (\delta \omega). \nonumber
\end{align}
As a consequence,
\begin{equation}
    e^{-2i\pi\Bigl(\frac{{\cal E}^s_P}{\omega_s} -\frac{{\cal E}^\omega_P}{\omega} \Bigr)} = 1-i \delta \omega \frac{2\pi}{\omega_s^2}{\text{\ae}}_P  +{\cal O}(\delta \omega^2).
\end{equation}
where ${\text{\ae}}_P=\frac{1}{4}\langle \phi^\omega_{P}  | H_{abcd}^s | \phi^\omega_{P}\rangle$.
Hence, for Floquet eigenstates $|\phi_n\big>$ belonging to the $P_n=P(|\phi_n\rangle)$ sub-band one gets:
\begin{equation}
    {\cal E}_n -\epsilon_{P_n} = -\frac{\delta\omega}{\omega_s} {\text{{\text{\ae}}}}_n \, 
              \label{eq:FloquetAE}
\end{equation}
establishing an interesting relation between the (sub-band shifted) Floquet spectrum and the average-energy spectrum in lowest order in $\delta\omega$. A very good agreement has been found numerically for $\omega=0.51$, as shown in figure~\ref{fig:CompCyl}.

\section{Floquet spectrum on a ladder geometry}
\label{app:ladder}

\begin{figure}
 \includegraphics[width=0.85\columnwidth]{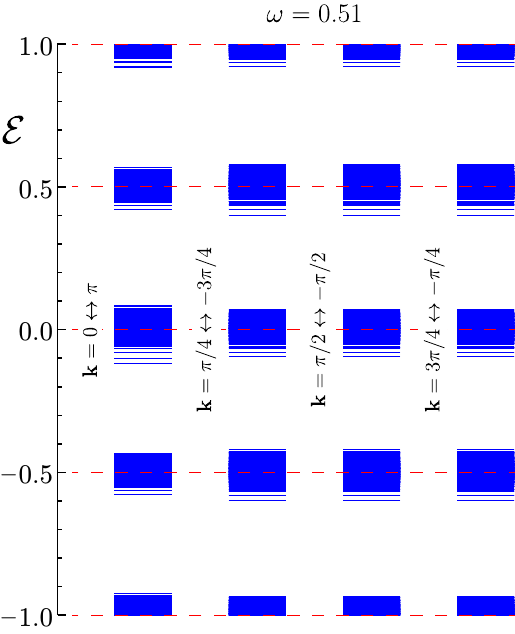}
 \caption{
 Floquet energy spectrum for 
 a $2\times 8$ periodic ribbon shown in the periodic FBZ $]-1,1]$ in units of $\omega/2$, at a detuned frequency $\omega=0.51$.
 Edge states at energies $0$, $\pm\omega/4$ and $\omega/2$ mod($\omega$) (shown as red dashed lines) are broaden into bands by the detuning. 
 The first-order Floquet Hamiltonian does not reproduce the sub-band nature of the Floquet spectrum. }
	\label{fig:FloquetSpectra2}
\end{figure}

We show here the Floquet energy spectrum of a $2\times 8$ periodic ribbon (ladder) at a small detuned frequency $\omega=0.51$. In that case, momenta $0$, $\pm\pi/4$ and $\pi/2$ mod($\pi$) are good quantum numbers. In the Floquet spectrum shown in figure~\ref{fig:FloquetSpectra2} one observes ${\cal L_{\rm cyl}}=4$ Floquet bands.

 \begin{figure*}
	\centering
\includegraphics[width=1.8\columnwidth]{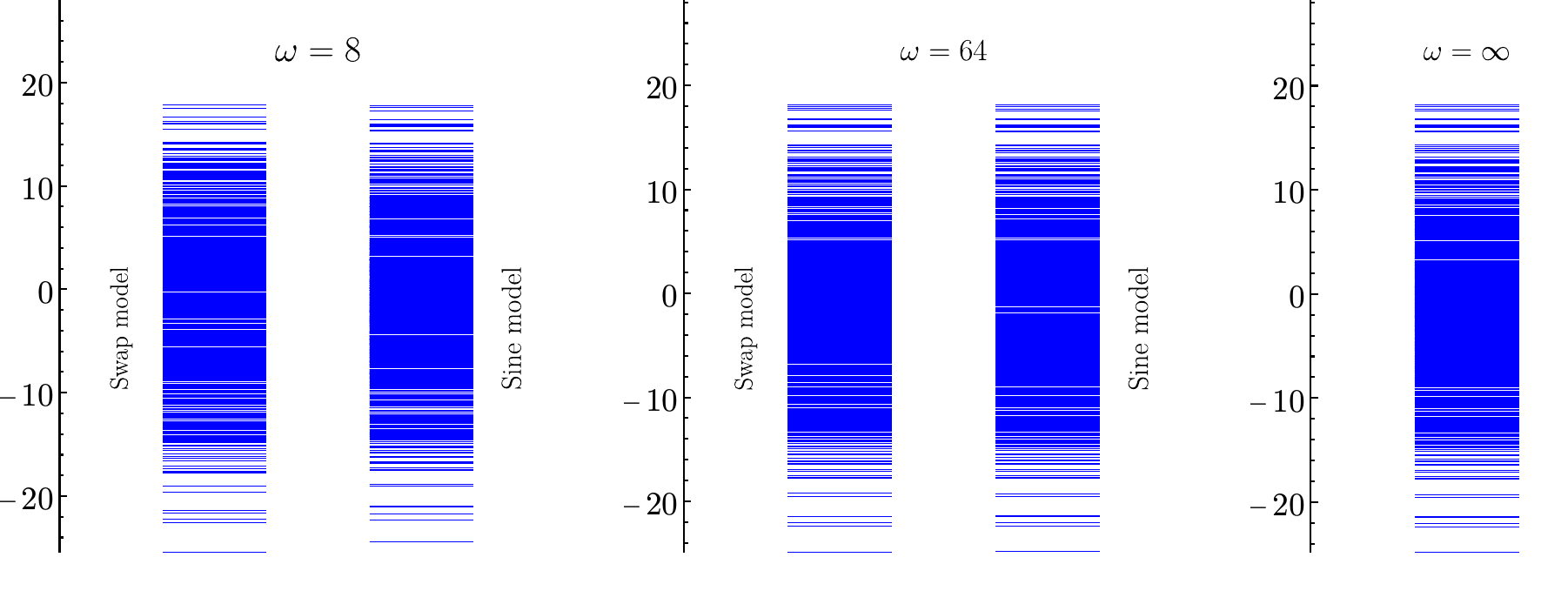}
 \caption{Comparison of Floquet quasi-energy spectra on a $4\times 4$ torus for two different types of time-periodic drives, a sinusoidal drive and a square drive, with $\omega=8$ (left panel) and $\omega=64$ (central panel). In the limit $\omega\rightarrow\infty$, the Floquet Hamiltonian is indeed given by equation~(\ref{eq:fullFloquet}) with $\lambda_\infty=2$ (right panel). Energy is in units of $J_F$ and the most symmetric sector (even total spin, $k=(0,0)-(\pi,\pi)$) is shown.}
	\label{fig:spectra_comp}
\end{figure*}

\section{Insertion of a magnetic flux: additional data}
\label{app:winding}

We shall give here additional results on a flux insertion, uniformly on a  $4\times 4$ cylinder. 

To compute ${\rm Tr} [H_F(\omega,\phi)]$, one can use the well-known identity ${\text{det} (e^A)} = e^{{\text{Tr}} (A)}$. Hence we first start by computing the determinant of $U_F(\omega,\phi)$:
\begin{align}
\label{eq:Det}
{\text{det}} \Bigl[ U_F(\omega,\phi) \Bigr] &= {\text{det}} \Bigl[ e^{-i \frac{2\pi}{\omega} H_F(\omega,\phi)}\Bigr] \nonumber \\
&= e^{-i \frac{2\pi}{\omega} {\text{Tr}}\Bigl[ H_F(\omega,\phi)\Bigr]}. 
\end{align}
As a consequence,
\begin{equation}
\label{eq:TrLogDet}
  \frac{1}{\omega}  {\text{Tr}}\Bigl[ H_F(\omega,\phi)\Bigr] = i \frac{1}{2\pi} 
  \ln\{ {\text{det}} \Bigl[ U_F(\omega,\phi) \Bigr] \} + k_\omega(\phi), 
\end{equation}
where $k_\omega (\phi) \in {\mathbb Z}$. It is important to note that the trace of $H_F$ is defined here as the sum of the quasi-energies {\it restrained to the first FBZ}, i.e. $\sum_n {\cal E}_n$, where all ${\cal E}_n\in ]-\omega/2,\omega/2]$. The choice of the first FBZ is equivalent in fact to the choice of the branch-cut of the log function. 

In our set-up $U_F=U_d U_c U_b U_a$, hence
\begin{equation}
\label{eq:DetProduct}
  \text{det} \Bigl[ U_F(\omega,\phi) \Bigr] = D_a(\omega) D_c(\omega) D_b(\omega,\varphi) D_d(\omega,\varphi)
\end{equation}
with $D_\alpha = {\text{det}} [ U_\alpha ]$ and $\varphi = 2\pi \phi / N$. Note that only  $b$ and $d$ terms (vertical bonds) contributions depend on $\varphi$.
The $D_\alpha(\omega)$ determinants can be evaluated easily since they involve the product of $N_{\rm bonds}\sim L_v L_h/2$ identical terms. Namely,
\begin{equation}
    D_{\alpha}(\omega) = \Bigl( {\text{det}} \; u_\alpha(\omega) \Bigr)^{N_{\rm bonds}},
\end{equation}
where $u_\alpha(\omega) =  e^{-\frac{i \pi}{2\omega} h_\alpha}$. It is easy to check that ${\rm Tr}\; h_\alpha=0$ i.e. ${\text{det}} \; u_\alpha(\omega) =1$ (even when $\varphi\ne 0$ on the $\alpha=b,d$ bonds) so that $\text{det} \Bigl[ U_F(\omega,\phi) \Bigr] =0$ and then,
\begin{equation}
    N_1(\omega,\phi) =    \frac{1}{\omega} {\text{Tr}}\Bigl[ H_F(\omega,\phi)\Bigr] = k_\omega (\phi)\, \in{\mathbb Z}\, . 
        \end{equation}
Note that the proof of quantization above is obtained by taking the trace over the full Hilbert space and is valid for all frequencies, even away from the fine-tuned values $\omega=\omega_s$. At fixed $S_z$, like for the unique spin-flip case $S_z=\pm (N/2-1)$ or the half-filled bosonic case $S_z=0$, the quantization of $N_1$ occurs only at fine-tuning $\omega=\omega_s$. However, $N_1(\phi)$ is always a staircase function of $\phi$ with integer steps, as seen in figure~\ref{fig:N1_vsPhi}.

\section{High-frequency limit}
\label{app:high}

First, let us provide here an alternative derivation of the high-frequency limit based on the decomposition of the drive Hamiltonian in terms of its Fourier harmonics
$H_{\rm drive}(t)=\sum_{n=0}^\infty H_n \exp{(in\omega t)} + h.c$ (see Appendix in Ref.~\cite{Goldman_Dalibard}).
In our case, each Fourier component can be separated in the four $H_{\cal B}$ Heisenberg terms operating on separate groups of bonds, 
\begin{equation}
H_n=c_n (H_a+i^n H_b + (-1)^n H_c + (-i)^n H_d)\, ,
\end{equation}
where $c_n=\frac{i}{2\pi n} (i^n - 1)$ is the Fourier component of the square function in the interval $[0,T/4]$.
Splitting the sum over $n$ in 4 groups, $n=4p$, $4p+1$, $4p+2$ and $4p+3$, $p\in \mathbb{N}$,
we obtain the 1st-order $1/\omega$ correction,
 \begin{eqnarray}
   H_{F}^{(1)}&=& \frac{1}{\omega} \sum_{n=1}^\infty\frac{1}{n} [H_n,H_{-n}]\nonumber\\
    &=& \frac{J^2}{\omega}\sum_{n=4p+1,4p+3}\!\! \frac{(-1)^{\frac{n-1}{2}}}{n} |c_n|^2 \,\, [h_x+ih_y,h_x-ih_y] \nonumber \\
    &=&  \frac{J^2}{\pi^2\omega}\,  \eta\, i [h_x,h_y] \nonumber\\
    \label{eq:FloquetHam}
    \end{eqnarray}    
    where we have introduced the static (staggered) Heisenberg couplings 
   $h_x=(H_a-H_c)/J$ and $h_y=(H_b-H_d)/J$ acting on the vertical and horizontal nearest-neighbor bonds, respectively, 
   and
   \begin{equation}
 \eta=\sum_{k=0}^\infty (-1)^k \frac{1}{(2k+1)^3}= \frac{\pi^3}{32}\, ,
   \end{equation}
in agreement with Eq.~(\ref{eq:H1}).

Secondly, we show in figure~\ref{fig:spectra_comp} Floquet spectra in the most symmetric sector, on a $4\times 4$ torus, for $\omega=8J$ and $\omega=64J$ with $\lambda_\omega=\lambda_\infty =2$, showing a perfect agreement with the spectrum of the effective Hamiltonian (\ref{eq:fullFloquet}) in the same sector. We have checked that a similar agreement also holds in all the other symmetry sectors.

\newpage

\bibliography{refs}

\end{document}